\newif\ifarxiv
\arxivtrue

\pdfoutput=1
\RequirePackage[T1]{fontenc}
\documentclass[12pt]{article}

\ifarxiv
\usepackage{float}
\usepackage{fullpage, multicol}
\usepackage{bm}
\usepackage{natbib}
\else
\usepackage[final]{template/neurips_2023}
\fi 

\usepackage{fancyhdr}
\newcommand{\runningtitle}{Quantum Variational Activation Functions Empower Kolmogorov-Arnold Networks}
\PassOptionsToPackage{round}{natbib}

\usepackage[height=8.85in,width=6.45in]{geometry}

\usepackage[utf8]{inputenc} %
\usepackage[T1]{fontenc}    %
\usepackage[pagebackref=true,breaklinks=true,colorlinks,hyperfootnotes=false]{hyperref}
\hypersetup{
  colorlinks,
  citecolor=citeblue,
  linkcolor=firebrick,
  urlcolor=firebrick
  }
\usepackage{url}            %
\usepackage{booktabs}       %
\usepackage{amsfonts}       %
\usepackage{nicefrac}       %
\usepackage{microtype}      %
\usepackage[table]{xcolor}
\usepackage{braket}

\usepackage{graphicx}
\usepackage{caption}
\usepackage{subcaption}
\usepackage{adjustbox}
\usepackage{rotating}
\usepackage{tikz}
\usetikzlibrary{positioning,calc,arrows.meta}

\usepackage[percent]{overpic}

\usepackage{longtable}   %

\usepackage{algorithm}
\usepackage{algorithmicx}
\usepackage{algpseudocode}
\usepackage{amsmath,amsthm,amssymb,bbm}
\usepackage{mathtools}
\numberwithin{equation}{section}
\usepackage{thmtools, thm-restate}

\DeclareFontFamily{OT1}{pzc}{}
\DeclareFontShape{OT1}{pzc}{m}{it}{<-> s * [1.10] pzcmi7t}{}
\DeclareMathAlphabet{\pzccal}{OT1}{pzc}{m}{it}

\usepackage{breakurl} %
\hypersetup{breaklinks=true}
\usepackage{enumitem}

\usepackage[nameinlink,capitalize,noabbrev]{cleveref}

\definecolor{lightyellow}{rgb}{1.0, 0.95, 0.7}
\definecolor{Blue}{rgb}{0, 0, 0.8}
\definecolor{blue}{rgb}{0,0,1}
\definecolor{mydarkblue}{rgb}{0,0.08,0.45}
\definecolor{mydarkblue2}{rgb}{0.133, 0.133, 0.698}
\definecolor{echodrk}{HTML}{0099cc}
\definecolor{mymauve}{rgb}{0.58,0,0.82}
\definecolor{darkgreen}{rgb}{0,0.40,0}
\definecolor{firebrick}{rgb}{0.698,0.133,0.133}
\definecolor{midnightblue}{rgb}{0.1,0.1,0.44}
\definecolor{citeblue}{RGB}{0, 113, 188}
\definecolor{oxfordblue}{rgb}{0.0,0.13,0.28}
\definecolor{prussianblue}{rgb}{0.0,0.19,0.33}
\definecolor{coolteal}{rgb}{0, 0.45, 0.45}
\definecolor{olive}{rgb}{0.1, 0.3, 0}
\definecolor{mypurple}{rgb}{0.5,0,0.5}
\definecolor{almond}{rgb}{0.94, 0.87, 0.8}

\definecolor{blue_ampEncoding}{HTML}{DAE8FC}
\definecolor{green_encoder}{HTML}{D5E8D4}
\definecolor{purple_decoder}{HTML}{E1D5E7}
\definecolor{yellow_measure}{HTML}{FFF2CC}
\definecolor{gray_block}{HTML}{F5F5F5}
\definecolor{pink_dru}{HTML}{FAD9D5}
\definecolor{orange_v}{HTML}{FAD7AC}
\definecolor{Lightblue}{HTML}{E7F4FC}

\DeclareDocumentCommand \norm { o m }{{\lVert #2 \rVert_#1}}

\newtheorem{theorem}{Theorem}[section]

\declaretheoremstyle[%
  spaceabove=10pt,%
  spacebelow=2pt,%
  headfont=\normalfont\itshape,%
  postheadspace=0em,%
  qed=%
]{prfstyle}

\theoremstyle{definition}
\newtheorem{remark}{Remark}[section]

\begin{document}

\begin{titlepage}
\vspace*{2.5em}
\begin{center}
{
\LARGE \rmfamily \bfseries
Quantum Variational Activation Functions Empower Kolmogorov-Arnold Networks
}
\vskip 1em
Jiun-Cheng Jiang$^{\dagger\ddagger}$\quad
Morris Yu-Chao Huang$^{\dagger\mathsection\diamond}$\quad
Tianlong Chen$^{\diamond}$\quad
Hsi-Sheng Goan$^{\dagger\ddagger\mathsection,*}$
\vspace{0.5em} %
{
\small
\begin{center}
\texttt{
\href{mailto:jcjiang@phys.ntu.edu.tw}{jcjiang@phys.ntu.edu.tw}, \{\href{mailto:morris@cs.unc.edu}{morris}, \href{mailto:tianlong@cs.unc.edu}{tianlong}\}@cs.unc.edu, \href{mailto:goan@phys.ntu.edu.tw}{goan@phys.ntu.edu.tw}
}
\end{center}
}

\begingroup
\renewcommand\thefootnote{}%
\footnotetext{%
\scriptsize
\begin{tabular}{@{}r p{0.90\linewidth}@{}}
$^\dagger$ & Department of Physics and Center for Theoretical Physics, National Taiwan University, Taipei 106319, Taiwan \\
$^\ddagger$ & Center for Quantum Science and Engineering, National Taiwan University, Taipei 106319, Taiwan \\
$^\mathsection$ & Physics Division, National Center for Theoretical Sciences, National Taiwan University, Taipei 106319, Taiwan \\
$^\diamond$ & Department of Computer Science, University of North Carolina at Chapel Hill, NC 27514, USA \\
$^*$ & Corresponding author. \\
\end{tabular}%
}
\endgroup

\end{center}

\begin{center}
\today
\end{center}
\noindent

\begin{abstract}
Variational quantum circuits (VQCs) are central to quantum machine learning, while recent progress in Kolmogorov-Arnold networks (KANs) highlights the power of learnable activation functions.
We unify these directions by introducing the quantum variational activation function (QVAF), a general framework in which parameterized quantum circuits serve as learnable activation functions; in this work we study an efficient single-qubit instantiation called DatA Re-Uploading ActivatioN (DARUAN).
We show that DARUAN with trainable data-preprocessing weights can realize an exponentially growing accessible frequency support with the number of re-uploading repetitions; for an explicit geometric choice of these weights, this gives a capacity-level exponential parameter reduction relative to independently parameterized Fourier activations.
Embedding DARUAN into KAN yields the quantum-inspired Kolmogorov-Arnold Network (QKAN), which retains the interpretability of the KAN architecture while improving parameter efficiency, expressivity, and generalization.
We further introduce layer extension and the hybrid QKAN (HQKAN) architecture to improve scalability and computational efficiency, enabling QKAN modules to act as compact replacements for multi-layer perceptrons (MLPs) in large-scale models.
We provide theoretical analysis and extensive experiments on function regression, image classification, and autoregressive generative language modeling, demonstrating the efficiency and scalability of QKANs.
Because the single-qubit circuits are efficiently simulable on classical quantum simulators, QKANs have quantum-inspired advantage in parameter efficiency and training stability; DARUANs and QKANs serve as present-day validation of the QVAF concept, and the trained DARUANs are directly executable and feasible on current noisy intermediate-scale quantum (NISQ) hardware for inference validation.
\begin{center}
\small Code available at: {\ttfamily\url{https://github.com/Jim137/qkan}}
\end{center}
\end{abstract}
\end{titlepage}

\tableofcontents
\thispagestyle{empty}
\clearpage

\section{Introduction}\label{sec:intro}

\begin{figure*}[t!]
  \centering
  \includegraphics[width=0.75\paperwidth]{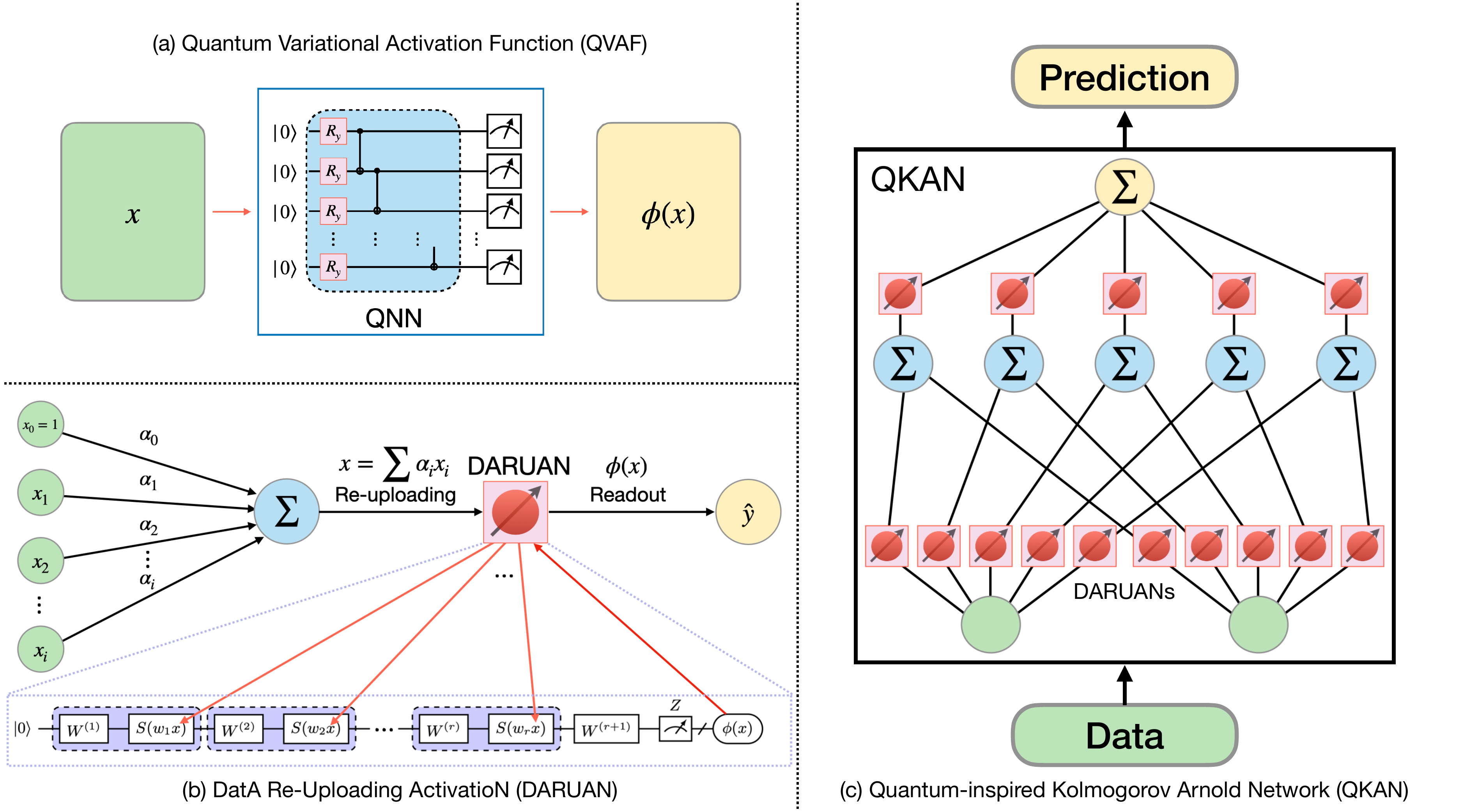}
  \caption{\textbf{Schematic overview of the proposed QVAF framework and QKAN architecture.}
    \textbf{(a)} The general concept of a quantum variational activation function (QVAF): a variational quantum circuit, including multi-qubit circuits that constitute QML models when entangling quantum resources are used, serves as a learnable activation function.
    \textbf{(b)} To instantiate an efficient and classically simulable QVAF, we adopt a single-qubit data re-uploading circuit (DARUAN) with trainable data pre-processing weights $w_\ell$ in the $\ell$-th encoding block $S(w_\ell x)$.
    The $\ell$-th parameterized unitary is $W^{(\ell)}=W^{(\ell)}(\bm{\theta}_\ell)$, where $\bm{\theta}_\ell$ is the set of trainable parameters in $W^{(\ell)}$.
    The aggregated input $x=\sum \alpha_ix_i$ is repeatedly uploaded into each data encoding block, and the measurement outcome of the parameterized circuit defines the activation function.
    \textbf{(c)} We further incorporate data re-uploading activations into the structure of Kolmogorov-Arnold networks (KANs), treating each edge's activation as the output of a distinct quantum circuit.
    Post-activation values are summed according to a predefined pattern to yield subsequent layer outputs, ultimately resulting in a quantum-inspired Kolmogorov-Arnold network (QKAN).
  }\label{fig:qkan}
\end{figure*}

Quantum computing (QC) and quantum machine learning (QML) represent rapidly evolving interdisciplinary research frontiers that leverage quantum mechanical principles to perform computation \citep{Biamonte_2017, PhysRevLett.117.130501, chen2025introductionquantummachinelearning}.
In QML, a central theme involves encoding classical data into high-dimensional Hilbert spaces using quantum states, harnessing the advantages of superposition, coherence, and entanglement to enable efficient learning from complex datasets \citep{Biamonte_2017, Ciliberto_2018, PhysRevLett.122.040504, phillipson2020quantum, Meyer_2023, liu2024quantumtrainrethinkinghybridquantumclassical, DEVADAS2025103318}.

Among the primary models in QML are variational quantum circuits (VQCs) which serve as quantum analogues of classical neural networks \citep{farhi2018classificationquantumneuralnetworks, 9144562}.
VQCs encode input data via structured ansatz and optimize parameterized gates using classical algorithms \citep{Schuld_2021, P_rez_Salinas_2020}.
VQCs are the foundation of the hybrid quantum-classical machine learning paradigms \citep{Mari_2020} and widely regarded as one of the most promising routes toward demonstrating quantum advantage in near-term applications \citep{Jerbi_2023}.

In parallel, advances in classical machine learning have spotlighted the use of learnable \textit{variational activation functions} (VAFs), particularly through the development of Kolmogorov-Arnold networks (KANs) \citep{liu2024kan}.
Inspired by the Kolmogorov-Arnold representation theorem (KART), KANs extend the concept of classical multilayer perceptrons (MLPs) by replacing fixed nonlinearities with learnable VAFs at each edge and performing summation at each node.
This approach has yielded improvements in both predictive accuracy and interpretability and has shown promising results across a broad range of tasks including computer vision \citep{li2024ukanmakesstrongbackbone, bodner2024convolutionalkolmogorovarnoldnetworks,
yang2025kolmogorovarnold} and time series forecasting \citep{vacarubio2024kolmogorovarnoldnetworkskanstime, xu2024kolmogorovarnoldnetworkstimeseries, genet2024tkantemporalkolmogorovarnoldnetworks}.
More researches have explored VAF implementations in KAN using Chebyshev polynomials \citep{ss2024chebyshevpolynomialbasedkolmogorovarnoldnetworks}, wavelets \citep{bozorgasl2024wavkanwaveletkolmogorovarnoldnetworks, seydi2024unveilingpowerwaveletswaveletbased}, Fourier series \citep{xu2024fourierkangcffourierkolmogorovarnoldnetwork}, radial basis functions \citep{li2024kolmogorovarnold}, and other function bases \citep{howard2024finitebasiskolmogorovarnoldnetworks, ta2024bsrbfkancombinationbsplinesradial, aghaei2024fkanfractionalkolmogorovarnoldnetworks, seydi2024exploringpotentialpolynomialbasis,yang2025kolmogorovarnold}.
Ref.~\cite{liu2024kan20kolmogorovarnoldnetworks} further enriched the expressivity of KANs by incorporating multiplicative interactions at the nodes.

Recent studies have established that VQCs can approximate any analytic function \citep{Mitarai_2018}, and under certain conditions, even arbitrary continuous functions \citep{Schuld_2021, yu2022powerlimitationssinglequbitnative, yu2024nonasymptoticapproximationerrorbounds}.
Nevertheless, much of the QML literature has focused on data encoding strategies rather than function approximation.
Notably, VQCs can provide compact parameterizations for learning single-variable functions $f(x)$ with enlarged accessible frequency spectra, motivating new approaches that leverage this parameter-efficient expressive potential \citep{Wach_2023, Schuld_2021}.

In this work, we propose to use VQCs not as standalone learners but as VAFs within classical or hybrid architectures, introducing a novel framework termed \textit{quantum variational activation functions} (QVAFs).
The QVAF framework is deliberately more general than the single-qubit model used in our experiments.
In principle, QVAFs may be instantiated by multi-qubit circuits with entangling gates, and such activations could become QML modules when future quantum hardware provides sufficiently accurate and scalable entangling operations.
At present, however, this route is limited by the NISQ hardware regime: although single-qubit gates have reached very high fidelities in leading trapped-ion and superconducting platforms~\citep{42w2-6ccy,PRXQuantum.5.040342}, two-qubit gates remain a major source of error for deep entangling circuits~\citep{Preskill2018quantumcomputingin,Singh_2024}.
Moreover, using a multi-qubit VQC for every KAN edge would require either many hardware executions or classically expensive multi-qubit simulations, and large multi-qubit VQCs may also suffer from barren-plateau trainability issues~\citep{mcclean2018barren}.

For this reason, we instantiate QVAF in this work using single-qubit data re-uploading circuits.
This single-qubit instantiation serves as a present-day validation platform for the QVAF idea: it allows us to test whether quantum-circuit-derived activation functions are effective, trainable, interpretable, and scalable under hardware and simulation constraints that are realistic today.
Because single-qubit data re-uploading circuits are classically efficiently simulable while still providing compact parameterizations of univariate functions~\citep{P_rez_Salinas_2020, Schuld_2021, yu2022powerlimitationssinglequbitnative}, they provide a practical route to evaluate the usefulness of QVAFs before large-scale high-fidelity multi-qubit quantum processors become available.
We name this single-qubit QVAF instantiation \textit{DatA Re-Uploading ActivatioN} (DARUAN).
The term ``daruan'' is derived from a Chinese traditional plucked string instrument renowned for its deep and coherent, mellow tone.

To validate the effectiveness of QVAFs in present-day computational settings, we further embed DARUAN into \textit{Quantum-inspired Kolmogorov-Arnold Networks} (QKANs), where DARUAN modules serve as quantum-inspired VAFs within the KAN framework.
This enables a controlled test of whether the QVAF principle can improve parameter efficiency, expressivity, and generalization using circuits that are already simulable on classical accelerators and executable on current quantum hardware for inference validation.
A schematic illustration of the QVAF, DARUAN, and QKAN architectures is provided in \cref{fig:qkan}, respectively.

We theoretically analyze the accessible frequency spectrum, Fourier approximation capacity, and parameter scaling of single-qubit data re-uploading circuits in QKANs with and without trainable data-preprocessing weights.
At the approximation-capacity level, we demonstrate that incorporating trainable weights into the data-preprocessing phase allows geometrically spaced frequency support to be reached with logarithmically many repetitions, yielding an exponential reduction in parameter size compared with a classical Fourier-series-based KAN that assigns independent coefficients up to the same maximum frequency.

We empirically validate our proposed models on a variety of tasks, including regression, classification, and generative modeling.
While QKANs and KANs show promising results, we notice that the number of parameters increases quadratically with the input and output dimensions, a challenge inherent to both architecture.
To mitigate this issue, we use Hybrid QKAN (HQKAN), a parameter-efficient architecture that adds classical compression and expansion layers around a QKAN latent core, thereby improving scalability.

To summarize the conceptual hierarchy in our paper: QVAF is a broad mathematical framework (\cref{eq:vap_map}) for using parameterized quantum circuits as learnable activation functions, and multi-qubit QVAFs with entangling gates can serve as QML models.
DARUAN is an efficiently simulable single-qubit instantiation of QVAF tailored to current hardware and simulation constraints.
QKAN embeds DARUAN as edge activations within the KAN topology, while HQKAN augments a QKAN latent core with two classical fully connected layers to handle high-dimensional inputs.
The resulting QKAN and HQKAN using single-qubit DARUAN can be understood as quantum-inspired learning frameworks.

QVAFs employing multiple entangled qubits can serve as QML models, yet their effectiveness in neural networks such as KAN remains inconclusive.
The single-qubit DARUAN used here is a deliberately practical instantiation of the broader QVAF framework: it validates quantum-circuit-derived activation design under present-day constraints, avoiding the two-qubit-gate fidelity bottlenecks of current NISQ devices and the unfavorable classical scaling of generic multi-qubit circuit simulation, while retaining a clear upgrade path to entangling multi-qubit QVAFs as hardware matures.
Because this instantiation is efficiently simulable, QKAN/HQKAN can be trained with standard distributed GPU infrastructure and scaled to large models, including large language models (LLMs).
Our successful demonstration of single-qubit DARUAN's efficiency and scalability in QKAN and HQKAN across different tasks suggests potential advantages for multi-qubit DARUAN implementations, where the introduction of more qubits and entanglement could increase success rates and reduce the number of layers required~\citep{Schuld_2021,P_rez_Salinas_2020}—though such benefits remain difficult to verify on current NISQ hardware or classical simulations.

\section{Results}\label{sec:res}

\subsection{The Quantum Variational Activation Function}

The concept of VAFs has recently gained attention in classical machine learning, where the activation functions within neural networks are no longer fixed but instead treated as trainable parameters $\bm{\theta} \in \mathbb{R}^d$.
\begin{align}\label{eq:vap_map}
   \phi_{\bm{\theta}} \colon \mathbb{R} \longrightarrow \mathbb{R}, \quad x \mapsto \phi_{\bm{\theta}}(x).
\end{align}
This approach enhances a network's expressivity by allowing it to learn the most suitable nonlinear transformations from data, leading to improvements in accuracy, convergence, and generalization performance~\citep{molina2019pade,Apicella_2021,liu2024kan}.
Parametric ReLU (PReLU)~\citep{7410480} and Swish~\citep{ramachandran2017searchingactivationfunctions} are two representative examples, where slope or gating parameters are learned during training.
More flexible formulations include adaptive piecewise linear (APL) units~\citep{agostinelli2015learningactivationfunctionsimprove}, kernel activation functions~\citep{SCARDAPANE201919}, and spline-based activations~\citep{liu2024kan}, all of which treat activation function as learnable entities.

Inspired by this classical paradigm, we propose a quantum analogue: the QVAF.
In this general framework, the role of an activation function is replaced with a variational quantum circuit. Multi-qubit QVAFs with entangling operations can be regarded as QML models when their circuit resources are used directly.
In general, a QVAF is defined as:
\begin{equation}
\phi_{\bm{\theta}}(x) = \langle \psi_0| U^\dagger(x; \boldsymbol{\theta}) \mathcal{M} U(x; \boldsymbol{\theta}) |\psi_0\rangle,
\end{equation}
where $|\psi_0\rangle$ is the initial state on the selected qubit register, the input $x \in \mathbb{R}$ is encoded via data re-uploading or angle encoding, $U(x; \boldsymbol{\theta})$ is a trainable unitary circuit, $\mathcal{M}$ is an Hermitian observable with norm $\lVert \mathcal{M} \rVert \le 1$, and the expectation value yields a bounded nonlinear function.

The expressive power of QVAFs stems from their ability to generate highly nonlinear, tunable transformations through quantum feature maps and trainable circuit structures~\citep{Schuld_2021, Mitarai_2018}.
Moreover, since these activations are inherently smooth and bounded, they are particularly well-suited for stable training.
Prior work has shown that VQCs are capable of approximating any analytic function~\citep{Mitarai_2018} and even arbitrary continuous functions under certain conditions~\citep{Schuld_2021}.
This positions QVAFs as universal approximators, analogous to classical VAFs but with parameterizations derived from quantum circuit structure.

Recent efforts such as variational quantum splines~\citep{10.1007/978-3-031-36030-5_14} and quantum-inspired activation circuits in hybrid convolutional networks~\citep{li2024quantuminspiredactivationfunctionsquantum} have demonstrated the potential of empirical viability of QVAFs on both synthetic and real-world data.

To implement this framework in a scalable form and facilitate integration into layered network architectures, we introduce DARUAN, a deliberately single-qubit instantiation of QVAF. DARUAN leverages the data re-uploading circuit framework~\citep{P_rez_Salinas_2020} to construct a quantum-inspired activation layer with multiple repetitions and trainable pre-processing weights, where the relevant structure is the SU(2) Lie algebra of rotations.
Each block consists of a data encoding alternated with trainable unitaries, forming a variational circuit capable of approximating smooth periodic and non-periodic functions.
The output is obtained via measurement of a Pauli observable (typically $\sigma_z$ in computational basis) and is used as the nonlinear transformation applied to the neuron output.
In \cref{fig:qkan}\textup{(b)}, DARUAN acts as VAF in a perceptron where the classical data is re-uploaded multiple times to a data re-uploading variational quantum circuit and readout the expectation value of the data re-uploading circuit as the final output.

Importantly, DARUAN supports architectural flexibility through a concept we term \textit{layer extension}, which progressively increases the number of re-uploading repetitions, where we discuss the details in \cref{sec:impl}.
In our latter part of the experiments, this design allows the model to scale its expressivity on demand while preserving previously learned features.
Layer extension addresses the practical challenge of optimizing large-repetition DARUANs by warm-starting from smaller circuits rather than training the full spectral bandwidth from scratch.

The simplicity of DARUAN, which relies solely on single-qubit circuits, makes it efficiently simulable on classical hardware, such as modern GPUs and multi-node high-performance computing (HPC) clusters.

At the same time, DARUAN circuits can be executed on quantum hardware without changing the trained parameters, which gives a direct inference-validation path rather than the main computational route. State-of-the-art trapped-ion platforms have achieved single-qubit gate error rates at the $10^{-7}$ level~\citep{42w2-6ccy}, while superconducting architectures have reached $10^{-5}$~\citep{PRXQuantum.5.040342}. For a simple-ansatz DARUAN with $r$ repetitions, the circuit depth is $3r$ single-qubit rotation gates; this empirically supports small-depth feasibility for $r \leq 30$, while statistical uncertainty from finite measurement shots and hardware errors remain practical constraints evaluated in \cref{sec:sup_real_device}.

Furthermore, the DARUAN instantiation benefits from a form of spectral compression: it can represent a large frequency support through a small set of coupled rotation parameters.
This suggests that single-qubit QVAF instantiations can serve not only as activation functions but also as compact structured feature extractors within quantum-inspired architectures, while the broader QVAF framework also accommodates quantum multi-qubit activations.

QVAFs open a promising path toward expressive, trainable nonlinearities in QML.
They provide both theoretical richness and practical flexibility, enabling efficient approximation capabilities in low-resource quantum settings.

\subsection{QKAN architecture}\label{sec:qkan_arch}

Building upon the insights from KANs \citep{liu2024kan} and the expressive power of DARUAN, we introduce the QKANs.
In this architecture, each activation function traditionally implemented via B-spline interpolation in KANs is replaced by DARUAN, a single-qubit data re-uploading variational quantum circuit, yielding a compact and trainable nonlinearity.

The central idea of QKANs is to harness the mathematical nature of Fourier-like expansion properties of data re-uploading quantum circuits, which approximate target functions via tunable superpositions of frequency components \citep{Schuld_2021}.
These circuits serve as QVAFs, enabling QKANs to learn highly expressive mappings using significantly fewer parameters than classical VAFs.
While KANs approximate activation functions through B-spline bases with grid size $G$, QKANs generate Fourier-like components through a compact set of coupled circuit parameters with a relatively small number of data re-uploading repetitions $r$.

Each QKAN layer is constructed from a collection of single-qubit DARUANs organized in a feedforward structure.
For a layer $l$ with $n_l$ input nodes and $n_{l+1}$ output nodes, the layer is defined as:
\begin{align}
  &\Phi_l = \{\phi_{l,j,i}\},~ i = 1,2,\ldots,n_{l},~j = 1,2,\ldots,n_{l+1}; \\
  &\phi_{l,j,i}(x_{l,i}) = \bra{0} U^\dagger(x_{l,i}, \bm{\theta}_{l,j,i}) \mathcal{M} U(x_{l,i}, \bm{\theta}_{l,j,i})\ket{0}; \\
  &x_{l+1,j} = \sum_{i=1}^{n_l} \phi_{l,j,i}(x_{l,i}),
\end{align}
where $i,j$ are indexes of input and output node respectively, $U(x; \bm{\theta})$ denotes the data re-uploading unitary circuit with trainable parameters $\bm{\theta}$, and $\mathcal{M}$ is the Pauli observable measured to obtain the circuit output.
The final model is obtained by composing these layers:
\begin{equation}
  \bm{y} = \text{QKAN}(\bm{x}) = (\Phi_{L}\circ\Phi_{L-1}\circ\cdots\circ\Phi_2\circ\Phi_1)(\bm{x}),
\end{equation}
where the output is bounded within $[-n_{L-1}, n_{L-1}]^{n_L}$ due to the nature of quantum expectation values.

QKANs offer both theoretical and practical advantages in terms of approximation capacity and parameter efficiency.
From a complexity perspective, the number of parameters required for a QKAN with depth $L$, width $N$, and repetition count $r$ scales as $\mathcal{O}(N^2 L r)$.
In contrast, KANs demand $\mathcal{O}(N^2 L G)$ parameters, where $G$ is the number of spline grid points.
As detailed in \cref{thm:qkan_linear}, trainable preprocessing weights can make the accessible Fourier frequencies grow exponentially with $r$, giving a capacity-level parameter advantage compared with classical grid-based and Fourier-based approaches.

Consequently, QKANs inherit the shallow architecture and structured interpretability of KANs while achieving parameter efficiency and enhanced expressivity through the quantum-inspired VAF DARUAN.
As such, they form a promising and scalable approach for compact, data-efficient, and interpretable quantum-inspired modeling.

\subsection{Theoretical Analysis of QKAN}\label{sec:theoretical_analysis}
We analyze the architectural design of QKAN and establish its parameter-efficiency advantages over classical KAN and Fourier-series-based KAN at the level of approximation capacity.
First, we present an approximation theory for KAN (\cref{thm:Approx_theory}) based on Fourier series, extending Theorem~2.1 of \citet{liu2024kan} from the case of B-spline basis functions.
We then investigate the accessible frequency spectrum and Fourier approximation capacity of single-qubit data re-uploading circuits within QKAN, as formalized in \cref{thm:qkan_linear}.

\begin{theorem}[Fourier-series approximation for KAN edge functions, adapted from Theorem~2.1 of \citet{liu2024kan}]
\label{thm:Approx_theory}
Let $\bm{x}=(x_1,\ldots,x_n)$ and suppose that $f$ admits a KAN representation
\begin{align}
f(\bm{x})=(\Phi_L\circ\Phi_{L-1}\circ\cdots\circ\Phi_1)(\bm{x}),
\end{align}
where each edge function $\phi_{l,j,i}$ is $(k+1)$-times continuously differentiable and, after the chosen input normalization, admits a $C^{k+1}$ periodic extension on the relevant compact interval. Then there exists a constant $C_f$ depending on $f$ and on the fixed representation, but not on the Fourier cutoff $K$, such that there are trigonometric-polynomial edge approximations $\phi^K_{l,j,i}$ with frequencies $|n|\le K$ satisfying, for any integer $m$ with $0\le m\le k$,
\begin{equation}
\begin{aligned}
\left\|
f-(\Phi^K_L\circ\Phi^K_{L-1}\circ\cdots\circ\Phi^K_1)(\bm{x})
\right\|_{C^m}
\le C_f K^{-(k+1-m)} .
\end{aligned}
\end{equation}
Here
\begin{align}
\|g\|_{C^m}
=
\max_{|\beta|\le m}
\sup_{\bm{x}\in[0,1]^n}
\left|D^\beta g(\bm{x})\right|,
\end{align}
and $D^\beta$ denotes the partial derivative of order $\beta$.
\end{theorem}
The proof is provided in \cref{sec:approx_theory}.
\begin{remark}[Fourier bandwidth as the analogue of spline grid resolution]
\cref{thm:Approx_theory} is the Fourier-series analogue of the spline approximation step used in classical KAN theory. In a spline-based KAN, approximation accuracy is controlled by the grid resolution $G$; in the Fourier formulation, it is controlled by the largest retained frequency $K$. Increasing $K$ increases the number of available Fourier modes and yields the $C^m$-error rate $\mathcal{O}(K^{-(k+1-m)})$ for edge functions with $C^{k+1}$ regularity.
\end{remark}

\begin{restatable}[Accessible spectrum and Fourier approximation capacity of QKAN with a trainable preprocessing layer]{theorem}{qkan}\label{thm:qkan_linear}
Fix an integer $k\!\ge\!0$ and let $f$ admit a $C^{k+1}$ periodic extension on the input domain after the chosen frequency normalization. For any depth $r\!\ge\!1$, consider two single-qubit data re-uploading circuits with generator $H=\sigma_j/2$. We measure frequencies in units of the eigenvalue gap of $H$, which equals one.

\medskip\noindent
\textbf{(A) Baseline (Un-weighted).}
\begin{align}
  U(x)&=W^{(r+1)}
        \bigl[S(x)\,W^{(r)}\bigr]\cdots
        \bigl[S(x)\,W^{(1)}\bigr], \\
  S(x)&=e^{-ixH},
\end{align}
where $W^{(\ell)}=W^{(\ell)}(\bm{\theta}_\ell)$ is the $\ell$-th parameterized unitary, $\bm{\theta}_\ell$ is its set of trainable parameters, and $H=\sigma_j/2$ with $\sigma_j\in\{\sigma_x,\sigma_y,\sigma_z\}$.

\noindent
\textbf{(B)  With a preprocessing layer.}
Let $\bm{\omega}=(w_1,\dots,w_r)^{\!\top}\in(0,\infty)^r$ and set
\begin{align}
  U_{\bm{\omega}}(x)&=W^{(r+1)}
          \prod_{\ell=r}^{1}\bigl[S(w_\ell x)\,W^{(\ell)}\bigr],
  \\
  S(w_\ell x)&=e^{-iw_\ell xH}.
\end{align}
Define the model outputs
\begin{align}
  f_A(x)&=\langle0|U^\dagger(x)\mathcal{M} U(x)|0\rangle,
  \\
  f_B(x)&=\langle0|U_{\bm{\omega}}^\dagger(x)\mathcal{M} U_{\bm{\omega}}(x)|0\rangle .
\end{align}

\begin{enumerate}
\item[\textup{(a)}] \textbf{Baseline accessible spectrum.}\;
For every choice of trainable unitary parameters,
\begin{equation}
  f_A(x)\in\operatorname{span}\{e^{i n x}:n=-r,\ldots,r\}.
\end{equation}
Equivalently, $\operatorname{supp}\widehat f_A\subseteq
\Omega_A(r):=\{-r,\ldots,-1,0,1,\ldots,r\}$. Thus the accessible signed nonzero frequency set has cardinality
\begin{equation}
  |\Omega_A(r)\setminus\{0\}|=2r .
\end{equation}
For special parameter choices some coefficients may vanish, so the actual support of a particular trained circuit can be a subset of this accessible set.

\item[\textup{(b)}] \textbf{Preprocessing-layer accessible spectrum.}\;
For any realized value of the trainable preprocessing weight vector $\bm{\omega}=(w_1,\ldots,w_r)^\top\in(0,\infty)^r$,
\begin{equation}
  f_B(x)\in\operatorname{span}\{e^{i\nu x}:\nu\in\Omega_B(\bm{\omega})\},
\end{equation}
where
\begin{equation}
  \Omega_B(\bm{\omega})
  :=
  \left\{\sum_{\ell=1}^{r}m_\ell w_\ell\;\middle|\;m_\ell\in\{-1,0,1\}\right\}.
\end{equation}
Hence the actual Fourier support of a particular circuit is contained in $\Omega_B(\bm{\omega})$. The signed nonzero accessible set satisfies
\begin{equation}
  2\le |\Omega_B(\bm{\omega})\setminus\{0\}|\le 3^r-1,
\end{equation}
with possible reductions caused by collisions among different signed sums. Its formal bandwidth is
\begin{equation}
  K_{\rm form}(\bm{\omega})
  :=\max_{\nu\in\Omega_B(\bm{\omega})}|\nu|
  =\sum_{\ell=1}^r w_\ell .
\end{equation}

\item[\textup{(c)}] \textbf{Fourier approximation benchmark.}\;
Let
\begin{equation}
  \mathcal T_K:=\operatorname{span}\{e^{i n x}: |n|\le K,\ n\in\mathbb Z\}.
\end{equation}
For $0\le m\le k$, the standard Jackson-type Fourier estimate gives
\begin{equation}
  \inf_{p\in\mathcal T_K}\|f-p\|_{C^m}
  \le C_f K^{-(k+1-m)},
\end{equation}
where $C_f$ depends on $f,k,m$ and the frequency normalization, but not on $K$. Consequently, the baseline accessible frequency spectrum gives
\begin{equation}
  \inf_{p\in\operatorname{span}\{e^{i n x}: n\in\Omega_A(r)\}}
  \|f-p\|_{C^m}
  \le C_f r^{-(k+1-m)} .
\end{equation}
For the weighted model, the corresponding Fourier approximation bound depends on the filled low-frequency block, not merely on the largest formal frequency. Define
\begin{equation}
  K_{\rm fill}(\bm{\omega})
  :=
  \max\Bigl\{K\in\mathbb N:\{-K,\ldots,-1,0,1,\ldots,K\}\subseteq\Omega_B(\bm{\omega})\Bigr\}.
\end{equation}
Whenever $K_{\rm fill}(\bm{\omega})\ge 1$,
\begin{equation}
  \inf_{p\in\operatorname{span}\{e^{i\nu x}: \nu\in\Omega_B(\bm{\omega})\}}
  \|f-p\|_{C^m}
  \le C_f K_{\rm fill}(\bm{\omega})^{-(k+1-m)} .
\end{equation}
This is a Fourier approximation capacity bound; it becomes a circuit approximation bound only after an additional coefficient-realization argument establishes that the DARUAN/QKAN parameterization can attain the corresponding Fourier approximation.

\item[\textup{(d)}] \textbf{Capacity comparison with Fourier-series-based KAN.}\;
Because the preprocessing weights are trainable, the hypothesis class contains the constructive geometric subfamily
\begin{equation}
  w_\ell=2^{\ell-1},\qquad \ell=1,\ldots,r .
\end{equation}
For this choice,
\begin{equation}
  K_{\rm form}(\bm{\omega})=2^r-1,
  \qquad
  \Omega_B(\bm{\omega})=\{-(2^r-1),\ldots,-1,0,1,\ldots,2^r-1\},
\end{equation}
and hence $K_{\rm fill}(\bm{\omega})=K_{\rm form}(\bm{\omega})=2^r-1$. The Fourier approximation benchmark therefore gives
\begin{equation}
  \inf_{p\in\operatorname{span}\{e^{i\nu x}:\nu\in\Omega_B(\bm{\omega})\}}
  \|f-p\|_{C^m}
  \le C_f(2^r-1)^{-(k+1-m)}
  \le C_f'2^{-r(k+1-m)} .
\end{equation}
Solving $C_f'2^{-r(k+1-m)}=\varepsilon$ yields
\begin{equation}
  r
  =
  \left\lceil
  \frac{\log_2(C_f'/\varepsilon)}{k+1-m}
  \right\rceil
  =
  \Theta\!\left(\log\frac1\varepsilon\right).
\end{equation}
Thus the geometric member of the trainable preprocessing-layer family provides an exponentially larger accessible Fourier frequencies in the number of re-uploading repetitions than the unweighted construction. A classical Fourier-series-based KAN with independently trainable coefficients requires $M=\Theta(\varepsilon^{-1/(k+1-m)})$ Fourier modes/coefficients to reach the same Jackson benchmark, whereas the weighted QKAN frequency spectrum reaches the same bandwidth benchmark with $\Theta(\log(1/\varepsilon))$ re-uploading repetitions. This comparison is a capacity statement: the Fourier coefficients of QKAN remain coupled through the shared circuit parameters and are not independent Fourier coefficients.
\end{enumerate}

\end{restatable}

The proof is provided in \cref{sec:reupload_fourier}.

\begin{remark}[Interpretation of \cref{thm:qkan_linear}]
\cref{thm:qkan_linear} concerns the \emph{accessible Fourier frequencies} of the single-qubit DARUAN/QKAN activation. Parts~\textup{(b)}--\textup{(c)} should be read conditionally: once the trainable preprocessing weight vector has taken a value $\bm{\omega}$, the corresponding circuit output has Fourier support contained in $\Omega_B(\bm{\omega})$, and the relevant Fourier approximation benchmark is governed by the filled low-frequency block $K_{\rm fill}(\bm{\omega})$.

The only place where a particular preprocessing weight vector is prescribed is part~\textup{(d)}. There, the geometric choice $w_\ell=2^{\ell-1}$ is used as a constructive witness inside the trainable preprocessing family. It is not an assumption that training must keep the preprocessing weights fixed at those values. The statement is therefore a capacity statement: the hypothesis class contains a member whose filled accessible Fourier bandwidth grows exponentially with the number of re-uploading repetitions.

\cref{thm:qkan_linear} is therefore an accessible-support and Fourier approximation capacity result. The complementary coefficient-level question is separated from the theorem and addressed next in \cref{prop:nonvanishing_coeff}.
\end{remark}

\cref{thm:qkan_linear} identifies the accessible frequency spectrum. We now record a complementary coefficient-accessibility statement showing that, in the single-qubit preprocessed setting, these algebraically accessible modes are not eliminated by an architecture-imposed cancellation.

\begin{restatable}[Coefficient accessibility of the single-qubit weighted spectrum]{proposition}{nonvanishingcoeff}\label{prop:nonvanishing_coeff}
Condition on any realized preprocessing weight vector $\bm{\omega}\in(0,\infty)^r$ and consider the single-qubit weighted circuit in \cref{thm:qkan_linear} with $H=\sigma_z/2$ and Pauli-$Z$ readout. Assume that each trainable one-qubit block can realize a nondegenerate mixer with independent $Z$-phase tags; for example, it contains gates of the form
\begin{equation}
  R_z(a_\ell)R_y(\pi/3)R_z(b_\ell)
\end{equation}
for independently variable phases $a_\ell,b_\ell$, $\ell=1,\ldots,r+1$. Let
\begin{equation}
  f_B(x;\Theta)=\sum_{\nu\in\Omega_B(\bm{\omega})}
  c_\nu(\Theta)e^{i\nu x}
\end{equation}
be the Fourier expansion, where coefficients with the same frequency are understood to be aggregated when different ternary strings produce the same signed sum. Then for every nonzero $\nu\in\Omega_B(\bm{\omega})$, the coefficient function $c_\nu(\Theta)$ is not identically zero.

Equivalently, no nonzero frequency in the accessible signed-sum set $\Omega_B(\bm{\omega})$ is removed by a parameter-independent cancellation imposed by the single-qubit architecture. For each $\nu\in\Omega_B(\bm{\omega})\setminus\{0\}$, there exist trainable unitaries such that $c_\nu(\Theta)\ne0$. Moreover, since $\Omega_B(\bm{\omega})$ is finite, generic choices of the unitary parameters make all the coefficients of the nonzero accessible frequencies simultaneously nonzero.
\end{restatable}
The proof is provided in \cref{sec:prop_nonvanishing_coeff}.

\begin{remark}[Scope of \cref{prop:nonvanishing_coeff} and relation to frequency-profile results]
\cref{prop:nonvanishing_coeff} is a coefficient-accessibility statement made after conditioning on an arbitrary realized preprocessing weight vector $\bm{\omega}$. This conditioning is not the geometric construction in \cref{thm:qkan_linear}\textup{(d)}; the latter is the only place where a specific preprocessing weight vector is prescribed, and only as a constructive capacity witness. \cref{prop:nonvanishing_coeff} instead applies to any chosen $\bm{\omega}$ and rules out architecture-imposed cancellation of the corresponding nonzero accessible frequencies.

The proposition does not assert optimization convergence, independent coefficient assignment, or a lower bound on the magnitude of trained coefficients. In particular, a frequency can be generically nonzero as a function of the trainable unitary parameters and still have small typical magnitude under a particular initialization, parameter ensemble, or training protocol.

This distinction explains why the result is not in conflict with the frequency-profile analysis of \citet{barthe2024gradients}.
Their result concerns the average behavior of Fourier-content profiles of quantum re-uploading models under ensemble assumptions in which each parameterized step (layer) is drawn from the Haar measure of unitaries and the data-encoding generator has no weighted, trainable preprocessing parameter.
For repeated identical data generators, successive convolutions of the same spectrum kernel produce a profile concentrated near low frequencies: the profile width grows as $O(\sqrt r)$, whereas the formal support grows as $O(r)$, leading to the phenomenon of vanishing high frequencies on average.
By contrast, \cref{thm:qkan_linear} identifies the algebraic frequency accessible with trainable data re-uploading of preprocessing weights, and \cref{prop:nonvanishing_coeff} shows only that the nonzero modes at those Fourier frequencies are not identically canceled by the single-qubit architecture.

The weighted DARUAN setting also differs from the repeated identical-generator setting because the effective data generator at layer $\ell$ is $w_\ell H$. The geometric vector used in \cref{thm:qkan_linear}\textup{(d)} is therefore closer to the powers-of-two single-qubit construction discussed by \citet{barthe2024gradients}, where the convolution support spreads over an exponentially large set. The relevant distinction is thus between average coefficient profiles under a parameter ensemble and the algebraic frequency available to a structured trainable architecture.
\end{remark}

Taken together, \cref{thm:Approx_theory}, \cref{thm:qkan_linear}, and \cref{prop:nonvanishing_coeff} provide a deliberately separated spectral picture. \cref{thm:Approx_theory} gives the Fourier approximation benchmark for KAN edge functions. \cref{thm:qkan_linear} shows that trainable preprocessing weights can enlarge the accessible Fourier frequencies, with the geometric subfamily giving an exponential filled-bandwidth witness. \cref{prop:nonvanishing_coeff} then shows that, under nondegenerate single-qubit mixing, these accessible nonzero modes are not structurally forced to vanish. The resulting theory is therefore a Fourier approximation capacity theory for a structured, coefficient-coupled parameterization. Its practical value is assessed empirically through the FourierKAN and GeometricFourierKAN baselines in \crefrange{sec:sup_fourierkan}{sec:sup_geofourier} and through the layer-extension experiments below.

Although single-qubit DARUAN avoids the multi-qubit barren-plateau setting, directly optimizing a large repetition number $r$ can still be difficult because increasing $r$ enlarges both the accessible frequency range and the parameter space. This motivates \emph{layer extension}, the QKAN analogue of grid extension in classical KANs. The model first learns lower-frequency global structure at small $r$, and higher-frequency refinements are then introduced by progressively increasing the number of re-uploading repetitions. Because $r$ controls both spectral capacity and computational cost, layer extension also enables adaptive stopping: training can terminate once the validation error reaches the desired tolerance, avoiding unnecessary parameters and computation for tasks that are already solved by smaller circuits.

\subsection{Numerical Results}
In this section, we assess the versatility and effectiveness of QKANs by simulating various tasks, including regression, classification, and generative modeling.
Overall, our experiments demonstrate that QKAN matches or improves upon classical KAN and MLP baselines in most evaluated settings.
In regression benchmarks, QKAN achieves up to an order-of-magnitude reduction in approximation error with fewer parameters in the main comparisons.
For classification tasks, QKANs and HQKANs provide comparable or better top-1 and top-5 accuracy under fixed backbones, while HQKANs achieve this with substantially fewer parameters.
In generative modeling, HQKAN integrated into GPT-2 obtains lower perplexity and reduced training time and computational resources compared to the corresponding MLP replacement.
These results establish QKAN as a scalable and computationally efficient quantum-inspired alternative to classical architectures, with a direct but optional pathway to quantum-hardware inference; the real-device inference validation is reported in \cref{sec:sup_real_device}.
Unless otherwise specified, the following results are obtained on a single NVIDIA RTX 4090 GPU, with software and system details summarized in \cref{sec:sup_system_info}.
The experiments conducted in the main context do not employ GPU-optimized techniques. Prior to the revision of the paper, we utilized a pure PyTorch~\citep{Ansel_PyTorch_2_Faster_2024} implementation during our initial development phase.
Detailed setups for each task are provided in \cref{sec:met}.
For regression experiments, reported metrics are mean $\pm$ standard deviation over 5 random seeds unless specified; other task-specific protocols are described below and in \cref{sec:met}.

We begin with a regression task focused on function fitting.
Previous studies have demonstrated that KANs outperform MLPs in various regression problems, including multivariate function fitting and solving partial differential equations (PDEs) \citep{liu2024kan, liu2024kan20kolmogorovarnoldnetworks}. In particular, KANs have shown strong performance in symbolic expression recovery \citep{yu2024kanmlpfairercomparison, liu2024kan20kolmogorovarnoldnetworks}.
To investigate the robustness of QKAN and its potential for real-world applications, we prepare a regression dataset with added noise in both training and testing labels.

\begin{figure}[t!]
    \centering
    \includegraphics[width=0.4\paperwidth]{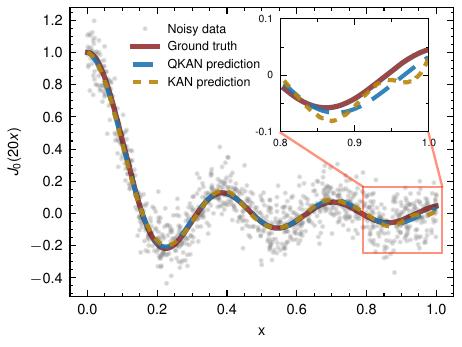}
    \caption{
    \textbf{Function fitting with noise using QKAN and KAN.}
    The target function is $f(x)=J_0(20x)$, fitted with noisy data.
    Both QKAN and KAN use a shape of $\left[1,1\right]$.
    The QKAN prediction exhibits smoother behavior compared to KAN which tends to overfit local noise features.
    }
    \label{fig:noise-plot}
\end{figure}

We start with heuristic function fitting, where the hidden layers and hidden nodes are manually specified.
As a simple example, we consider fitting the function $f(x)=J_0(20x)$, where $J_0$ is the Bessel function of the first kind, using QKAN and KAN with a heuristic QKAN (KAN) shape $\left[1,1\right]$.
The training label is subject to a Gaussian error with a standard deviation of 0.1.
The results are shown in \cref{fig:noise-plot}, where QKAN yields a smoother approximation to the noisy data, while KAN captures more localized features, which in this case correspond to the noise.

To further assess model performance, we conduct more complex heuristic function fitting experiments using the QKAN (KAN) shapes suggested in ref.~\citep{liu2024kan}.
We report mean $\pm$ standard deviation over 5 random seeds in \cref{tab:main_table}, and summarize the corresponding test loss and parameter sizes for QKANs and KANs in \cref{tab:reg-param}.
We also report the visualization of each VAF in learned QKANs in \cref{fig:nodes,sec:sup_activation_visualization}. For a concrete qualitative interpretability demonstration in the style of Fig.~2.4 of~\citet{liu2024kan}, \cref{sec:sup_interpretability,fig:qkan_symbolic,tab:qkan_symbolic} additionally show per-edge symbolic regression on a QKAN$([2,5,1])$ trained on $\exp(\sin(\pi x)+y^2)$: after pruning and symbolic fitting, the recovered primitives cleanly match the target's building blocks ($\sin$ on the $x$ channel, $y^2$ on the $y$ channel, and exponential on the output) with $R^2=1.000$ on the three active edges and a pruned-QKAN test MSE of $3.7\times10^{-6}$, confirming that the KAN interpretability workflow is preserved in QKAN settings.

\begin{table*}[t!]
\centering
\caption{
\textbf{Heuristic noisy function regression.}
We select 10 functions in Feynman dataset  \citep{Udrescu:2019mnk,10.5555/3495724.3496132} and each function is represented by a dimensionless formula with its variables.
Moreover, we randomly sample the data with 10 \% noise level and 1000 training, 1000 testing data points for each function.
Each function has inputs within the range of $\left[0, 1\right]$.
The shapes of QKAN/KAN are represented by the number of hidden nodes in each layer.
The performance is measured by the root mean square error (RMSE) of the predicted outputs compared to the test data.
The best performance is highlighted in \textbf{bold} while the second is labeled with \underline{underline}.
Results are reported as mean $\pm$ standard deviation over 5 random seeds.
The results show that QKAN outperforms KAN and MLP in most cases, demonstrating the effectiveness of the QKAN architecture in noisy function regression tasks.
}\label{tab:main_table}
\resizebox{\textwidth}{!}{%
\begin{tabular}{lcc|ccc|c}
\toprule
Feynman eq. & Dimensionless formula & Variables  & QKAN/KAN shape & QKAN & KAN & MLP \\
\midrule
I.12.11&$1+a\sin{\theta}$ & $a,\theta$ &[2, 2, 1]& $\mathbf{1.192\times10^{-1}}$ $\pm 8.5\times10^{-5}$ & $\underline{1.210\times10^{-1}}$ $\pm 1.6\times10^{-4}$ & $1.388\times10^{-1}$ $\pm 3.9\times10^{-2}$ \\
I.29.16&$\sqrt{1+a^2-2a\cos{(\theta_1-\theta_2)}}$ & $a,\theta_1,\theta_2$ &[3, 2, 3, 1]& $\mathbf{1.444\times10^{-1}}$ $\pm 3.4\times10^{-4}$ & $\underline{1.466\times10^{-1}}$ $\pm 5.4\times10^{-4}$ & $3.192\times10^{-1}$ $\pm 1.4\times10^{-1}$\\
I.40.1&$n_0e^{-a}$ & $n_0,a$ &[2, 2, 1, 1, 1, 2, 1]& $\mathbf{3.085\times10^{-2}}$ $\pm 1.1\times10^{-4}$ & $\underline{3.821\times10^{-2}}$ $\pm 7.4\times10^{-3}$ & $7.958\times10^{-1}$ $\pm 2.9\times10^{-2}$ \\
I.50.26&$\cos{a}+\alpha\cos^2{a}$ & $a,\alpha$ &[2, 2, 3, 1]& $\underline{1.171\times10^{-1}}$ $\pm 1.2\times10^{-4}$ & $1.186\times10^{-1}$ $\pm 7.5\times10^{-4}$ & $\mathbf{1.135\times10^{-1}}$ $\pm 5.4\times10^{-2}$ \\
II.2.42&$(a-1)b$ & $a,b$ &[2, 2, 1]& $\mathbf{2.405\times10^{-2}}$ $\pm 2.3\times10^{-5}$ & $\underline{2.442\times10^{-2}}$ $\pm 4.2\times10^{-5}$ & $4.320\times10^{-1}$ $\pm 3.1\times10^{-1}$\\
II.6.15a&$\frac{1}{4\pi}c\sqrt{a^2+b^2}$ & $a,b,c$ &[3, 2, 1, 1]& $\underline{3.115\times10^{-3}}$ $\pm 2.3\times10^{-5}$ & $\mathbf{3.080\times10^{-3}}$ $\pm 2.4\times10^{-5}$ & $7.930\times10^{-3}$ $\pm 2.6\times10^{-3}$\\
II.35.18&$\frac{n_0}{\exp{(a)}+\exp{(-a)}}$ & $n_0,a$ &[2, 1, 1]& $\mathbf{2.114\times10^{-2}}$ $\pm 1.2\times10^{-5}$ & $\underline{2.128\times10^{-2}}$ $\pm 1.0\times10^{-5}$ & $1.109\times10^{-1}$ $\pm 1.2\times10^{-1}$ \\
II.36.38&$a+\alpha b$ & $a,b,\alpha$ &[3, 2, 1]& $\mathbf{7.331\times10^{-2}}$ $\pm 1.6\times10^{-4}$ & $\underline{8.014\times10^{-2}}$ $\pm 6.2\times10^{-3}$ & $1.687\times10^{-1}$ $\pm 2.5\times10^{-1}$\\
III.10.19&$\sqrt{1+a^2+b^2}$ & $a,b$ &[2, 1, 1]& $\mathbf{1.242\times10^{-1}}$ $\pm 9.8\times10^{-5}$ & $\underline{1.244\times10^{-1}}$ $\pm 9.1\times10^{-5}$ &$1.485\times10^{-1}$ $\pm 3.1\times10^{-2}$\\
III.17.37&$\beta(1+\alpha\cos{\theta})$ & $\alpha,\beta,\theta$ &[3, 3, 1]& $\mathbf{6.903\times10^{-2}}$ $\pm 2.7\times10^{-4}$ & $\underline{7.066\times10^{-2}}$ $\pm 4.9\times10^{-4}$ & $3.367\times10^{-1}$ $\pm 2.8\times10^{-1}$ \\
\bottomrule
\end{tabular}
}
\end{table*}

\begin{table*}[ptbh]
    \centering
    \caption{\textbf{
            Comparison of QKAN and KAN in heuristic noisy function regression.
        }
        Continue from \cref{tab:main_table}, we report the number of parameters and RMSE loss of QKAN and KAN models.
        The best performance is highlighted in \textbf{bold}.
        Results are reported as mean $\pm$ standard deviation over 5 random seeds.
        The results show that QKAN outperforms KAN while requiring 24\,\% fewer parameters in average.
    }\label{tab:reg-param}
    \resizebox{\textwidth}{!}{
    \begin{tabular}{lcccc}
        \toprule
          Feynman          & \multicolumn{2}{@{}c@{}}{QKAN} & \multicolumn{2}{@{}c@{}}{KAN}                                          \\
        \cmidrule{2-3} \cmidrule{4-5}
         equation & RMSE loss                      & \# Params                     & RMSE loss                  & \# Params \\
        \midrule
        I.12.11     & $\bm{1.192\times10^{-1}}$ $\pm 8.5\times10^{-5}$     &  102                           & $1.210\times10^{-1}$ $\pm 1.6\times10^{-4}$       & 135       \\
        I.29.16     & $\bm{1.444\times10^{-1}}$ $\pm 3.4\times10^{-4}$     & 255                           & $1.466\times10^{-1}$ $\pm 5.4\times10^{-4}$      & 336       \\
        I.40.1      & $\bm{3.085\times10^{-2}}$ $\pm 1.1\times10^{-4}$     & 204                           & $3.821\times10^{-2}$ $\pm 7.4\times10^{-3}$ & 272       \\
        I.50.26     & $\bm{1.171\times10^{-1}}$ $\pm 1.2\times10^{-4}$     & 221                           & $1.186\times10^{-1}$ $\pm 7.5\times10^{-4}$      & 292       \\
        II.2.42     & $\bm{2.405\times10^{-2}}$ $\pm 2.3\times10^{-5}$     & 102                           & $2.442\times10^{-2}$ $\pm 4.2\times10^{-5}$      & 135       \\
        II.6.15a    & $3.115\times10^{-3}$ $\pm 2.3\times10^{-5}$          & 153                           & $\bm{3.080\times10^{-3}}$ $\pm 2.4\times10^{-5}$ & 202       \\
        II.35.18    & $\bm{2.114\times10^{-2}}$ $\pm 1.2\times10^{-5}$     & 51                            & $2.128\times10^{-2}$ $\pm 1.0\times10^{-5}$      & 68        \\
        II.36.38    & $\bm{7.331\times10^{-2}}$ $\pm 1.6\times10^{-4}$     & 136                           & $8.014\times10^{-2}$ $\pm 6.2\times10^{-3}$      & 179       \\
        III.10.19   & $\bm{1.242\times10^{-1}}$ $\pm 9.8\times10^{-5}$     & 51                            & $1.244\times10^{-1}$ $\pm 9.1\times10^{-5}$      & 68        \\
        III.17.37   & $\bm{6.903\times10^{-2}}$ $\pm 2.7\times10^{-4}$     & 204                           & $7.066\times10^{-2}$ $\pm 4.9\times10^{-4}$      & 268       \\
        \bottomrule
    \end{tabular}
    }
\end{table*}

As shown in \cref{tab:main_table}, QKAN consistently outperforms both KAN and MLP baselines in the majority of cases, achieving the lowest mean test RMSE on 8 out of the 10 benchmark equations.
This trend demonstrates that QVAF in QKAN effectively captures complex patterns in noisy regression tasks.
In the remaining two cases, QKAN still ranks second, with performance closely matching or slightly trailing that of the best-performing baseline.
The results suggest that QKAN's expressive feature mappings are especially well-suited for capturing high-frequency or compositional structures, where classical models tend to underperform.

\cref{tab:reg-param} further reveals that QKAN models not only deliver superior or comparable performance but also achieve this with significantly fewer parameters.
On average, QKAN uses about 24\% fewer parameters than KAN while maintaining or improving generalization accuracy.
This parameter efficiency is especially advantageous in scenarios with limited model capacity or deployment constraints.

However, in most scenarios the underlying functional form is unknown a priori.
To evaluate model robustness under such uncertainty, we extend our study to a diverse suite of 66 symbolic expressions drawn from the Feynman dataset \citep{Udrescu:2019mnk,10.5555/3495724.3496132}, each input normalized to the unit hypercube and output subject to 10\% additive noise.
For each equation, we optimize QKAN, KAN and MLP architectures over hidden-layer depths as described in \cref{sec:numerica_method}, and record the lowest test RMSE attained across five random seeds.
The reason for reporting only the best test RMSE is that in real-world machine learning for science (ML4Science) scenarios, where the exact equation is unknown beforehand, we rely solely on the best model we have obtained.

\cref{fig:loss} displays these best‐case losses on a logarithmic scale, with the total number of trainable parameters annotated for both QKANs and KANs (see \cref{tab:com_reg}).
Notably, QKAN achieves the lowest RMSE on over 80\% of the benchmark equations, despite employing on average 30\% fewer parameters than classical KAN.
In the minority of cases where KAN marginally outperforms, the QKAN remains competitive, often within the same order of magnitude, while retaining its parameter‐efficiency advantage.
By contrast, standard MLPs exhibit rapidly deteriorating generalization as equation complexity increases, underscoring the critical role of structured feature embeddings in noisy symbolic regression.

These results substantiate the strength of DARUAN, the QVAF instantiation used in QKAN: its structured circuit parameterization enhances expressivity in the absence of exact analytic priors while also delivering substantial parameter savings.
Consequently, QKAN represents a compelling approach for data‐driven discovery and modeling in scientific applications where both accuracy and resource constraints are paramount.  
The quantum-inspired architecture provides both improved generalization and model compactness, highlighting its potential for broader applications in data-driven scientific modeling.

To evaluate the expressive power and scalability of QKANs beyond function regression, we investigate their performance on image classification tasks.
We consider three standard benchmarks—MNIST~\citep{mnist}, CIFAR-10, and CIFAR-100~\citep{krizhevsky2009learning}—and use a hybrid architecture where a convolutional neural network (CNN) is followed by a fully connected network (FCN).
In our setup, the FCN is instantiated using either an MLP, a KAN, or a QKAN, enabling a direct comparison across model families with identical convolutional backbones.

\begin{table*}[t]
\centering
\caption{\textbf{Performance of different models on MNIST, CIFAR-10, and CIFAR-100 datasets.}
The top-1, top-5 test accuracy and the parameter size of each model.
The second column indicates the parameter size of CNN shared by all models.
}
\resizebox{\textwidth}{!}{%
\begin{tabular}{l|c|c|c|c|c|c|c|c|c|c|c|c|c}
\toprule
Model & (CNN) & \multicolumn{3}{c|}{CNN+MLP} & \multicolumn{3}{c|}{CNN+KAN(G=10)} & \multicolumn{3}{c|}{CNN+QKAN(r=3)} & \multicolumn{3}{c}{CNN+HQKAN}\\
\midrule
Training & CNN & Top-1 & Top-5 & MLP & Top-1 & Top-5 & KAN & Top-1 & Top-5 & QKAN & Top-1 & Top-5 & HQKAN\\
dataset & \# Params & \multicolumn{2}{c|}{Accuracy (\%)} & \# Params & \multicolumn{2}{c|}{Accuracy (\%)} & \# Params & \multicolumn{2}{c|}{Accuracy (\%)} & \# Params & \multicolumn{2}{c|}{Accuracy (\%)} & \# Params\\
\midrule
MNIST & 1,084 & 97.9 & \textbf{100.0} & 850 & 97.7 &  \textbf{100.0} & 1,500 & \textbf{98.0} & \textbf{100.0} & 800 & 95.9 & 99.7 & 222 \\
CIFAR-10 & 56,320 & 71.4 & 97.8 & 41,802 & 68.4 & 97.4 & 39,900 & 68.8 & 97.0 & 21,280 & \textbf{71.6} & \textbf{97.9} & 14,370 \\
CIFAR-100 & 56,320 & 39.8 & 69.4 & 86,948 & 40.6 & 70.4 & 384,000 & \textbf{41.2} & 70.0 & 204,800 & 39.9 & \textbf{70.6} & 32,636 \\
\bottomrule
\end{tabular}
}
\label{tab:combined_performance}
\end{table*}

\cref{tab:combined_performance} reports the top-1 and top-5 classification accuracies together with the parameter counts of the FCN components. 
Top-1 accuracy measures the proportion of test samples for which the model’s single most confident prediction coincides with the true label, whereas top-5 accuracy considers a prediction correct if the ground-truth label is contained within the top five highest probabilities.

On MNIST, all models achieve high accuracy, with CNN+QKAN attaining the highest top-1 accuracy of 98.0\% and perfect top-5 accuracy using only 800 parameters in the FCN.
In contrast, CNN+MLP and CNN+KAN require significantly more parameters (850 and 1,500, respectively) to achieve comparable accuracy.
This suggests that QKAN can achieve strong performance with minimal parameter overhead, even on simple tasks.

The advantage of QKAN becomes more evident on CIFAR-10, where CNN+QKAN achieves the comparable top-1 and top-5 accuracies (68.8\% and 97.0\%, respectively) while requiring nearly half the parameters of CNN+KAN (21,280 vs.\ 39,900).
CNN+MLP, despite having a similar parameter count to KAN, doesn't outperform in both metrics.
This highlights QKAN’s superior parameter efficiency and generalization capacity.

For the more challenging CIFAR-100 dataset, CNN+QKAN achieves the highest top-1 accuracy (41.2\%), outperforming CNN+KAN and CNN+MLP while using fewer parameters than CNN+KAN.
Notably, CNN+KAN and CNN+QKAN require a substantial number of parameters due to their linear scaling with output size, 384k and 205k respectively, posing limitations for practical deployment.

To address the scalability constraints of QKAN and KAN on high-dimensional tasks, we introduce \textit{Hybrid QKAN} (HQKAN), which incorporates two additional fully connected layers to form an autoencoder-like structure around a compact QKAN core.
By compressing features into a small latent space, HQKAN leverages QKAN’s expressivity while significantly reducing parameter count.

As demonstrated in \cref{tab:combined_performance}, HQKAN achieves competitive accuracy with an order-of-magnitude reduction in parameters.
For example, on CIFAR-100, HQKAN attains a top-5 accuracy of 70.6\% using only 32,636 parameters, compared to 86,948 and 384,000 required by MLP and KAN models, respectively.
On CIFAR-10, HQKAN not only outperforms both MLP and KAN in top-1 and top-5 accuracy but does so with the fewest parameters (14,370).
This result underscores the practicality of HQKAN for memory- and compute-constrained environments.

Together, these results validate the efficacy and versatility of QKANs and HQKANs across classification tasks of varying complexity.
They demonstrate that QVAFs, particularly the single-qubit data re-uploading circuit DARUAN, provide an expressive, scalable, and parameter-efficient quantum-inspired approach to deep learning.

\begin{figure}[t!]
    \centering
    \includegraphics[width=0.4\paperwidth]{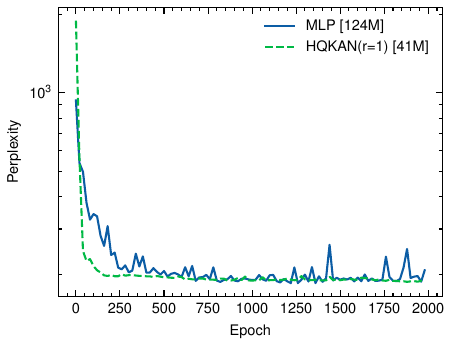}
    \caption{
        \textbf{GPT-2 model trained on the WebText dataset~\citep{webtext}, incorporating HQKAN and MLP layers.}
        To further reduce parameter count and improve efficiency, we adopt the hybrid QKAN (HQKAN) strategy, where the input and output dimensions fed into QKAN are compressed via fully connected layers, forming an autoencoder-like bottleneck.
        This compression enables QKAN to operate effectively in a low-dimensional latent space with reduced overhead.
        Parameter sizes are indicated in square brackets.
        HQKAN achieves better perplexity performance than MLP while using only one-third of its parameters and the same training time, demonstrating the effectiveness of QKAN-based architectures for generative modeling on classical hardware.
    }\label{fig:gen-hqkan}
\end{figure}

We further evaluate the generative modeling capability of QKANs by integrating them into autoregressive language models, specifically the GPT-2 architecture.
For preliminary testing, we utilize an open-source \texttt{kan-gpt} module \citep{GANESH2024KANGPT}.
In this approach, we replace all the linear layers within the transformer blocks with HQKANs we introduced earlier in classification tasks to address the scaling problem in KANs architecture.
Subsequently, we pretrain the resulting models on the WebText dataset \citep{webtext}.

\cref{fig:gen-hqkan} presents the perplexity curves during training of GPT-2 models equipped with MLP and HQKAN modules.
Despite operating in a reduced-dimensional latent space, HQKAN consistently achieves superior convergence and reduced final perplexity compared to the MLP baseline.
This is achieved while taking only one-third of the parameters and 30\% less memory during training time.

To complement the generative modeling results presented in the main text, we provide detailed runtime and memory benchmarks for GPT-2 models incorporating QKAN-based components.
In particular, we aim to assess the scalability of these models under large-batch training regimes, as large batch sizes have been shown to significantly improve performance in LLMs with scaling laws~\citep{brown2020languagemodelsfewshotlearners,shuai2024scalinglawlanguagemodels}.

\cref{tab:app-gpt2} summarizes the performance of various GPT-2 configurations on the WebText dataset.
The top section of the table (above the second double midrule) evaluates models using the \texttt{kan-gpt} framework, where the linear layers in GPT-2 are replaced by HQKANs.
Models marked with an asterisk (\textasteriskcentered) correspond to those visualized in \cref{fig:gen-hqkan} of the main text.
We measure the iteration time and memory consumption for each methods.

The lower section of the table reports results for the \texttt{KANsformer} architecture~\citep{xie2024kansformerscalablebeamforming}, 
which integrates HQKAN modules directly into the feedforward layers of transformer blocks, with flash attention~\citep{dao2022flashattentionfastmemoryefficientexact}.
Flash attention significantly reduces memory usage and improves speed.
We report performance at various batch sizes to demonstrate the scalability and hardware compatibility of QKAN-based architectures under realistic training regimes.

\begin{table*}[ptbh]
\centering
\caption{
\textbf{Performance comparison of GPT-2 models on the WebText dataset \citep{webtext}.}
This table presents the runtime and memory usage for various GPT-2 configurations equipped with QKAN-based modules.
The top section (above the second double midrule) corresponds to models using the \texttt{kan-gpt} framework introduced in the main text, where all linear layers are replaced by HQKAN.
The models marked with an asterisk (\textasteriskcentered) are those displayed in \cref{fig:gen-hqkan}.
Results include training speed (ms per iteration) and memory consumption.
The lower section (below the second double midrule) reports large-scale results using the \texttt{KANsformer} \citep{xie2024kansformerscalablebeamforming} architecture with flash attention \citep{dao2022flashattentionfastmemoryefficientexact}, which integrates HQKAN directly into the feed-forward network of  transformer block instead of all linear layers.
We benchmark across batch sizes and hardware configurations, including single- and multi-GPU setups with NVIDIA RTX 4090, V100S, H100, and H200 GPUs.
These configurations demonstrate the scalability and practical feasibility of HQKAN-based models in both single- and distributed-training regimes.
}
\label{tab:app-gpt2}
\resizebox{\textwidth}{!}{%
\begin{tabular}{l|c|c|r|r}
\toprule
Method& \# Params & Batch size & Time/iter [GPU $\cdot$ ms] & Device memory \\
\midrule
\midrule
MLP-based GPT  & 124 M & *1 & 63 [RTX 4090 $\cdot$ ms] & 6.5 GB \\
w/o flash atten.& & 48 & 4,536 [V100S $\cdot$ ms] & 111 GB \\
\midrule
HQKAN-gpt & 41 M & *1 & 60 [RTX 4090 $\cdot$ ms] & 4.1 GB\\
w/o flash atten. & & 48 & 4,648 [V100S $\cdot$ ms] & 120 GB \\
\midrule
\midrule
MLP-based GPT & 124 M & 1 & 40 [RTX 4090 $\cdot$ ms] & 4.7 GB \\
w/ flash atten.& & 10 & 252 [RTX 4090 $\cdot$ ms] & 17.3 GB \\
& & 48 & 3,884 [V100S $\cdot$ ms] & 81.2 GB \\
&  & 320 & 6,400 [H100 $\cdot$ ms] & 656 GB \\
&  & 800 & 14,784 [H200 $\cdot$ ms] & 1,340 GB \\
\midrule
\textbf{HQKANsformer} & 67 M & 1 & 39.5 [RTX 4090 $\cdot$ ms] & 3.8 GB \\
\textbf{w/ flash atten.} & & 10 & 240 [RTX 4090 $\cdot$ ms] & 15.6 GB \\
& & 48 & 3,540 [V100S $\cdot$ ms] & 73.6 GB \\
&  & 320 & 5,760 [H100 $\cdot$ ms] & 592 GB \\
&  & 800 & 13,232 [H200 $\cdot$ ms] & 1,224 GB \\
\bottomrule
\end{tabular}
}
\end{table*}

\cref{tab:app-gpt2} highlights the runtime and memory efficiency of HQKAN-based GPT-2 models. 
On a single RTX 4090, HQKAN achieves comparable speed to standard MLPs (60\,ms vs.\ 63\,ms per iteration), while reducing memory consumption by over 35\% (4.1\,GB vs.\ 6.5\,GB). 
When scaled to 4$\times$V100S GPUs at batch size 48, HQKAN maintains efficiency with only marginal overhead (4,648\,ms vs.\ 4,536\,ms), despite a roughly threefold reduction in model size (41M vs.\ 124M parameters).
Nevertheless, the device memory consumption and training time reveal scaling limitations for the \texttt{kan-gpt} method, which restricts its practicality for very large models.

To address this issue, we turn to \texttt{KANsformer} architecture with flash attention, and the results are shown in the lower section of \cref{tab:app-gpt2}. 
This architecture provides further memory savings and speed improvements, enabling scalability to larger training regimes. 
At large batch sizes (e.g., 320), HQKANsformer reduces device memory from 656\,GB to 592\,GB compared to MLP, while also lowering per-iteration time (6,400\,ms vs.\ 5,760\,ms on total H100 times). 
Even at batch size 800 on 16$\times$H200, HQKANsformer completes iterations with both 10\% less memory and 10\% less time, demonstrating its efficiency at scale. 
In summary, HQKANsformer effectively overcomes the scalability encoder-decoder bottleneck of the \texttt{kan-gpt} approach, making it a practical solution for training foundation models at large scale.

Overall, the results demonstrate that HQKAN can serve as an effective, efficient, and scalable \textit{plug-and-play replacement} for classical fully connected layers in generative transformer architectures.
Its parameter efficiency and faster convergence open the door to practical quantum-inspired modeling in large-scale natural language processing tasks.

Additionally, to investigate the compatibility between quantum-inspired and classical KAN, we present a knowledge distillation method to transfer learned parameters from a trained QKAN to a corresponding KAN (see \cref{sec:trans} for details).
This process enables QKANs to serve as a pretraining mechanism for KANs, potentially accelerating convergence and improving generalization in the classical regime.

\begin{figure}
    \centering
    \includegraphics[width=0.4\paperwidth]{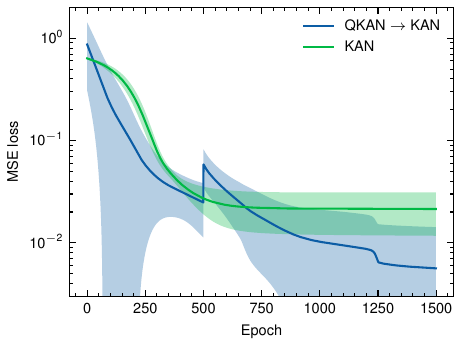}
    \caption{
        \textbf{Knowledge distillation from QKAN to KAN.}  
        We consider the regression task \(f(x, y) = \sin(e^x + y^2)\).
        A QKAN is first trained for 500 epochs, after which its learned variational parameters are converted into B-spline coefficients and transferred into a KAN of matching architecture.
        Due to approximation errors during the conversion, a small loss shift is observed, which is corrected through continued training.
        The resulting QKAN-initialized KAN achieves a 70\% reduction in test loss compared to a KAN trained from scratch, demonstrating the effectiveness of QKAN as a pretraining strategy for classical KAN.
    }\label{fig:q2c}
\end{figure}

We illustrate this approach with a toy function, \(f(x, y) = \sin(e^x + y^2)\), a nonlinear expression with compositional structure.
As shown in \cref{fig:q2c}, the QKAN is first trained close to convergence.
Learned parameters are then converted into B-spline basis representations and transferred into a structurally matched KAN.
Owing to spline approximation errors during the conversion, the KAN continues training for a limited number of epochs to fine-tune the inherited parameters.

This transfer approach yields a substantial performance gain: the transferred KAN achieves a 70\% lower test loss compared to a baseline KAN trained from scratch under identical conditions. 
These results suggest that QKAN can serve as a powerful initialization strategy for KAN, effectively bootstrapping their learning via QVAFs.

\section{Discussion}\label{sec:dis}

In this work, we introduce QKANs, a novel framework that integrates QVAFs into the classical KAN architecture.
QVAF is the general quantum activation concept and can be instantiated by multi-qubit circuits as QML; the concrete QKAN studied here uses DARUAN, a single-qubit data re-uploading instantiation with trainable data-preprocessing weights.
DARUANs enable QKANs to approximate complex nonlinear mappings using significantly fewer parameters than traditional methods.

QVAFs offer a compelling perspective on variational quantum circuits, not only as data encoders or feature maps, but also as expressive, learnable nonlinearities embedded directly within neural architectures.
By systematically organizing DARUAN into KAN layers, QKANs obtain a compact structured parameterization whose accessible Fourier frequencies can grow rapidly with the number of re-uploading repetitions. Its mathematical form arises naturally from the Lie algebra of SU(2) rotations and remains efficiently simulable classically. The quantum circuit formalism provides: (i) a principled derivation of the activation form from physical symmetry; (ii) structured coupling of the accessible Fourier coefficients through shared rotation parameters, which can act as an implicit regularizer; and (iii) a direct pathway to hardware inference validation as quantum devices mature.

Our architecture preserves the structured interpretability of classical KANs while achieving improved parameter efficiency through quantum-inspired DARUAN.
Empirical evaluations across regression, classification, and generative modeling tasks consistently show that QKANs outperform or rival both MLPs and traditional KANs, despite using fewer parameters and reduced training overhead.

To improve scalability and adaptability, we further introduced layer extension and HQKANs, allowing QKANs to flexibly accommodate various task complexities.
Layer extension is particularly useful as an adaptive capacity-control mechanism: by progressively increasing $r$, QKAN can stop at the smallest circuit depth that reaches a desired validation error, reducing unnecessary parameter and computational overhead; the corresponding ablation is reported in \cref{sec:sup_layerext}.
These extensions broaden the model's applicability without compromising its foundational design.

Our classification and language-modeling studies are component-level comparisons: the CNN or transformer backbone is fixed, and only the MLP/KAN/QKAN replacement block is varied. Therefore, the relevant question is not whether a QKAN head outperforms full SOTA vision or language architectures with different feature extractors, but whether DARUAN provides a more compact drop-in activation module under the same surrounding architecture.
Under this controlled setting, QKANs serve as versatile modules across the spectrum of classical architectures and consistently improve parameter efficiency with comparable or better performance.

To clarify the source of QKAN's advantage, we additionally compare against a Fourier-based KAN (FourierKAN; a KAN with classical Fourier basis activations) and a GeometricFourierKAN using geometric frequency allocation $\{1,2,4,\ldots,2^{K-1}\}$ matched to the DARUAN support construction. These baselines test whether the empirical gains arise merely from Fourier-type modes or from geometric frequency placement. Across the reported benchmark equations, QKAN obtains the lowest or tied-lowest mean RMSE in the GeometricFourierKAN comparison and outperforms FourierKAN in the direct Fourier baseline comparison (\crefrange{sec:sup_fourierkan}{sec:sup_geofourier}). This supports the interpretation that the SU(2)-induced coefficient coupling, rather than support size alone, is important in practice. In addition, the ablation of the DARUAN contribution weight $w_d$ versus the base activation weight $w_b$ in \cref{sec:sup_ablation} shows that DARUAN is the dominant contributor, with learned contribution fractions of approximately $63$--$69\%$ and a 2--10$\times$ RMSE degradation when it is removed.

Importantly, QKANs are efficiently simulable on classical hardware: in the performance benchmark of \cref{sec:sup_runtime,tab:runtime}, the GPU-optimized FlashQKAN block gives lower mean forward and training-step latency than the corresponding Fourier-based KAN (FourierKAN) block and substantially lower latency than the B-spline-based KAN (SplineKAN) block, while the parameter and simulation precision analyses are provided in \cref{sec:sup_precision}. The single-qubit quantum circuits of QKAN are also executable on quantum hardware without modification for NISQ-era inference experiments \citep{42w2-6ccy,PRXQuantum.5.040342}; we provide a direct real-device validation on \texttt{ibm\_aachen} with a one-dimensional spherical Bessel $j_0(20x)=\sin(20x)/(20x)$ function-fit measurement-shot sweep through 100,000 shots and CIFAR-10/CIFAR-100 QKAN inference (\cref{sec:sup_real_device,fig:ibm_aachen_hqkan,fig:ibm_aachen_funcfit_curve}). The function-fit test MSE decreases from $3.9439\times10^{-2}$ at 1,000 shots to $5.392\times10^{-4}$ at 100,000 shots; because these values were acquired under different calibrations and layouts, the decrease is consistent with convergence but is not a controlled estimate of shot scaling alone. At 100 shots, the HQKAN runs reach $73.0\%$/$40.0\%$ top-1 accuracy on CIFAR-10/CIFAR-100 (exact-solver $75.0\%$/$43.0\%$). For the single-qubit case studied here, classical simulation remains sufficient and preferred for all practical purposes.
While the possibility of transferring trained QKAN parameters to classical KANs enables a seamless hybridization of quantum and classical pipelines, offering practical benefits even before full-scale quantum hardware becomes widely accessible.

During the final stage of our manuscript revision, we became aware of several recent studies that also explore similar idea on QVAFs in KANs \citep{werner2025qukanquantumcircuitborn, wakaura2025enhancedvariationalquantumkolmogorovarnold, wakaura2025adaptivevariationalquantumkolmogorovarnold}.
These works propose alternative VQC ansatz to serve as activation functions within the KAN framework, aligning with our formulation of QVAFs.
However, a key distinction lies in the quantum resource requirements of these approaches.
The aforementioned studies utilize multi-qubit circuits for their VQCs, which presents three major scalability challenges.
First, the current limitations of quantum hardware, as well as the computational cost of classical quantum simulators, restrict the practical deployment of models with large qubit counts.
Second, and more fundamentally, scaling up multi-qubit VQCs often leads to the emergence of barren plateaus, regions in the optimization landscape where the gradient vanishes exponentially with system size, making the training of large-scale quantum neural networks exceedingly difficult \citep{mcclean2018barren}.
Moreover, two-qubit gate fidelities on NISQ devices still make deep entangling circuits difficult to execute reliably \citep{Preskill2018quantumcomputingin, Singh_2024, 42w2-6ccy}, which limits the scalability of multi-qubit QVAF designs.
These limitations significantly hinder the scalability and practicality of multi-qubit QVAF-based KANs.
Third, existing models still depend on access to real quantum hardware when scaling up, both during training and inference.
Such devices remain scarce and must typically be accessed through remote servers, introducing latency that is particularly detrimental for real-time tasks.
In contrast, our proposed architecture, QKAN, employs only single-qubit data re-uploading circuits for its VAFs.
These circuits are lightweight and efficient enough to be simulated on classical quantum simulators, thereby removing the reliance on real quantum devices during inference.
This single-qubit design supports direct execution on near-term quantum devices for inference validation, while keeping classical simulation as the preferred practical, scalable, and robust route for quantum-inspired learning.

In summary, QKANs instantiate the general QVAF idea inside KANs using the scalable and computationally efficient single-qubit DARUAN design.
Crucially, this design is already executable on classical hardware with optimized GPU kernels and demonstrates readiness for large-scale deployment, while preserving a direct pathway to quantum-hardware inference as devices mature.
Our work thus establishes QKANs as a bridge between interpretable classical architectures and quantum-circuit-derived structured activations.
The successful demonstration of the efficiency and scalability of single-qubit DARUAN within QKAN and HQKAN architectures across multiple tasks indicates potential advantages for multi-qubit DARUAN implementations with trainable preprocessing weights, where increasing the number of qubits and incorporating entanglement may enhance success probabilities and reduce the depth (i.e., the number of layers) required \citep{Schuld_2021, P_rez_Salinas_2020}, though systematically validating these anticipated benefits on a large scale remains challenging with current NISQ-era hardware and classical simulation capabilities.

\section{Methods}\label{sec:met}

This section provides a detailed overview of the core methods proposed and employed in this work. Additional implementation details and software/system information are provided in \crefrange{sec:sup_software_impl}{sec:sup_system_info} and \cref{tab:version}.

\subsection{Kolmogorov-Arnold Networks}\label{sec:KAN_intro}
KANs are designed to approximate multivariate functions using compositions of univariate functions and addition, establishing themselves as universal approximators with high interpretability and efficiency.

\paragraph{Kolmogorov-Arnold representation theorem.}
KART \citep{kolmogorov1957representation} states that any multivariate continuous function $f(x_1, \ldots, x_N)$ can be represented as a finite composition of univariate functions and addition:
\begin{equation}
  f(\bm{x}) = \sum_{q=1}^{2N+1} \Phi_q \left( \sum_{p=1}^{N} \phi_{q,p}(x_p) \right),
\end{equation}
where $\phi_{q,p} : [0,1] \rightarrow \mathbb{R}$ and $\Phi_q : \mathbb{R} \rightarrow \mathbb{R}$ are continuous functions.
While KART provides theoretical guarantees of universal approximation, the original functions can be non-smooth or hard to learn in practice \citep{10.1162/neco.1989.1.4.465,poggio2020theoretical, liu2024kan}, hence the need for parameterized and trainable alternatives.

\paragraph{Formalism of Kolmogorov-Arnold networks.}
\citet{liu2024kan} introduced KANs as a practical realization of the KART, generalizing it to deep and wide architectures.
Each activation in KANs is modeled as a learnable VAF parameterized by B-splines, which are piecewise polynomial functions capable of approximating any continuous function with arbitrary precision \citep{liu2024kan, douzette2017b}.

Formally, a KAN layer maps the output of the $l$-th layer to the $(l+1)$-th layer via:
\begin{equation}
  x_{l+1,j} = \sum_{i=1}^{n_l} \phi_{l,j,i}(x_{l,i}),
\end{equation}
where $\phi_{l,j,i}$ is the learnable univariate VAF connecting input node $i$ to output node $j$.

This can be expressed in matrix notation as:
\begin{align}
  &\bm{x}_{l+1}  = \Phi_l(\bm{x}_l),                                                                            \\
  &\Phi_l        = \begin{pmatrix}
                     \phi_{l,1,1}(\cdot)       & \phi_{l,1,2}(\cdot)       & \cdots & \phi_{l,1,n_l}(\cdot)       \\
                     \phi_{l,2,1}(\cdot)       & \phi_{l,2,2}(\cdot)       & \cdots & \phi_{l,2,n_l}(\cdot)       \\
                     \vdots                    & \vdots                    & \ddots & \vdots                      \\
                     \phi_{l,n_{l+1},1}(\cdot) & \phi_{l,n_{l+1},2}(\cdot) & \cdots & \phi_{l,n_{l+1},n_l}(\cdot)
                   \end{pmatrix}.
\end{align}

A KAN with depth $L$, width $N$, spline order $k$, and $G$ grid intervals requires $\mathcal{O}(N^2L(G+k)) \sim \mathcal{O}(N^2LG)$ parameters \citep{liu2024kan}.
This is comparable to MLPs with $\mathcal{O}(\mathcal{N}^2L)$ parameters for depth $L$ and width $\mathcal{N}$, but KANs typically require a significantly smaller width $N$, resulting in improved efficiency and generalization.

However, the computational cost of evaluating B-spline bases and rescaling grids can become a computational bottleneck in practice \citep{li2024kolmogorovarnold}.
This motivates our exploration of quantum-inspired methods to replace spline-based activations with more efficient quantum circuit approximators.

\subsection{Data Re-Uploading Circuits}\label{sec:data_reuplaod}
To address the computational challenges associated with B-spline activations in classical KANs, we leverage the expressive power of VQCs, particularly through data re-uploading circuits \citep{P_rez_Salinas_2020}.

Data re-uploading circuits encode classical inputs via repeated parameterized embeddings into quantum states, alternating with trainable unitaries.
In general, for an implementation with $r$ repetitions with $n$-dimensional input $\bm{x}\in\mathbb{R}^n$, the circuit can be expressed as:
\begin{equation}
  U(\bm{x}, \bm{\theta}) = W^{(r+1)} S(\bm{x}) W^{(r)} \cdots S(\bm{x}) W^{(1)},
\end{equation}
where each $W^{(\ell)}$ is a trainable unitary and $S(\bm{x})$ is the data-encoding operation.
A common instantiation is $S(\bm{x} \circ \bm{\omega} + \bm{\beta})$ with trainable parameters $\bm{\omega}, \bm{\beta} \in \mathbb{R}^n$ \citep{zhao2024quantum}, where $\circ$ denotes a Hadamard product.
In the DARUAN instantiation used for QKAN, each edge activation receives a single scalar input, which allows us to reduce the circuit to
\begin{equation}
  U(x, \bm{\theta}) = W^{(r+1)} S(w_rx+b_r) W^{(r)} \cdots S(w_1x+b_1) W^{(1)}.
\end{equation}

\subsection{Knowledge Distillation from QKANs to KANs}\label{sec:trans}
One of the key advantages of our quantum-inspired KANs is their ability to transfer learned VAFs to classical KANs after training, enabling deployment on classical hardware.

Since each DARUAN unit approximates an univariate function, we can estimate its functional form and reparameterize it using classical spline or Fourier bases. The transfer procedure involves three steps:
\begin{enumerate}
  \item Forward evaluation on a quantum machine: after training, each DARUAN activation is evaluated across a discretized input domain $\bm{x}$ to sample its function output.
  \item Classical or quantum coefficient estimation: using the sampled outputs, we fit spline coefficients either classically or using quantum linear solvers such as the Harrow–Hassidim–Lloyd (HHL) algorithm \citep{Harrow_2009}.
  \item VAF replacement: the estimated coefficients are used to define classical B-spline or Fourier basis functions that replace the DARUAN activations in the classical KAN.
\end{enumerate}

This strategy enables training on quantum hardware and inference on classical machines, facilitating hybrid deployment.
Additionally, for Fourier-based replacements, coefficients can be estimated using either the discrete Fourier transform (DFT) or quantum Fourier transform (QFT), enabling flexible and efficient post-training adaptation.

\subsection{Distributed Training of QKANs}
If deployed on quantum hardware for inference or gradient evaluation, the architecture of QKANs is naturally parallelizable because each DARUAN activation is independent.

Each activation is implemented as a single-qubit data re-uploading circuit, requiring no entanglement between qubits.
For a QKAN layer with $m$ input nodes and $n$ output nodes, a total of $(m \cdot n)$ independent qubits are required to compute the layer outputs in parallel.
To process mini-batches of size $b$, we consider two distributed strategies:
\begin{enumerate}[leftmargin=*,label=\roman*]
  \item \textbf{Synchronous parallelism:} Execute $(b \cdot m \cdot n)$ qubits simultaneously to process all batches in parallel.
  \item \textbf{Asynchronous parallelism:} Run $b$ quantum systems independently, each with $(m \cdot n)$ qubits, and perform asynchronous gradient updates.
\end{enumerate}

These distributed approaches are compatible with current quantum cloud infrastructures, which often impose qubit count limitations per device.
As such, QKANs can be deployed on multiple quantum machines with classical communication links for parameter synchronization.
Moreover, it is compatible to quantum federated learning (QFL) \citep{e23040460}, each quantum machine can independently compute gradients for its local data batch, and the results can be aggregated to update the global model parameters.

This architecture aligns well with the emerging paradigm of quantum-centric supercomputing and enables \textit{quantum-accelerated learning} via parallelized quantum computation, similar to tensor parallelism in classical deep learning frameworks.
This also enables us to perform classical simulations efficiently on both personal computers (PCs) and HPCs, particularly for multi-node clusters that can handle large-scale training tasks, such as training LLMs with our QKAN.

\subsection{QKAN Implementation Methods}\label{sec:impl}

\paragraph{Layer extension.}

Analogous to the grid extension strategy used in KANs to refine approximation accuracy~\citep{liu2024kan}, QKANs leverage a technique termed \textit{layer extension}, wherein the number of data re-uploading repetitions $r$ is progressively increased.
This extension enriches the model's frequency spectrum and improves its capacity to approximate complex univariate functions.

Practically, layer extension for data re-uploading circuit can be interpreted as a transition from a shallower, pre-trained model to a deeper one.
Specifically, the learned parameters from the original model are retained, while the newly added parameterized unitaries in the extended layers are initialized to identity (i.e., their corresponding parameters are set to zero).
In this configuration, the new encoding blocks $S(x)$ do not contribute to the output, as the new parameterized unitaries $W$ act trivially on the quantum state.
And it is easy to see that the results are remained the same after layer extension.

As training proceeds, these newly introduced layers begin to adjust, allowing the extended DARUAN to fine-tune the overall function approximation without disrupting the performance already achieved by the shallower model.
In practice, layer extension also enables an adaptive resource-allocation strategy.
Since each extension stage increases the repetition number $r$, and hence the number of trainable circuit parameters and the cost of evaluating the DARUAN activation, one may monitor the validation loss after each stage and stop once a target error tolerance is achieved.
This avoids training unnecessarily large-$r$ models when a smaller repetition number already provides sufficient accuracy, thereby reducing both computational and parameter overhead.

\paragraph{Post-activation process.}

The output of a QKAN layer is intrinsically bounded due to the nature of quantum measurement.
Specifically, the expectation value of a single-qubit observable, such as $\langle \sigma_z \rangle$, is confined to the interval $[-1, 1]$.
Consequently, for $l$-th QKAN layer receiving $n_{l}$ inputs and producing $n_{l+1}$ outputs, the output domain is constrained to $\bm{y} \in [-n_{l}, n_{l}]^{n_{l+1}}$, assuming summation over independent channels.
While this bounded output range is beneficial for stability, it can be limiting when modeling functions that require outputs beyond this interval.
Increasing the number of hidden units to expand the range is not always practical due to parameter overhead and architectural constraints.

To overcome this limitation, we explore two post-activation strategies:

First, a lightweight output mapping network, such as a shallow MLP, can be appended to the QKAN.
Given the universal approximation capabilities of MLPs, such a model can flexibly rescale or reshape the QKAN output to meet task-specific requirements.
This hybrid setup preserves the advantages of QKAN, including reduced parameter count and quantum-inspired expressivity, while enabling output adaptation for broader applications.

Second, we incorporate learnable scaling and bias parameters directly into the QKAN output.
These parameters act multiplicatively and additively on the expectation values produced by the quantum circuits, allowing direct rescaling without architectural expansion.
This approach is both efficient and seamlessly integrates with the QKAN framework.
Importantly, it keeps the compatibility of parameter transfer mechanisms from QKAN to classical KAN, preserving interoperability between quantum and classical VAFs.

Together, these methods ensure that QKANs remain scalable and adaptable across a variety of tasks.
Moreover, they highlight QKAN’s flexibility as a modular component that can be combined with classical models to enhance performance or meet specific output constraints.
These strategies enhance the practical use of QKAN and pave the way for integrating QVAFs with conventional learning pipelines.

\paragraph{Base activation function.}

To further enhance training stability and optimization in QKAN, we incorporate a \emph{base activation function}, inspired by a similar mechanism in the original KAN framework \citep{liu2024kan}.
This base activation acts analogously to a residual connection, providing a direct and smooth signal path during learning.

The output with the residual activation function is defined as:
\begin{equation}
  \phi(x) = w_b b(x) + w_d \langle 0 | U(x, \boldsymbol{\theta})^\dagger \mathcal{M} U(x, \boldsymbol{\theta}) | 0 \rangle,
\end{equation}
where $w_b$ and $w_d$ are learnable weights, and $b(x)$ denotes the base activation, chosen here as the SiLU function, i.e., $b(x) = x \cdot \text{sigmoid}(x)$.

This residual pathway ensures that even if in the early stages of training, when VAFs and QVAFs may be poorly initialized, the model still has access to a smooth nonlinear function.
As a result, the learning landscape is improved, and optimization becomes more robust.
This method is particularly beneficial in deeper QKANs or settings with limited data, where maintaining gradient flow is critical.

\subsection{Numerical Methods}\label{sec:numerica_method}

\paragraph{Function regression with heuristic architectures.}
Ten multivariate equations are drawn from the Feynman dataset \citep{Udrescu:2019mnk,10.5555/3495724.3496132}.
For each function \(f\), inputs \(\bm{x}_i\in(-1,1)^{d}\) are sampled uniformly and noisy targets are generated as
\begin{equation}
  y_i \;=\; f(\bm{x}_i)\;+\;\epsilon,\qquad \epsilon\sim\mathcal{N}(0,\,0.1\,\mu_f)\,,
\end{equation}
where $\mu_f$ is the mean value of $f(\bm{x})$ over the input domain.
Each dataset comprises 1,000 training and 1,000 test points.

Three model classes are compared:
\begin{enumerate}[label=\roman*.]
  \item \textbf{KAN.}
        Hidden‐layer shapes are adopted from the best settings reported in Table~2 of ref.~\cite{liu2024kan}.
        The grid number $G$ is extended over $\{5, 10,\dots,50\}$.
  \item \textbf{QKAN.}
        The same layer counts and widths as in the corresponding KANs are employed.
        Data re-uploading repetition number $r$ is extended over $\{3,6,\dots,30\}$.
  \item \textbf{MLP.}
        A fixed hidden-layer width of 5 is used, with depths chosen from $\{3,5,10\}$.
\end{enumerate}
All parameters are initialized at random and optimized via L-BFGS \citep{Liu1989OnTL} for 200 epochs.
For each combination of architecture and extension hyperparameter, five independent runs are performed; \cref{tab:main_table,tab:reg-param} report mean $\pm$ standard deviation of the selected setting.

\paragraph{Function regression with empirical architectures.}
To further assess performance without prior knowledge of layer shapes, a broader benchmark of 66 equations is selected from the Feynman dataset (see \cref{sec:sup_fig3_details,tab:com_reg}).
Datasets are constructed as above, with inputs restricted to avoid singularities and 10\% additive noise.

Model families and hyperparameter sweeps are defined as follows:
\begin{enumerate}[label=\roman*.]
  \item \textbf{KAN.}
        Five hidden-layer depths $\{2,3,4,5,6\}$ are tested with fixed width 5, and grid numbers $G\in\{5,10,15,20,25\}$.
  \item \textbf{QKAN.}
        Depths match those of KAN, with width 5; data-re-uploading repetitions $r\in\{3,6,9,12,15\}$.
  \item \textbf{MLP.}
        Width is fixed to 5, and depths vary over $\{2,3,4,5,6\}$.
\end{enumerate}
Training is performed as above. For each hyperparameter combination, five random seeds are used and the best test RMSE is recorded.
Results are presented in \cref{fig:loss}.

\paragraph{Image classification.}

We evaluate the classification performance of QKAN, KAN, and MLP modules on three datasets: MNIST~\citep{mnist}, CIFAR-10, and CIFAR-100~\citep{krizhevsky2009learning}.
MNIST consists of $28\times28$ grayscale digits, while CIFAR-10 and CIFAR-100 consist of $32\times32$ RGB natural images.
All datasets are normalized such that pixel values are centered at 0.5 with a standard deviation of 0.5.

For each dataset, we construct a CNN backbone comprising either two (for MNIST) or three layers (for CIFAR-10 and CIFAR-100), each consisting of a 2D convolutional layer, followed by a ReLU activation and a 2D max-pooling layer.
The resulting feature maps are then flattened and passed to a FCN, instantiated as either an MLP, KAN, or QKAN.

All models are trained for 100 epochs using the Adam optimizer~\citep{Kingma2014AdamAM} with a learning rate of $10^{-3}$ and a batch size of 1000.
The baseline CNN+MLP uses a fixed hidden width of 5 and serves as a standard reference.
For QKAN, we sweep over multiple CNN output sizes and QKAN configurations, fixing the re-uploading repetition number $r=3$ and selecting the best performing combination.
KAN models are then configured to match the chosen CNN features and evaluated with a fixed grid number of $G=10$.

The total parameter count of the convolutional backbone is reported separately from the FCN to clearly illustrate the contribution of the representation module.
We conduct all evaluations using the same initialization and training setup for consistency.

In addition to standard QKAN architectures, we design a parameter-efficient variant referred to as HQKAN.
HQKAN introduces two auxiliary fully connected layers before and after the QKAN block, respectively, functioning as feature compressor and expander.
This design mimics an autoencoder structure, wherein the high-dimensional feature vector is downscaled to a latent representation whose size is logarithmic in the original dimension.
The compressed features are processed by the QKAN layer, which benefits from its high expressivity even in small latent spaces, before being upscaled again to match the required output dimension.

During training, the latent dimension is chosen based on the logarithm of the original input size and output size to QKAN.
All other training configurations, including optimizer, learning rate, and batch size, are kept consistent with those used for the main QKAN and KAN experiments.
This setup enables direct comparisons in both performance and model compactness.

\paragraph{Natural language generation.}

To assess the generative modeling capabilities of QKANs and their variants, we utilize the open-source \texttt{kan-gpt} module~\citep{GANESH2024KANGPT}, which replaces all linear layers in the GPT architecture with either QKAN modules.
For the implementation of \textit{QKANsformer}, we modified an open-source \texttt{nanoGPT}~\citep{nanogpt} project and customized the feed-forward network to utilize HQKANs.
Our experiments are conducted on the WebText dataset~\citep{webtext}, with GPT-2 as the backbone model.
The GPT-2 architecture consists of 12 transformer layers, each comprising 12 attention heads, with an embedding dimension of 768.
All models are evaluated using perplexity as the primary metric.

Training is performed over 2000 epochs using the AdamW optimizer~\citep{Loshchilov2017DecoupledWD} with a batch size of 1, a learning rate of $5 \times 10^{-3}$, and momentum parameters $\beta_1 = 0.9$, $\beta_2 = 0.95$.
A weight decay of $0.1$ is applied to all matrix multiplication weights in both linear and QKAN-based layers.

For large-scale batch size, models are benchmarked using multi-GPU clusters setups, including 4×V100S (1 node), 32×H100 (4 nodes), and 16×H200 (2 nodes) configurations.
GPUs in each node are interconnected in pairs using NVLink bridges or NVSwitch technologies, facilitating efficient data transfer within nodes.
In the multi-node training setup, each node is interconnected via InfiniBand (IB), enabling high-throughput and low-latency communication essential for efficient distributed learning.

\paragraph{Transferring VAFs from QKAN to KAN.}

We design a two-phase training strategy: (1) QKAN is trained on a target regression task for 500 epochs, and (2) its learned activation parameters are mapped to a classical B-spline basis and loaded into a KAN of equivalent depth and width.
Due to mismatch between the QKAN’s Fourier-based activations and KAN’s B-spline basis, a small loss offset appears upon transfer.
To mitigate this, the KAN continues training from the transferred parameters for an additional 1000 epochs.

For comparison, we train a KAN from scratch under the same architecture and with a fixed total training epochs.
The grid size is set to 5 in both models, while data re-uploading repetition number is set to 3 for QKAN.
We evaluate mean squared error over a held-out test set of 1,000 samples drawn from the input domain \((x, y) \in [-1, 1]^2\).
Training is performed using the L-BFGS optimizer with identical learning settings across all models.

\section*{Data and Code Availability}
The data used and generated in this study is available at: \url{https://github.com/Jim137/qkan}~\citep{jiang2025qkan_github}.
The code used in this study, implemented using PyTorch~\citep{Ansel_PyTorch_2_Faster_2024}, is available at: \url{https://github.com/Jim137/qkan}~\citep{jiang2025qkan_github}.

\ifarxiv
\section*{Acknowledgements}
J.-C. Jiang would like to thank Qian-Rui Lee, Damien Jian, Chen-Yu Liu, and Dr. Samuel Yen-Chi Chen for helpful discussions and comments.
J.-C. Jiang also thanks the National Center for High-Performance Computing (NCHC), National Institutes of Applied Research (NIAR), Taiwan, for providing computational and storage resources supported by NSTC, Taiwan, under Grants No. NSTC 114-2119-M-007-013.
H.-S. Goan acknowledges support from the NSTC, Taiwan, under Grants No. NSTC 113-2112-M-002-022-MY3, No. NSTC 113-2119M-002-021, No. NSTC 114-2119-M-002-018, No. NSTC 114-2119-M-002-017-MY3, and No. NSTC 115-2119-M-002-005 and from the National Taiwan University under Grants No. NTU-CC-115L8937, No. NTU-CC-115L893704 and No. NTU-CC-115L8512.  H.-S. Goan is also grateful for the support of the “Center for Advanced Computing and Imaging in Biomedicine (NTU-115L900702)” through the Featured Areas Research Center Program within the framework of the Higher Education Sprout Project by the Ministry of Education (MOE), Taiwan, the support of Taiwan Semiconductor Research Institute (TSRI) through the Joint Developed Project (JDP) and the support from the Physics Division, National Center for Theoretical Sciences, Taiwan.
The authors acknowledge the National Taiwan University--IBM Quantum Hub (NTU--IBM Q Hub) for providing IBM Quantum system, and NVIDIA AI Technology Center (NVAITC), NVIDIA for their technical support in accelerating computing.
\else
\begin{ack}
J.-C. Jiang would like to thank Qian-Rui Lee, Damien Jian, Chen-Yu Liu, and Dr. Samuel Yen-Chi Chen for helpful discussions and comments.
J.-C. Jiang also thanks the National Center for High-Performance Computing (NCHC), National Institutes of Applied Research (NIAR), Taiwan, for providing computational and storage resources supported by NSTC, Taiwan, under Grants No. NSTC 114-2119-M-007-013.
H.-S. Goan acknowledges support from the NSTC, Taiwan, under Grants No. NSTC 113-2112-M-002-022-MY3, No. NSTC 113-2119M-002-021, No. NSTC 114-2119-M-002-018, No. NSTC 114-2119-M-002-017-MY3, and No. NSTC 115-2119-M-002-005 and from the National Taiwan University under Grants No. NTU-CC-115L8937, No. NTU-CC-115L893704 and No. NTU-CC-115L8512.  H.-S. Goan is also grateful for the support of the “Center for Advanced Computing and Imaging in Biomedicine (NTU-115L900702)” through the Featured Areas Research Center Program within the framework of the Higher Education Sprout Project by the Ministry of Education (MOE), Taiwan, the support of Taiwan Semiconductor Research Institute (TSRI) through the Joint Developed Project (JDP) and the support from the Physics Division, National Center for Theoretical Sciences, Taiwan.
The authors acknowledge the National Taiwan University--IBM Quantum Hub (NTU--IBM Q Hub) for providing IBM Quantum system, and NVIDIA AI Technology Center (NVAITC), NVIDIA for their technical support in accelerating computing.
\end{ack}
\fi

\section*{Author Contribution}
The project was conceived by J.-C.J.
The theoretical aspects of this work were developed by M.Y.-C.H.
The numerical experiments were conducted by J.-C.J.
T.C. discussed the results and reviewed the manuscript.
The project is supervised by H.-S.G.
All authors contributed to technical discussions and the writing of the manuscript and have read and approved it.

\section*{Conflict of Interest}
Author Hsi-Sheng Goan is an Associate Editor of EPJ: Quantum Technology, International Journal of Quantum Information, and Chinese Journal of Physics.
Author Tianlong Chen is an Associate Editor of npj Artificial Intelligence.
Hsi-Sheng Goan and Tianlong Chen were not involved in the journal's review of, or decisions related to, this manuscript.

\bibliographystyle{plainnat}
\bibliography{ref}

\newpage
\appendix

\section{Supplementary Theoretical Backgrounds}\label{sec:sup_theory}

This section provides the theoretical details supporting the main text. QVAF is the general activation-function framework and may be instantiated by genuine multi-qubit QML circuits; the results below focus on DARUAN, the single-qubit QVAF used in QKAN. \cref{thm:Approx_theory} establishes the Fourier-series approximation benchmark for KAN edge functions, adapting the approximation step in the KAN theory of~\citet{liu2024kan}. \cref{thm:qkan_linear} combines the standard spectral representation of data re-uploading circuits~\citep{Schuld_2021} with trainable data-preprocessing weights, showing that a geometric choice of these weights gives an exponentially large filled accessible frequency block with only linearly many re-uploading repetitions. This is a capacity statement about accessible support and Fourier approximation; the Fourier coefficients remain coupled through shared single-qubit rotation parameters. \cref{prop:nonvanishing_coeff} addresses the complementary coefficient-accessibility question by showing that, under a nondegenerate single-qubit mixing assumption, no nonzero accessible frequency is forced to vanish by a parameter-independent cancellation. The empirical consequences of this structured parameterization are examined in \cref{sec:sup_baselines}.

\subsection{Proof of \texorpdfstring{\cref{thm:Approx_theory}}{Theorem}}\label{sec:approx_theory}

\begin{proof}[Proof of \cref{thm:Approx_theory}]
The argument follows the layer-replacement proof used for KAN approximation theory, with the spline approximation step replaced by a trigonometric-polynomial approximation. For each edge function $\phi_{l,j,i}$, classical Fourier approximation theory gives a trigonometric polynomial $\phi^K_{l,j,i}$ with frequencies $|n|\le K$ such that, for $0\le m\le k$,
\begin{equation}
  \|\phi_{l,j,i}-\phi^K_{l,j,i}\|_{C^m}
  \le C_{l,j,i}K^{-(k+1-m)} ,
\end{equation}
on the compact interval containing the corresponding edge input. The constant depends on the edge function and the chosen normalization, but not on $K$.

Because the KAN representation is fixed and all relevant functions are smooth on compact domains, composition by the remaining layers is locally Lipschitz in the $C^m$ norm. Thus replacing one layer at a time introduces a residual
\begin{equation}
\begin{aligned}
R_l
=&\,
(\Phi^K_L\circ\cdots\circ\Phi^K_{l+1}\circ\Phi_l\circ\Phi_{l-1}\circ\cdots\circ\Phi_1)(\bm{x})\\
&-
(\Phi^K_L\circ\cdots\circ\Phi^K_{l+1}\circ\Phi^K_l\circ\Phi_{l-1}\circ\cdots\circ\Phi_1)(\bm{x})
\end{aligned}
\end{equation}
satisfying
\begin{equation}
  \|R_l\|_{C^m}\le C_lK^{-(k+1-m)} .
\end{equation}
Summing the finitely many layer residuals gives
\begin{equation}
\begin{aligned}
\left\|
f-(\Phi^K_L\circ\cdots\circ\Phi^K_1)(\bm{x})
\right\|_{C^m}
&\le
\sum_{l=1}^L\|R_l\|_{C^m}  \\
&\le
\left(\sum_{l=1}^L C_l\right)K^{-(k+1-m)} .
\end{aligned}
\end{equation}
Since the representation depth $L$ is fixed, the sum of constants can be absorbed into a single constant $C_f$, yielding
\begin{equation}\label{eqn:KAN-approx}
  \left\|
  f-(\Phi^K_L\circ\cdots\circ\Phi^K_1)(\bm{x})
  \right\|_{C^m}
  \le
  C_fK^{-(k+1-m)} .
\end{equation}
This completes the proof.
\end{proof}

\subsection{Proof of \texorpdfstring{\cref{thm:qkan_linear}}{Theorem}}\label{sec:reupload_fourier}

We restate \cref{thm:qkan_linear} from the main paper for completeness and provide the detailed proof. Relative to prior spectral analyses of data re-uploading circuits, the additional step here is the trainable preprocessing weight vector $\bm{\omega}$, which expands the accessible support from the arithmetic spectrum of repeated identical encodings to subset-sum spectra controlled by $\{w_\ell\}_{\ell=1}^r$.
\qkan*

\begin{proof}[Proof of \cref{thm:qkan_linear}]

The proof of part~\textup{(a)} follows the spectral representation of data-encoding circuits~\citep{Schuld_2021}, specialized to one qubit. Consider
\begin{equation}\label{eq:quantum_model_app}
  f_A(x)=\langle0|U^\dagger(x)\mathcal M U(x)|0\rangle,
\end{equation}
where
\begin{equation}
  U(x)=W^{(r+1)}S(x)W^{(r)}\cdots S(x)W^{(1)},
  \qquad
  S(x)=e^{-ixH}.
\end{equation}
For a single qubit, $H=\sigma_j/2$ has eigenvalues $\pm 1/2$. Since $H$ is Hermitian, it can be unitarily diagonalized as
\begin{equation}
  H=V^\dagger \frac{\sigma_z}{2}V .
\end{equation}
The corresponding $V$ and $V^\dagger$ factors can be absorbed into the adjacent trainable unitaries. Thus, without loss of generality, we may take the data encoder to be
\begin{equation}
  S(x)=e^{-ix\sigma_z/2}.
\end{equation}

Next, we expand the quantum state $U(x)|0\rangle$ in terms of the eigenvalues of $H$.  
Denote multi-indices $\bm{j}=(j_1,j_2,\ldots,j_r)\in\{1,2\}^r$ and define the sums of eigenvalues
\begin{align}\label{eqn:eigens}
  \Lambda_{\bm{j}} \;=\; \lambda_{j_1} + \lambda_{j_2} + \cdots + \lambda_{j_r},\\
  \quad
  \lambda_1 = \tfrac{1}{2},\;
  \lambda_2 = -\tfrac{1}{2}.
\end{align}
The components of the state are
\begin{align}
[U(x)|0\rangle]_i
  = \sum_{\bm{j} \in \{1,2\}^r} 
     e^{-i \Lambda_{\bm{j}} x}\,
     W^{(r+1)}_{i j_r}\cdots W^{(2)}_{j_2 j_1}W^{(1)}_{j_1 1}.
\end{align}
Similarly, the adjoint operation gives $\langle 0|U^\dagger(x) = [U(x)|0\rangle]^\dagger$.  
Substituting into \eqref{eq:quantum_model_app} yields
\begin{align}
f_A(x) = \sum_{\bm{k},\bm{j} \in \{1,2\}^r} 
        e^{i (\Lambda_{\bm{k}} - \Lambda_{\bm{j}}) x}\,
        a_{\bm{k},\bm{j}},
\end{align}
where $a_{\bm{k},\bm{j}}$ are coefficients involving the unitaries $W^{(\ell)}$ and the measurement observable $\mathcal{M}$:
\begin{align}
a_{\bm{k},\bm{j}} 
  = \sum_{i,i'}
      \bigl(W^{(1)}_{1 k_1}\bigr)^{\!*}
      \bigl(W^{(2)}_{k_1 k_2}\bigr)^{\!*}
      \cdots
      \bigl(W^{(r+1)}_{k_r i}\bigr)^{\!*}
      \mathcal{M}_{i i'}
      W^{(r+1)}_{i' j_r}
      \cdots
      W^{(2)}_{j_2 j_1}
      W^{(1)}_{j_1 1}.
\end{align}
We group terms with the same frequency $\omega = \Lambda_{\bm{k}} - \Lambda_{\bm{j}}$ and define the frequency spectrum
\begin{align}\label{eqn:freq_spec-app}
\Omega_\text{freq} = \{\,\omega = \Lambda_{\bm{k}} - \Lambda_{\bm{j}} \mid \bm{k},\bm{j} \in \{1,2\}^r\}.
\end{align}
Thus, the quantum model $f_A(x)$ can be rewritten as
\begin{align}
f_A(x) = \sum_{\omega \in \Omega_\text{freq}} c_\omega e^{i \omega x},
\end{align}
where the coefficients $c_\omega$ are given by
\begin{align}
c_\omega 
  = \sum_{\substack{\bm{k},\bm{j} \in \{1,2\}^r \\ \Lambda_{\bm{k}} - \Lambda_{\bm{j}} = \omega}}
      a_{\bm{k},\bm{j}}.
\end{align}
Because each $\Lambda_{\bm{j}}$ is a sum of $r$ eigenvalues and the eigenvalue gap is normalized to one, the differences $\omega=\Lambda_{\bm{k}}-\Lambda_{\bm{j}}$ are integer frequencies:
\begin{align}
  \Omega_A 
  = \{\,m \mid m = -r,-(r-1),\ldots,r\,\}.
\end{align}
Accounting for the zero frequency separately, the maximum number of distinct nonzero frequencies is therefore
\begin{align}
    |\Omega_A\setminus\{0\}| = 2r.
\end{align}
This completes the proof of (a).

For part~\textup{(b)}, the generator remains $H=\sigma_j/2$, but the $\ell$-th data encoder receives the scaled input $w_\ell x$:
\begin{equation}
  S(w_\ell x)=e^{-iw_\ell x\sigma_j/2}.
\end{equation}
In the state expansion, a path accumulates an eigenphase
\begin{equation}
  \frac12\sum_{\ell=1}^r \tau_\ell w_\ell x,
  \qquad
  \tau_\ell\in\{+1,-1\}.
\end{equation}
The frequency difference between two paths is therefore
\begin{equation}
  \sum_{\ell=1}^r
  \frac{\tau_\ell-\tau_\ell'}{2}w_\ell
  =
  \sum_{\ell=1}^r m_\ell w_\ell,
  \qquad
  m_\ell\in\{-1,0,1\}.
\end{equation}
Hence
\begin{equation}
  \Omega_B(\bm{\omega})
  =
  \left\{
  \sum_{\ell=1}^r m_\ell w_\ell
  \;\middle|\;
  m_\ell\in\{-1,0,1\}
  \right\}.
\end{equation}
There are at most $3^r$ ternary strings, one of which gives the zero sum, so
\begin{equation}
  |\Omega_B(\bm{\omega})\setminus\{0\}|
  \le 3^r-1 .
\end{equation}
Since $w_1>0$, the two frequencies $\pm w_1$ are accessible, giving the lower bound
\begin{equation}
  |\Omega_B(\bm{\omega})\setminus\{0\}|\ge 2 .
\end{equation}
Finally,
\begin{equation}
  K_{\rm form}(\bm{\omega})
  =
  \max_{\nu\in\Omega_B(\bm{\omega})}|\nu|
  =
  \sum_{\ell=1}^r w_\ell ,
\end{equation}
attained by taking all $m_\ell=1$ or all $m_\ell=-1$. This proves part~\textup{(b)}.

For (c), let
\begin{equation}
  \mathcal T_K=\operatorname{span}\{e^{i n x}: |n|\le K,\ n\in\mathbb Z\}.
\end{equation}
The standard Jackson-type Fourier estimate for a function with a $C^{k+1}$ periodic extension gives, for $0\le m\le k$,
\begin{equation}
  \inf_{p\in\mathcal T_K}\|f-p\|_{C^m}
  \le C_fK^{-(k+1-m)}.
\end{equation}
Applying this estimate to the filled integer block contained in the accessible frequency spectrum proves the two capacity bounds in part (c): the baseline has filled bandwidth $r$, and the weighted model has filled bandwidth $K_{\rm fill}(\bm{\omega})$. This step is a Fourier approximation benchmark; it does not assume that the coupled DARUAN parameters independently set every Fourier coefficient.

For (d), choose the constructive geometric weights $w_\ell=2^{\ell-1}$. Every integer $n$ with $|n|\le 2^r-1$ has a signed binary representation $n=\sum_{\ell=1}^r m_\ell2^{\ell-1}$ with $m_\ell\in\{-1,0,1\}$. Hence
\begin{equation}
  \Omega_B(\bm{\omega})=\{-(2^r-1),\ldots,-1,0,1,\ldots,2^r-1\},
\end{equation}
and $K_{\rm fill}(\bm{\omega})=K_{\rm form}(\bm{\omega})=2^r-1$. Substituting this filled bandwidth into part (c) gives
\begin{equation}
  \inf_{p\in\operatorname{span}\{e^{i\nu x}:\nu\in\Omega_B(\bm{\omega})\}}
  \|f-p\|_{C^m}
  \le C_f(2^r-1)^{-(k+1-m)}
  \le C_f'2^{-r(k+1-m)}.
\end{equation}
Solving $C_f'2^{-r(k+1-m)}=\varepsilon$ gives $r=\Theta(\log(1/\varepsilon))$. In contrast, setting the Fourier-series benchmark \eqref{eqn:KAN-approx} equal to $\varepsilon$ requires $K=\Theta(\varepsilon^{-1/(k+1-m)})$ modes for an independently parameterized Fourier-series-based KAN. Thus part (d) is a capacity comparison between accessible frequency spectrums, while the QKAN coefficients remain coupled through the circuit parameters.
This completes the proof.
\end{proof}

\subsection{Proof of \texorpdfstring{\cref{prop:nonvanishing_coeff}}{Proposition}}\label{sec:prop_nonvanishing_coeff}

The preceding theorem proves the support and Fourier approximation capacity statements. We next give the independent coefficient-level construction used in the main text to show that the accessible frequencies are not merely formal in the single-qubit weighted case.

\nonvanishingcoeff*

\begin{proof}[Proof of \cref{prop:nonvanishing_coeff}]
Fix the realized weights $\bm{\omega}$. We work in the Heisenberg picture and use the traceless Pauli-transfer representation. The identity component is invariant under conjugation and contributes only to the zero-frequency part, so it can be omitted for the present nonzero-frequency claim. Let
\begin{equation}
  E_+=\frac{\sigma_x+i\sigma_y}{2},\qquad
  E_-=\frac{\sigma_x-i\sigma_y}{2},\qquad
  E_0=\sigma_z,
\end{equation}
and define
\begin{equation}
  \eta(+)=1,\qquad \eta(0)=0,\qquad \eta(-)=-1.
\end{equation}
For the data-encoding gate $S(w_\ell x)=e^{-i w_\ell x\sigma_z/2}$, we have
\begin{equation}
  S(w_\ell x)^\dagger E_s S(w_\ell x)
  =
  e^{i\eta(s)w_\ell x}E_s,
  \qquad s\in\{+,0,-\}.
\end{equation}
Thus, whenever the Heisenberg-evolved observable passes through basis element $E_{s_\ell}$ at the $\ell$-th data encoding, that encoding contributes the phase $e^{i\eta(s_\ell)w_\ell x}$.

For any single-qubit unitary $V$, let $T(V)$ denote its transfer matrix in the basis $\{E_+,E_0,E_-\}$, defined by
\begin{equation}
  V^\dagger E_t V=\sum_{s\in\{+,0,-\}}T(V)_{s,t}E_s .
\end{equation}
Expanding the full circuit in this transfer basis gives a path expansion
\begin{equation}
  f_B(x;\Theta)
  =
  \sum_{\bm{s}\in\{+,0,-\}^r}
  A_{\bm{s}}(\Theta)
  \exp\!\left(
     i\sum_{\ell=1}^r \eta(s_\ell)w_\ell x
  \right),
\end{equation}
where $\bm{s}=(s_1,\ldots,s_r)$ records which transfer-basis element passes through each data-encoding gate. The coefficient $A_{\bm{s}}(\Theta)$ is a product of transfer-matrix entries from the trainable interleaving unitaries, including the two boundary blocks and the overlaps with the initial state and the Pauli-$Z$ readout.

We now show that the sum of all paths producing any fixed nonzero frequency cannot vanish identically. It is enough to exhibit a parameter subfamily on which the path contributions are nonzero and carry distinct phase characters. Choose each trainable block to be of the tagged-mixer form
\begin{equation}
  W^{(\ell)}=R_z(a_\ell)R_y(\beta)R_z(b_\ell),
  \qquad \beta=\pi/3,
\end{equation}
with independently variable phases $a_\ell,b_\ell$. In the transfer basis,
\begin{equation}
  R_z(\phi)^\dagger E_s R_z(\phi)
  =
  e^{i\eta(s)\phi}E_s .
\end{equation}
Moreover, for $\beta=\pi/3$, the transfer matrix of $R_y(\beta)$ has no zero entries in the basis $\{E_+,E_0,E_-\}$. Indeed, its entries are combinations of
\begin{equation}
  \frac{1+\cos\beta}{2},\qquad
  \frac{1-\cos\beta}{2},\qquad
  \frac{\sin\beta}{2},\qquad
  \cos\beta,
\end{equation}
all of which are nonzero at $\beta=\pi/3$.

Let $s_0=s_{r+1}=0$ denote the boundary Pauli-$Z$ components. For a path $\bm{s}=(s_1,\ldots,s_r)$, the corresponding prefactor along the tagged-mixer subfamily has the form
\begin{equation}
  A_{\bm{s}}(\Phi)
  =
  B_{\bm{s}}
  \exp\!\left(
    i\sum_{\ell=1}^{r+1}
    \bigl[
      \eta(s_\ell)a_\ell+\eta(s_{\ell-1})b_\ell
    \bigr]
  \right),
\end{equation}
where $\Phi$ denotes all phase tags and $B_{\bm{s}}\ne0$ because every transition entry of the fixed $R_y(\pi/3)$ mixer is nonzero. If two paths $\bm{s}\ne\bm{s}'$ differ at some layer $j$, then $\eta(s_j)\ne\eta(s'_j)$, and hence the exponent of the independent phase $a_j$ differs. Therefore the corresponding phase characters are distinct.

Now fix any nonzero frequency $\nu\in\Omega_B(\bm{\omega})$. By definition of $\Omega_B(\bm{\omega})$, there is at least one path $\bm{s}$ such that
\begin{equation}
  \sum_{\ell=1}^r \eta(s_\ell)w_\ell=\nu .
\end{equation}
The aggregated coefficient of $e^{i\nu x}$, restricted to the tagged-mixer subfamily, is
\begin{equation}
  c_\nu(\Phi)
  =
  \sum_{\substack{\bm{s}\in\{+,0,-\}^r:\\
  \sum_{\ell=1}^r \eta(s_\ell)w_\ell=\nu}}
  B_{\bm{s}}\,
  \chi_{\bm{s}}(\Phi),
\end{equation}
where each $B_{\bm{s}}\ne0$ and the characters $\chi_{\bm{s}}$ are distinct. Distinct characters on a torus are linearly independent, so this finite trigonometric polynomial is not identically zero. Thus $c_\nu$ is not the zero function of the trainable unitary parameters.

Finally, $c_\nu(\Theta)$ is a complex-valued real-analytic function of the single-qubit unitary parameters in any Euler-angle chart. Since it is not identically zero, at least one of its real or imaginary parts is a nonzero real-analytic function; the zero set of such a function has measure zero. Therefore the set on which $c_\nu(\Theta)=0$ has measure zero. Because $\Omega_B(\bm{\omega})$ is finite, the union of these zero sets over all nonzero accessible frequencies also has measure zero. Hence generic choices of the unitary parameters make all the coefficients of the nonzero accessible frequencies simultaneously nonzero.

The proposition proves coefficient accessibility only. It does not imply independent coefficient control, a lower bound on coefficient magnitudes, or optimization convergence to a desired Fourier series.
\end{proof}

\paragraph{Relation to frequency-profile analyses.}
The coefficient-accessibility result should be distinguished from the average frequency-profile results of \citet{barthe2024gradients}. In \cref{thm:qkan_linear}, the preprocessing weight vector is trainable. The only place where a specific vector is prescribed is \cref{thm:qkan_linear}\textup{(d)}, where $w_\ell=2^{\ell-1}$ is used as a constructive witness inside the trainable family. In \cref{prop:nonvanishing_coeff}, by contrast, we condition on an arbitrary realized value of $\bm{\omega}$ in order to ask whether any nonzero frequency in $\Omega_B(\bm{\omega})$ is forced to vanish by the single-qubit architecture.

The two statements answer different questions. Barthe and P\'erez-Salinas analyze typical Fourier-content profiles under ensemble assumptions. For repeated identical data generators, the profile is governed by repeated convolution of the same spectrum kernel and concentrates near low frequencies, so high-frequency coefficients can be small on average even when the formal support is larger. \cref{prop:nonvanishing_coeff} instead proves that, for any realized weighted single-qubit encoding and nondegenerate mixing blocks, each nonzero accessible coefficient is a nonzero analytic function of the trainable unitary parameters. A coefficient can therefore be structurally accessible but still small for typical parameter draws.

The geometric construction in \cref{thm:qkan_linear}\textup{(d)} changes the relevant profile setting because it uses layer-dependent effective generators $w_\ell H$, rather than repeated copies of the same generator. It is therefore aligned with the powers-of-two single-qubit example discussed by \citet{barthe2024gradients}, where the convolution support spreads over an exponentially large set. The theorem uses this construction only as a capacity witness; the empirical sections then test whether the resulting structured coefficient coupling is useful in optimization and generalization.

\section{Numerical Results Details}\label{sec:sup_numerical_details}

\begin{sidewaysfigure*}[ptbh]
  \centering
  \includegraphics[width=1\paperwidth]{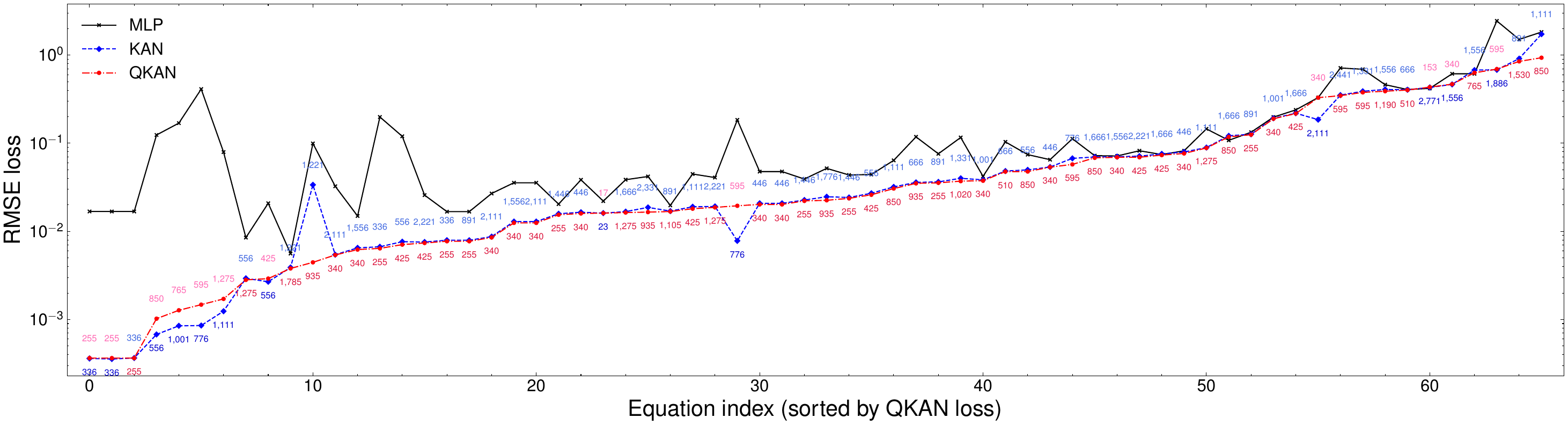}
  \caption{\textbf{Empirical noisy function fitting.}
    Test root-mean-square error (RMSE) on the 66-equation Feynman regression benchmark.
    The vertical axis shows the test RMSE on a logarithmic scale (lower is better), and the horizontal axis indexes the equations, ordered by increasing quantum-inspired Kolmogorov-Arnold network (QKAN) loss.
    For each equation, the plotted value is the average test RMSE of the model at its best hidden-layer depth.
    Three models are compared: the multilayer perceptron (MLP) is drawn as a black solid line with cross-shaped markers; the Kolmogorov-Arnold network (KAN) is drawn as a blue dashed line with diamond-shaped markers; and the QKAN is drawn as a red dash-dotted line with circular markers.
    The number printed next to each marker is the parameter count of the corresponding model, given in blue for the KAN and in red for the QKAN.
    QKAN attains the best test RMSE on more than 80\% of the equations, and with far fewer parameters than KAN in most cases.
    See \cref{tab:com_reg} for details.
  }\label{fig:loss}
\end{sidewaysfigure*}

\subsection{Complementary results of \texorpdfstring{\cref{fig:loss}}{Figure}}\label{sec:sup_fig3_details}

In addition to the summary presented in \cref{fig:loss}, we provide the complete set in \cref{tab:com_reg}.
This table includes the full list of 66 equations used in the regression benchmark, which were not explicitly shown in the main text.
For each equation, we report the RMSE loss and the number of trainable parameters for both QKAN and classical KAN models.
These results further illustrate that QKANs can achieve comparable or superior approximation accuracy while maintaining significantly fewer parameters.

\begin{center}
\tiny %
\setlength{\tabcolsep}{3pt} %
\renewcommand{\arraystretch}{0.9} %

\begin{longtable}{cccccc}
\caption{
\textbf{Detailed regression results corresponding to \cref{fig:loss}.}
This table presents a comparison between QKAN and classical KAN across 66 symbolic regression expressions.
For each expression, we report the root mean square error (RMSE) loss and the total number of trainable parameters.
\textbf{Bold values} highlight the better-performing model in each row (lower RMSE).
The results demonstrate that QKAN consistently achieves comparable or superior approximation accuracy with significantly fewer parameters, validating the expressive efficiency of DARUAN.
}
\label{tab:com_reg} \\
\toprule
\multicolumn{2}{@{}c@{}}{Equation} & \multicolumn{2}{@{}c@{}}{QKAN} & \multicolumn{2}{@{}c@{}}{KAN} \\
\cmidrule{1-2} \cmidrule{3-4} \cmidrule{5-6}
Idx & Expression & RMSE Loss & \# Params & RMSE Loss & \# Params \\
\midrule
\endfirsthead

\multicolumn{6}{c}{{\bfseries \tablename\ \thetable{} -- continued from previous page}} \\
\toprule
\multicolumn{2}{@{}c@{}}{Equation} & \multicolumn{2}{@{}c@{}}{QKAN} & \multicolumn{2}{@{}c@{}}{KAN} \\
\cmidrule{1-2} \cmidrule{3-4} \cmidrule{5-6}
Idx & Expression & RMSE Loss & \# Params & RMSE Loss & \# Params \\
\midrule
\endhead

\midrule \multicolumn{6}{r}{{Continued on next page}} \\ \bottomrule
\endfoot

\bottomrule
\endlastfoot

1 & $\hbar \omega$ & $3.652\times 10^{-04}$ & 255 & $\mathbf{3.619\times 10^{-04}}$ & $\mathbf{336}$\\
2 & $N_{n} \mu$ & $3.652\times 10^{-04}$ & 255 & $\mathbf{3.568\times 10^{-04}}$ & \textbf{336}\\
3 & $E_{f} q_{2}$ & $\mathbf{3.652\times 10^{-04}}$ & \textbf{255} & $3.662\times 10^{-04}$ & 336\\
4 & $0.5 m \left(u^{2} + v^{2} + w^{2}\right)$ & $1.023\times 10^{-03}$ & 850 & $\mathbf{6.779\times 10^{-04}}$ & \textbf{556}\\
5 & $F r \sin{(\theta)}$ & $1.274\times 10^{-03}$ & 765 & $\mathbf{8.490\times 10^{-04}}$ & \textbf{1,001}\\
6 & $x_{1} y_{1} + x_{2} y_{2} + x_{3} y_{3}$ & $1.476\times 10^{-03}$ & 595 & $\mathbf{8.556\times 10^{-04}}$ & \textbf{776}\\
7 & $m r v \sin{\theta}$ & $1.715\times 10^{-03}$ & 1,275 & $\mathbf{1.244\times 10^{-03}}$ & \textbf{1,111}\\
8 & $0.25 m x^{2} \left(\omega^{2} + \omega_{0}^{2}\right)$ & $\mathbf{2.818\times 10^{-03}}$ & \textbf{1,275} & $2.933\times 10^{-03}$ & 556\\
9 & ${p_{d} \cos{\theta}}/{4 \pi \epsilon r^{2}}$ & $2.915\times 10^{-03}$ & 425 & $\mathbf{2.683\times 10^{-03}}$ & \textbf{556}\\
10 & ${\hbar \omega^{3}}/{\pi^{2} c^{2} \left(\exp\left({\hbar \omega}/{T k_{b}}\right) - 1\right)}$ & $\mathbf{3.800\times 10^{-03}}$ & \textbf{1,785} & $3.898\times 10^{-03}$ & 1,221\\
11 & $q \left(B v \sin{\theta} + E_{f}\right)$ & $\mathbf{4.455\times 10^{-03}}$ & \textbf{935} & $3.377\times 10^{-02}$ & 1,221\\
12 & $2 U \left(1 - \cos{\left(d k \right)}\right)$ & $\mathbf{5.404\times 10^{-03}}$ & \textbf{340} & $5.442\times 10^{-03}$ & 2,111\\
13 & ${q r v}/{2}$ & $\mathbf{6.219\times 10^{-03}}$ & \textbf{340} & $6.503\times 10^{-03}$ & 1,556\\
14 & $\left(x + y\right) \sin{\left(\exp({2 y}) \right)}$ & $\mathbf{6.419\times 10^{-03}}$ & \textbf{255} & $6.693\times 10^{-03}$ & 336\\
15 & ${3 p_{d} \sin{\theta} \cos{\theta}}/{4 \pi \epsilon r^{3}}$ & $\mathbf{7.082\times 10^{-03}}$ & \textbf{425} & $7.648\times 10^{-03}$ & 556\\
16 & ${E_{f} \alpha \epsilon n}/({- {\alpha n}/{3} + 1})$ & $\mathbf{7.388\times 10^{-03}}$ & \textbf{425} & $7.572\times 10^{-03}$ & 2,221\\
17 & $0.5 k_{s} x^{2}$ & $\mathbf{7.729\times 10^{-03}}$ & \textbf{255} & $7.942\times 10^{-03}$ & 336\\
18 & $0.5 E_{f}^{2} \epsilon$ & $\mathbf{7.729\times 10^{-03}}$ & \textbf{255} & $7.937\times 10^{-03}$ & 891\\
19 & $E_{f}^{2} c \epsilon$ & $\mathbf{8.531\times 10^{-03}}$ & \textbf{340} & $8.722\times 10^{-03}$ & 2,111\\
20 & $g m z$ & $\mathbf{1.249\times 10^{-02}}$ & \textbf{340} & $1.293\times 10^{-02}$ & 1,556\\
21 & $T k_{b} \mu$ & $\mathbf{1.249\times 10^{-02}}$ & \textbf{340} & $1.294\times 10^{-02}$ & 2,111\\
22 & $E_{f}^{2} \epsilon$ & $\mathbf{1.546\times 10^{-02}}$ & \textbf{255} & $1.586\times 10^{-02}$ & 1,446\\
23 & ${I_{0} \sin^{2}{\left({n \theta}/{2} \right)}}/{\sin^{2}{\left({\theta}/{2} \right)}}$ & $\mathbf{1.598\times 10^{-02}}$ & \textbf{340} & $1.650\times 10^{-02}$ & 446\\
24 & ${\sqrt{2} \exp({- {\theta^{2}}/{2}})}/{2 \sqrt{\pi}}$ & $1.611\times 10^{-02}$ & 17 & $\mathbf{1.611\times 10^{-02}}$ & \textbf{23}\\
25 & ${4 I_{0} \sin^{2}{\left({\alpha}/{2} \right)} \sin^{2}{\left({N \delta}/{2} \right)}}/{\alpha^{2} \sin^{2}{\left({\delta}/{2} \right)}}$ & $\mathbf{1.637\times 10^{-02}}$ & \textbf{1,275} & $1.676\times 10^{-02}$ & 1,666\\
26 & ${n_{0}}/\left({\exp({{B \mu}/{T k_{b}}}) + \exp({- {B \mu}/{T k_{b}}})}\right)$ & $\mathbf{1.655\times 10^{-02}}$ & \textbf{935} & $1.872\times 10^{-02}$ & 2,331\\
27 & ${Y}/({2 \sigma + 2})$ & $\mathbf{1.681\times 10^{-02}}$ & \textbf{1,105} & $1.700\times 10^{-02}$ & 891\\
28 & ${l}/{2 \pi c^{2} \epsilon r}$ & $\mathbf{1.810\times 10^{-02}}$ & \textbf{425} & $1.903\times 10^{-02}$ & 1,111\\
29 & ${\hbar \omega}/({\exp({{\hbar \omega}/{T k_{b}}}) - 1})$ & $\mathbf{1.872\times 10^{-02}}$ & \textbf{1,275} & $1.920\times 10^{-02}$ & 2,221\\
30 & $(- {H^{2} c^{2} \cdot \left(1 - 2 \alpha\right) + {c^{4} k_{f}}/{a_{f}^{2}}})/{8 \pi G}$ & $1.952\times 10^{-02}$ & 595 & $\mathbf{7.809\times 10^{-03}}$ & \textbf{776}\\
31 & $- B \mu_{M} \cos{\theta}$ & $\mathbf{2.023\times 10^{-02}}$ & \textbf{340} & $2.084\times 10^{-02}$ & 446\\
32 & $- E_{f} p_{d} \cos{\theta}$ & $\mathbf{2.023\times 10^{-02}}$ & \textbf{340} & $2.082\times 10^{-02}$ & 446\\
33 & $\operatorname{asin}{\left(n \sin{\theta_{2}} \right)}$ & $\mathbf{2.219\times 10^{-02}}$ & \textbf{255} & $2.270\times 10^{-02}$ & 1,446\\
34 & $\mu n \tanh{\left({B \mu}/{T k_{b}} \right)}$ & $\mathbf{2.254\times 10^{-02}}$ & \textbf{935} & $2.473\times 10^{-02}$ & 1,776\\
35 & $\hbar n$ & $\mathbf{2.372\times 10^{-02}}$ & \textbf{255} & $2.427\times 10^{-02}$ & 1,446\\
36 & ${d \left(1 - \alpha^{2}\right)}/({\alpha \cos{\left(\theta_{1} - \theta_{2} \right)} + 1})$ & $\mathbf{2.603\times 10^{-02}}$ & \textbf{425} & $2.709\times 10^{-02}$ & 556\\
37 & ${E}/{({E \left(1 - \cos{\theta}\right)}/{c^{2} m} + 1)}$ & $\mathbf{3.062\times 10^{-02}}$ & \textbf{850} & $3.198\times 10^{-02}$ & 1,111\\
38 & $T k_{b} n \log{\left({V_{2}}/{V_{1}} \right)}$ & $\mathbf{3.509\times 10^{-02}}$ & \textbf{935} & $3.597\times 10^{-02}$ & 666\\
39 & $1.5 V p_{F}$ & $\mathbf{3.558\times 10^{-02}}$ & \textbf{255} & $3.645\times 10^{-02}$ & 891\\
40 & $n_{0} \exp({- {g m x}/{T k_{b}}})$ & $\mathbf{3.716\times 10^{-02}}$ & \textbf{1,020} & $4.005\times 10^{-02}$ & 1,331\\
41 & $B \mu \left(\chi + 1\right)$ & $\mathbf{3.761\times 10^{-02}}$ & \textbf{340} & $3.845\times 10^{-02}$ & 1,001\\
42 & ${3 \left(H^{2} + {c^{2} k_{f}}/{a_{f}^{2}}\right)}/{8 \pi G}$ & $\mathbf{4.767\times 10^{-02}}$ & \textbf{510} & $4.834\times 10^{-02}$ & 666\\
43 & $\mu \sqrt{B_{x}^{2} + B_{y}^{2} + B_{z}^{2}}$ & $\mathbf{4.778\times 10^{-02}}$ & \textbf{850} & $4.992\times 10^{-02}$ & 556\\
44 & ${1}/({{n}/{d_{2}} + {1}/{d_{1}}})$ & $\mathbf{5.355\times 10^{-02}}$ & \textbf{340} & $5.372\times 10^{-02}$ & 446\\
45 & $V_{e} q + \sqrt{c^{4} m^{2} + c^{2} \left(- A q + p\right)^{2}}$ & $\mathbf{5.753\times 10^{-02}}$ & \textbf{595} & $6.745\times 10^{-02}$ & 776\\
46 & ${1}/({\exp({{\hbar \omega}/{T k_{b}}}) - 1})$ & $\mathbf{6.836\times 10^{-02}}$ & \textbf{850} & $6.983\times 10^{-02}$ & 1,666\\
47 & $\beta \left(\alpha \cos{\theta} + 1\right)$ & $\mathbf{6.930\times 10^{-02}}$ & \textbf{340} & $7.042\times 10^{-02}$ & 1,556\\
48 & $x_{1} \left(\alpha \cos^{2}{\left(\omega t \right)} + \cos{\left(\omega t \right)}\right)$ & $\mathbf{6.938\times 10^{-02}}$ & \textbf{425} & $7.132\times 10^{-02}$ & 2,221\\
49 & $({m_{1} r_{1} + m_{2} r_{2}})/({m_{1} + m_{2}})$ & $\mathbf{7.342\times 10^{-02}}$ & \textbf{425} & $7.578\times 10^{-02}$ & 1,666\\
50 & $\sqrt{{\gamma p}/{\rho}}$ & $\mathbf{7.662\times 10^{-02}}$ & \textbf{340} & $7.965\times 10^{-02}$ & 446\\
51 & $\sqrt{x_{1}^{2} - 2 x_{1} x_{2} \cos{\left(\theta_{1} - \theta_{2} \right)} + x_{2}^{2}}$ & $\mathbf{8.846\times 10^{-02}}$ & \textbf{1,275} & $8.931\times 10^{-02}$ & 1,111\\
52 & $\sqrt{\left(- x_{1} + x_{2}\right)^{2} + \left(- y_{1} + y_{2}\right)^{2}}$ & $\mathbf{1.181\times 10^{-01}}$ & \textbf{850} & $1.215\times 10^{-01}$ & 1,666\\
53 & ${\alpha n}/({- {\alpha n}/{3} + 1}) + 1$ & $\mathbf{1.252\times 10^{-01}}$ & \textbf{255} & $1.265\times 10^{-01}$ & 891\\
54 & $I_{1} + I_{2} + 2 \sqrt{I_{1} I_{2}} \cos{\left(\delta \right)}$ & $\mathbf{1.896\times 10^{-01}}$ & \textbf{340} & $1.920\times 10^{-01}$ & 1,001\\
55 & ${k_{b} v}/{A \left(\gamma - 1\right)}$ & $\mathbf{2.172\times 10^{-01}}$ & \textbf{425} & $2.192\times 10^{-01}$ & 1,666\\
56 & $\sin^{2}{\left({E t}/{\hbar} \right)}$ & $3.291\times 10^{-01}$ & 340 & $\mathbf{1.853\times 10^{-01}}$ & \textbf{2,111}\\
57 & ${3 p_{d} z \sqrt{x^{2} + y^{2}}}/{4 \pi \epsilon r^{5}}$ & $\mathbf{3.467\times 10^{-01}}$ & \textbf{595} & $3.517\times 10^{-01}$ & 2,441\\
58 & ${k_{G} m \left(\sqrt{{2 E L^{2}}/{k_{G}^{2} m} + 1} \cos{\left(\theta_{1} - \theta_{2} \right)} + 1\right)}/{L^{2}}$ & $\mathbf{3.777\times 10^{-01}}$ & \textbf{595} & $3.887\times 10^{-01}$ & 1,331\\
59 & ${\sqrt{2} \exp({- {\left(\theta - \theta_{1}\right)^{2}}/{2 \sigma^{2}}})}/{2 \sqrt{\pi} \sqrt{\sigma^{2}}}$ & $\mathbf{3.895\times 10^{-01}}$ & \textbf{1,190} & $4.078\times 10^{-01}$ & 1,556\\
60 & $\sqrt{{8 \pi G \rho}/{3} - {c^{2} k_{f}}/{a_{f}^{2}}}$ & $\mathbf{4.017\times 10^{-01}}$ & \textbf{510} & $4.050\times 10^{-01}$ & 666\\
61 & ${G m_{1} m_{2}}/({\left(- x_{1} + x_{2}\right)^{2} + \left(- y_{1} + y_{2}\right)^{2} + \left(- z_{1} + z_{2}\right)^{2}})$ & $4.331\times 10^{-01}$ & 153 & $\mathbf{4.295\times 10^{-01}}$ & \textbf{2,771}\\
62 & ${V p_{F}}/({\gamma - 1})$ & $4.681\times 10^{-01}$ & 340 & $\mathbf{4.655\times 10^{-01}}$ & \textbf{1,556}\\
63 & ${h q}/{4 \pi m}$ & $\mathbf{6.304\times 10^{-01}}$ & \textbf{765} & $6.766\times 10^{-01}$ & 1,556\\
64 & $({m^{2} \omega^{2} x^{2} \left({\alpha y}/{x} + 1\right) + p^{2}})/{2 m}$ & $6.937\times 10^{-01}$ & 595 & $\mathbf{6.810\times 10^{-01}}$ & \textbf{1,886}\\
65 & ${\sqrt{2} \exp({- {\theta^{2}}/{2 \sigma^{2}}})}/{2 \sqrt{\pi} \sqrt{\sigma^{2}}}$ & $\mathbf{8.487\times 10^{-01}}$ & \textbf{1,530} & $9.130\times 10^{-01}$ & 891\\
66 & ${q}/{\sqrt{d^{2} - 2 d r \cos{\alpha} + r^{2}}}$ & $\mathbf{9.339\times 10^{-01}}$ & \textbf{850} & $1.731\times 10^{+00}$ & 1,111\\

\end{longtable}

\end{center}

\subsection{Visualization of QKAN activations on noisy function regression}\label{sec:sup_activation_visualization}

To further analyze the expressive power and interpretability of QKANs, we visualize the learned activation functions from selected regression tasks involving noisy symbolic expressions.
This follows a similar methodology to KANs, where each activation function is a learnable one-dimensional function, enabling insight into the internal structure of the model.

\begin{figure*}
    \centering
    \includegraphics[width=0.8\textwidth]{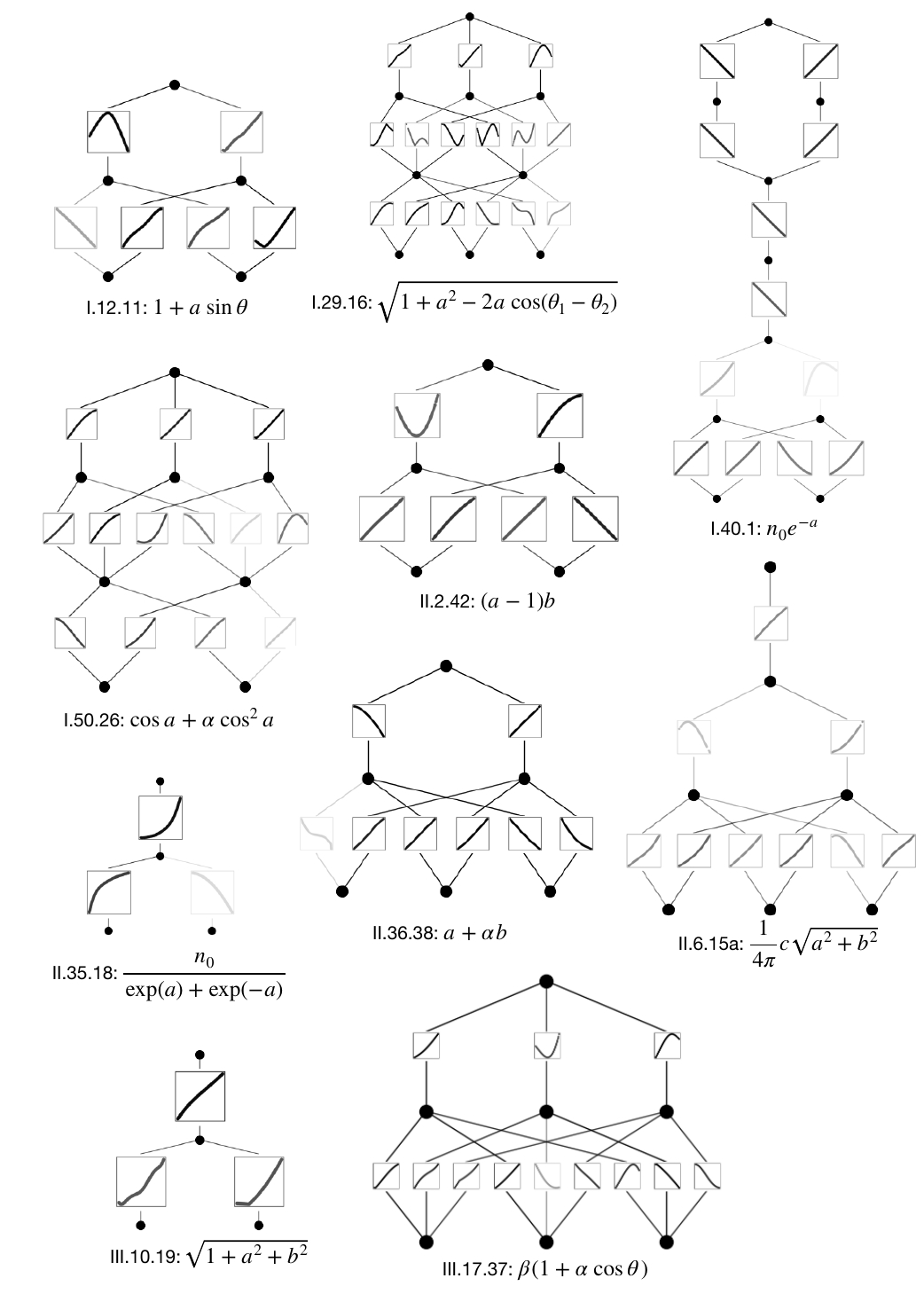}
    \caption{
        \textbf{Visualization of learned QKAN activations for heuristic noisy regression equations.}
        This figure illustrates the per-node activation functions learned by QKANs, similar to the interpretability offered by KANs. 
        We display the QKAN models that achieved the lowest RMSE for a subset of representative symbolic regression equations, as listed in \cref{tab:main_table}. 
        The transparency of each node is proportional to output range divided by input range; darker therefore denote stronger connections.
        Each sub-panel corresponds to a distinct equation, demonstrating that QKANs can learn smooth and structured nonlinear transformations despite the presence of noise.
    }
    \label{fig:nodes}
\end{figure*}

\cref{fig:nodes} presents the per-node activation functions learned by QKANs that achieved the lowest RMSE for representative equations drawn from the benchmark in \cref{tab:main_table}.
These visualizations demonstrate that QKANs are capable of learning smooth, structured nonlinearities even in the presence of input noise, highlighting their robustness and alignment with symbolic structure.

\subsection{Real-device QKAN/HQKAN inference on IBM Aachen}\label{sec:sup_real_device}

To assess the direct hardware pathway of DARUAN-based activations, we evaluated trained QKAN/HQKAN models on the \texttt{ibm\_aachen} superconducting processor.
We emphasize that this subsection is intended as a hardware-inference feasibility and robustness validation, not as a hardware-speedup claim.
The main controlled error source in the measurement-shot sweep is the finite number $S$ of measurement shots used to estimate each single-qubit Pauli expectation value.
For a fixed trained circuit, a Pauli-$Z$ measurement returns samples $Y_s\in\{-1,+1\}$ with mean $\mu$.
The empirical estimator
\begin{equation}
  \widehat{\mu}_S=\frac{1}{S}\sum_{s=1}^S Y_s
\end{equation}
has variance
\begin{equation}
  \operatorname{Var}(\widehat{\mu}_S)=\frac{1-\mu^2}{S}\le \frac{1}{S}.
\end{equation}
Thus the statistical expectation-value uncertainty scales as $S^{-1/2}$, and its contribution to mean-square error scales as $S^{-1}$.
The single-qubit DARUAN circuits used here are shallow: with $r$ data re-uploading repetitions, each edge activation has depth $3r$ single-qubit rotation gates before measurement and coherent circuit time
\begin{equation}
  T_{\rm circ}\approx 3r\,\tau_{\rm 1q}.
\end{equation}
Within a fixed calibration window and physical-qubit layout, after selecting calibrated qubits with short circuit time relative to $T_1$ and $T_2$, the remaining gate, decoherence, and readout effects act approximately as a fixed hardware bias across a measurement-shot sweep.
Comparisons made within such a window therefore primarily probe the reduction of finite-shot statistical error, whereas a residual large-shot gap to the exact solver can reflect hardware and mitigation bias that increasing $S$ alone does not remove.

For a function-fit validation we trained a single-qubit QKAN$([1,1])$ with $r=3$ on the spherical Bessel function $J_0(20x)=\sin(20x)/(20x)$ (defined by continuity at $x=0$), using the exact solver with LBFGS, transferred the trained parameters to a Qiskit-backed QKAN layer, and swept $\{10^0, 10^1, 10^2, 10^3, 10^4, 10^5\}$ shots per circuit over 1,000 held-out test points $x\in[0,1]$.
For image classification, the CIFAR-10 model uses CNN $\rightarrow$ Linear$(256,40)$ $\rightarrow$ QKAN$([40,24])$ $\rightarrow$ Linear$(24,10)$ and the CIFAR-100 model uses CNN $\rightarrow$ Linear$(256,56)$ $\rightarrow$ QKAN$([56,56])$ $\rightarrow$ Linear$(56,100)$; both use 100 held-out test images and transfer the exact-solver QKAN parameters directly to the Qiskit-backed QKAN layer under the original $\{10^0,10^1,10^2,10^3\}$-shot schedule.
We deployed the asynchronous parallelism distributed method introduced in the main context.
This method parallelizes the execution of independent single-qubit circuits on the same quantum processor unit (QPU), using as many calibrated physical qubits as possible.
The QPU then executes the submitted circuit batches asynchronously until all circuits are completed.
In all function-fit runs we select calibrated physical qubits with single-qubit \texttt{sx} gate error no larger than $10^{-3}$ and readout error below $1\%$, and pack circuits in parallel across the qualifying qubits.
The hardware evaluations use \texttt{resilience\_level=2}; no gate-twirling option is enabled explicitly, while measurement twirling follows the IBM Runtime default.

\begin{figure}[t!]
    \centering
    \includegraphics[width=\textwidth]{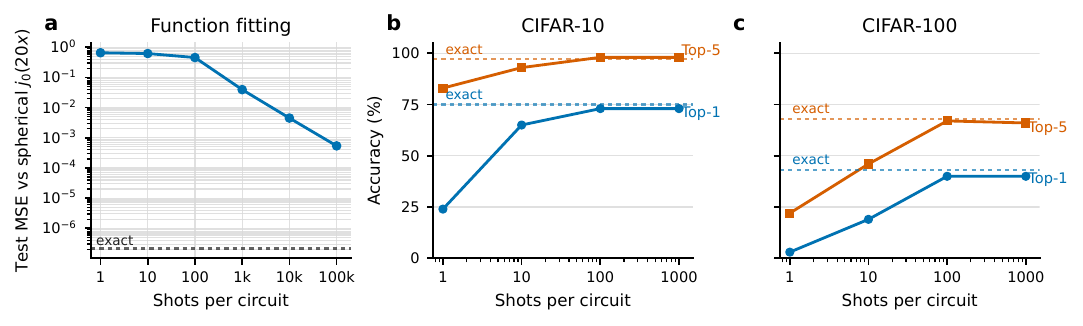}
    \caption{
        \textbf{Real-device measurement-shot sweep on IBM Aachen.}
        Panel (a) plots the test MSE of the QKAN$([1,1], r{=}3)$ fit to the spherical Bessel function $J_0(20x)=\sin(20x)/(20x)$ over 1,000 points for $\{10^0, 10^1, 10^2, 10^3, 10^4, 10^5\}$ measurement shots per circuit; the dashed line marks the exact-solver reference, MSE $\approx 2.1\times10^{-7}$.
        Panels (b,c) show top-1 (blue) and top-5 (orange) accuracies for the CIFAR-10 and CIFAR-100 HQKAN models over 100 test images under the original $\{10^0,10^1,10^2,10^3\}$-shot schedule; dashed lines mark the corresponding exact-solver references.
        The function fit measures an analog expectation value at every test point and its observed error continues to decrease at the two added shot counts.
        By contrast, classification only requires preserving the correct discrete decision, so the accuracy saturates once the hardware-induced logit perturbation is smaller than the relevant class margins for most samples.
    }
    \label{fig:ibm_aachen_hqkan}
\end{figure}

\begin{figure}[t!]
    \centering
    \includegraphics[width=\textwidth]{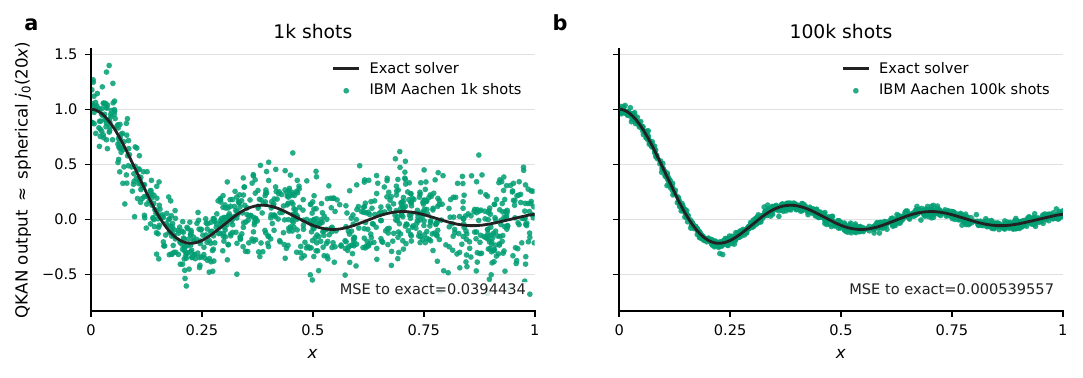}
    \caption{
        \textbf{Real-device QKAN function fit on IBM Aachen.}
        The black curves are the outputs of the trained QKAN$([1,1], r{=}3)$ evaluated with the exact solver over 1,000 test points $x\in[0,1]$, and the green markers are the outputs of the QKAN layer on \texttt{ibm\_aachen}.
        Panels (a,b) show 1,000 and 100,000 shots, respectively.
        Each panel reports the pointwise hardware-to-exact-solver MSE: $0.039443$ at 1,000 shots and $0.00053956$ at 100,000 shots; the corresponding target-function test MSEs are $0.039439$ and $0.0005392$.
    }
    \label{fig:ibm_aachen_funcfit_curve}
\end{figure}

These results provide a direct real-device validation of the trained QKAN/HQKAN inference path.
For the function fit (\cref{fig:ibm_aachen_hqkan}(a)), the test MSE drops from $6.56\times10^{-1}$ at 1 shot to $6.19\times10^{-1}$ at 10 shots, $4.56\times10^{-1}$ at 100 shots, $3.9439\times10^{-2}$ at 1000 shots, $4.510\times10^{-3}$ at 10,000 shots, and $5.392\times10^{-4}$ at 100,000 shots (exact-solver reference $\approx 2.1\times10^{-7}$).
Because the four lower-shot points and two higher-shot points were collected in different calibration windows and with different layouts, this overall decrease is consistent with convergence as the shot count increases but is not a controlled estimate of shot scaling alone.
The remaining gap at 100,000 shots can reflect finite-shot error together with residual single-qubit gate error, decoherence, readout error, and mitigation bias.
For CIFAR-10, the 100-shot hardware run reaches $73.0\%$ top-1 and $98.0\%$ top-5 accuracy, close to the exact-solver reference of $75.0\%$ and $97.0\%$; 1,000 shots give the same top-1/top-5 accuracy.
For CIFAR-100, 100 shots reaches $40.0\%$ top-1 and $67.0\%$ top-5 accuracy (exact reference $43.0\%$ and $68.0\%$), and 1,000 shots reaches $40.0\%$/$66.0\%$.
\cref{fig:ibm_aachen_funcfit_curve} compares the 1,000- and 100,000-shot outputs point-by-point against the exact-solver reference, confirming that the hardware captures the main oscillatory structure of $J_0(20x)$ and showing the reduction in pointwise analog error at the higher shot count.

The different behavior of regression and classification is expected.
In regression, the circuit output itself is the target signal.
If the exact-solver prediction is $f_{\rm ex}(x)$ and the hardware-evaluated output is
\begin{equation}
  \widehat f(x)=f_{\rm ex}(x)+\delta_{\rm stat}(x)+\delta_{\rm hw}(x),
\end{equation}
then the test MSE directly accumulates the pointwise statistical measurement term $\delta_{\rm stat}(x)$ and the residual hardware-bias term $\delta_{\rm hw}(x)$ across the full test grid.
Increasing the number of measurement shots reduces the statistical contribution $\delta_{\rm stat}$, but coherent gate error, decoherence, and readout bias can remain after mitigation.
Therefore, high-precision regression requires accurate analog expectation values at each input point, and the MSE continues to improve with shots while still showing a residual gap relative to the exact solver.

Classification is less sensitive because the task only requires the correct discrete decision rather than pointwise reconstruction of every analog expectation value.
Let $\bm z$ be the exact-solver logit vector and $\widehat{\bm z}=\bm z+\Delta\bm z$ be the hardware-perturbed logit vector.
For top-1 classification, if the exact top-class margin is
\begin{equation}
  \gamma_1=z_{(1)}-z_{(2)},
\end{equation}
where $z_{(1)}$ and $z_{(2)}$ are the largest and second-largest logits, then the predicted label is unchanged whenever the hardware perturbation is smaller than the margin, for example when $\|\Delta\bm z\|_\infty<\gamma_1/2$.
For top-5 accuracy the tolerance is even larger: the true class only needs to remain inside the five largest logits, so perturbations are tolerated as long as they do not push the true-class logit below the fifth/sixth-rank decision boundary.
Moreover, HQKAN classification aggregates many single-qubit activation outputs through subsequent linear layers, which can partially average independent statistical measurement fluctuations.
Thus moderate per-circuit expectation-value errors can be absorbed by class margins and downstream aggregation, explaining why the CIFAR-10 and CIFAR-100 accuracies saturate already at 100 shots while the continuous regression MSE continues to benefit from increased shots.

This distinction should not be interpreted as saying that classification is insensitive to all hardware errors.
Samples close to a decision boundary can still flip under small logit perturbations, and the robustness depends on the trained model's class margins, the dataset, and the calibration quality of the selected qubits.
Rather, the real-device results show a task-dependent robustness pattern: margin-based classification can remain close to the exact-solver reference under present-day single-qubit NISQ noise, whereas high-precision function regression remains substantially more demanding because it penalizes analog pointwise deviations.

A point worth emphasizing is that our \texttt{sx} gate-error cutoff of $10^{-3}$ and readout cutoff of $1\%$ on \texttt{ibm\_aachen} sit roughly two orders of magnitude above the $\sim 10^{-5}$ single-qubit gate-error figure cited in the main text for optimized superconducting benchmarks~\citep{PRXQuantum.5.040342}.
The residual appears in \cref{fig:ibm_aachen_hqkan}(a) as the remaining offset between the 100,000-shot Aachen MSE of $5.392\times10^{-4}$ and the exact-solver reference of $2.1\times10^{-7}$.
Crucially for the practical viability of the single-qubit DARUAN pathway, the end-to-end classification tasks remain usable without best-case hardware: on \texttt{ibm\_aachen} the HQKAN models reach $73.0\%/40.0\%$ top-1 and $98.0\%/67.0\%$ top-5 on CIFAR-10/CIFAR-100 at just 100 shots per circuit, within $2$--$3$ percentage points of the exact-solver reference ($75.0\%/43.0\%$ top-1, $97.0\%/68.0\%$ top-5).
These results indicate that inference of single-qubit DARUAN in HQKAN degrades gracefully from optimized single-qubit benchmark regimes to public-access NISQ hardware for margin-based classification tasks, while continuous function regression remains more sensitive to residual hardware error and statistical measurement error from finite shots.
For the single-qubit setting studied in this work, classical GPU simulation therefore remains the preferred practical route, and quantum hardware execution is best understood as a direct inference-validation pathway.

\subsection{Interpretability: the full symbolic-regression workflow}\label{sec:sup_interpretability}

To provide a concrete qualitative interpretability analysis, we reproduce \emph{all six steps} of the symbolic-regression workflow shown in Fig.~2.4 of~\citet{liu2024kan} on a QKAN. The target is $f(x,y) = \exp(\sin(\pi x) + y^2)$ with 1,000 training / 1,000 test points sampled uniformly from $[-1,1]^2$.

The six steps are: (1)~\emph{train with sparsification} --- fit a QKAN$([2,5,1])$ under L1+entropy regularisation with incremental DARUAN \emph{layer extension} $r{:}3{\to}24$;
(2)~\emph{prune} --- take the most-important hidden neuron and retrain a fresh QKAN$([2,1,1])$ from scratch with layer extension $r{:}3{\to}48$ (seed~$1$; the optimisation landscape of QKAN$([2,1,1])$ is seed-sensitive, but with layer extension, seeds $0$ and $1$ both reach MSE $<10^{-5}$ with sin/$x^2$/exp fits, matching KAN Fig.~2.4 Step~2 exactly);
(3)~\emph{set symbolic functions} --- for each of the three surviving edges run over the primitive library $\{x, x^2, \dots, e^x, \sin, \cos, \mathrm{gaussian}, \dots\}$; the best fits are $\sin$ ($R^2=1.000$), $x^2$ ($R^2=1.000$), and $\exp$ ($R^2=1.000$);
(4)~\emph{train affine parameters} --- LBFGS refines each edge's $(a,b,c,d)$ to machine precision;
(5)~\emph{output symbolic formula} --- \texttt{sympy} composes the edges into $0.999\cdot\exp\!\bigl(0.0613\,(4.04y)^2 + 1.001\,\sin(3.142 x)\bigr) + 5.9{\times}10^{-4}$;
(6)~\emph{number snap} --- after expanding inner products and rounding near-integer, near-$\pi$, and near-small-rational floats, the composed expression collapses exactly to $\exp(y^2 + \sin(\pi x))$, recovering the ground-truth target.

\begin{figure}[ht!]
    \centering
    \resizebox{0.86\textwidth}{!}{%
    \begin{tikzpicture}[
        nd/.style={circle, draw=black, fill=black, inner sep=1.25pt, outer sep=0pt},
        edgeplot/.style={inner sep=0pt, outer sep=0pt, fill=white},
        faintconn/.style={-, draw=black!18, line width=0.34pt, line cap=round},
        conn/.style={-, draw=black!78, line width=0.58pt, line cap=round},
        symconn/.style={-, draw=red!62!black, line width=0.58pt, line cap=round},
        flow/.style={-{Stealth[length=2.8mm,width=2.4mm]}, draw=black, line width=0.85pt, line cap=round},
        finalflow/.style={-{Stealth[length=3.4mm,width=3.0mm]}, draw=black, line width=1.10pt, line cap=round},
        steplbl/.style={font=\bfseries\sffamily\small, align=center, text=black},
        smallnote/.style={font=\sffamily\scriptsize, align=center, text=black},
        sym/.style={font=\scriptsize, color=red!72!black, align=center},
        snapbox/.style={
            draw=black,
            double=black,
            double distance=0.55pt,
            line width=0.55pt,
            rounded corners=1pt,
            inner xsep=8pt,
            inner ysep=5pt,
            font=\Large
        },
    ]
        \path[use as bounding box] (-10.15,-5.75) rectangle (9.05,6.75);

        \begin{scope}[shift={(-6.50,0.72)}]
            \def\yout{3.80}\def\yLone{3.03}\def\yhid{2.22}\def\yLzero{1.08}\def\yin{0.16}
            \def\ha{-2.40}\def\hb{-1.20}\def\hc{0}\def\hd{1.20}\def\he{2.40}
            \def\xEa{-3.10}\def\xEb{-2.35}\def\xEc{-1.60}\def\xEd{-0.85}\def\xEe{-0.10}
            \def\yEa{0.10}\def\yEb{0.85}\def\yEc{1.60}\def\yEd{2.35}\def\yEe{3.10}

            \coordinate (s1_xin) at (-2.15,\yin);
            \coordinate (s1_yin) at ( 2.15,\yin);
            \coordinate (s1_out) at (0,\yout);

            \foreach \col/\id in {\ha/a,\hb/b,\hc/c,\hd/d,\he/e} {
                \coordinate (s1_L1\id) at (\col,\yLone);
                \coordinate (s1_h\id)  at (\col,\yhid);
            }
            \foreach \col/\id in {\xEa/a,\xEb/b,\xEc/c,\xEd/d,\xEe/e} {
                \coordinate (s1_L0x\id) at (\col,\yLzero);
            }
            \foreach \col/\id in {\yEa/a,\yEb/b,\yEc/c,\yEd/d,\yEe/e} {
                \coordinate (s1_L0y\id) at (\col,\yLzero);
            }

            \foreach \id in {a,b,c,d,e} {
                \draw[faintconn] (s1_L1\id) -- (s1_out);
                \draw[faintconn] (s1_h\id) -- (s1_L1\id);
                \draw[faintconn] (s1_L0x\id) -- (s1_h\id);
                \draw[faintconn] (s1_L0y\id) -- (s1_h\id);
                \draw[faintconn] (s1_xin) -- (s1_L0x\id);
                \draw[faintconn] (s1_yin) -- (s1_L0y\id);
            }

            \foreach \id/\n in {a/0,b/1,c/2,d/3,e/4} {
                \node[edgeplot, opacity=0.72] at (s1_L1\id)
                    {\includegraphics[width=0.64cm]{fig/interp_steps/step1_L1_j0_i\n.pdf}};
                \node[edgeplot, opacity=0.72] at (s1_L0x\id)
                    {\includegraphics[width=0.57cm]{fig/interp_steps/step1_L0_j\n_i0.pdf}};
                \node[edgeplot, opacity=0.72] at (s1_L0y\id)
                    {\includegraphics[width=0.57cm]{fig/interp_steps/step1_L0_j\n_i1.pdf}};
            }

            \node[nd, label={[font=\normalsize, yshift=0.7mm]above:$\exp(\sin(\pi x)+y^2)$}] at (s1_out) {};
            \foreach \id in {a,b,c,d,e} {\node[nd] at (s1_h\id) {};}
            \node[nd, label={[font=\scriptsize]below:$x$}] at (s1_xin) {};
            \node[nd, label={[font=\scriptsize]below:$y$}] at (s1_yin) {};

            \node[smallnote, anchor=north] at (0,-0.24)
                {QKAN$([2,5,1])$, layer-ext.\ $r{:}3{\to}24$};
        \end{scope}

        \begin{scope}[shift={(0.40,0.72)}]
            \def\yout{3.80}\def\yLone{3.03}\def\yhid{2.22}\def\yLzero{1.08}\def\yin{0.16}
            \coordinate (s2_xin) at (-1.00,\yin);
            \coordinate (s2_yin) at ( 1.00,\yin);
            \coordinate (s2_out) at (0,\yout);
            \coordinate (s2_L1a) at (0,\yLone);
            \coordinate (s2_ha) at (0,\yhid);
            \coordinate (s2_L0x) at (-1.00,\yLzero);
            \coordinate (s2_L0y) at ( 1.00,\yLzero);

            \draw[conn] (s2_L1a) -- (s2_out);
            \draw[conn] (s2_ha) -- (s2_L1a);
            \draw[conn] (s2_L0x) -- (s2_ha);
            \draw[conn] (s2_L0y) -- (s2_ha);
            \draw[conn] (s2_xin) -- (s2_L0x);
            \draw[conn] (s2_yin) -- (s2_L0y);

            \node[edgeplot] at (s2_L1a) {\includegraphics[width=0.88cm]{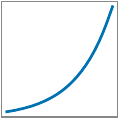}};
            \node[edgeplot] at (s2_L0x) {\includegraphics[width=0.88cm]{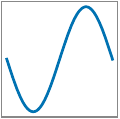}};
            \node[edgeplot] at (s2_L0y) {\includegraphics[width=0.88cm]{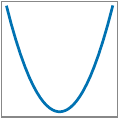}};

            \node[nd] at (s2_out) {};
            \node[nd] at (s2_ha) {};
            \node[nd, label={[font=\scriptsize]below:$x$}] at (s2_xin) {};
            \node[nd, label={[font=\scriptsize]below:$y$}] at (s2_yin) {};

            \node[smallnote, anchor=north] at (0,-0.24)
                {$[2,1,1]$, MSE $3.7{\times}10^{-6}$};
        \end{scope}

        \begin{scope}[shift={(5.65,0.72)}]
            \def\yout{3.80}\def\yLone{3.03}\def\yhid{2.22}\def\yLzero{1.08}\def\yin{0.16}
            \coordinate (s3_xin) at (-1.00,\yin);
            \coordinate (s3_yin) at ( 1.00,\yin);
            \coordinate (s3_out) at (0,\yout);
            \coordinate (s3_L1a) at (0,\yLone);
            \coordinate (s3_ha) at (0,\yhid);
            \coordinate (s3_L0x) at (-1.00,\yLzero);
            \coordinate (s3_L0y) at ( 1.00,\yLzero);

            \draw[symconn] (s3_L1a) -- (s3_out);
            \draw[symconn] (s3_ha) -- (s3_L1a);
            \draw[symconn] (s3_L0x) -- (s3_ha);
            \draw[symconn] (s3_L0y) -- (s3_ha);
            \draw[symconn] (s3_xin) -- (s3_L0x);
            \draw[symconn] (s3_yin) -- (s3_L0y);

            \node[edgeplot] at (s3_L1a) {\includegraphics[width=0.88cm]{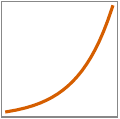}};
            \node[edgeplot] at (s3_L0x) {\includegraphics[width=0.88cm]{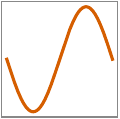}};
            \node[edgeplot] at (s3_L0y) {\includegraphics[width=0.88cm]{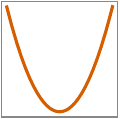}};

            \node[sym, anchor=west] at (0.58,\yLone) {$\exp(\cdot)$};
            \node[sym, anchor=east] at (-1.48,\yLzero) {$\sin$};
            \node[sym, anchor=west] at (1.48,\yLzero) {$x^2$};

            \node[nd] at (s3_out) {};
            \node[nd] at (s3_ha) {};
            \node[nd, label={[font=\scriptsize]below:$x$}] at (s3_xin) {};
            \node[nd, label={[font=\scriptsize]below:$y$}] at (s3_yin) {};
        \end{scope}

        \begin{scope}[shift={(5.65,-5.05)}]
            \def\yout{3.80}\def\yLone{3.03}\def\yhid{2.22}\def\yLzero{1.08}\def\yin{0.16}
            \coordinate (s4_xin) at (-1.00,\yin);
            \coordinate (s4_yin) at ( 1.00,\yin);
            \coordinate (s4_out) at (0,\yout);
            \coordinate (s4_L1a) at (0,\yLone);
            \coordinate (s4_ha) at (0,\yhid);
            \coordinate (s4_L0x) at (-1.00,\yLzero);
            \coordinate (s4_L0y) at ( 1.00,\yLzero);

            \draw[symconn] (s4_L1a) -- (s4_out);
            \draw[symconn] (s4_ha) -- (s4_L1a);
            \draw[symconn] (s4_L0x) -- (s4_ha);
            \draw[symconn] (s4_L0y) -- (s4_ha);
            \draw[symconn] (s4_xin) -- (s4_L0x);
            \draw[symconn] (s4_yin) -- (s4_L0y);

            \node[edgeplot] at (s4_L1a) {\includegraphics[width=0.88cm]{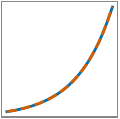}};
            \node[edgeplot] at (s4_L0x) {\includegraphics[width=0.88cm]{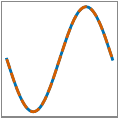}};
            \node[edgeplot] at (s4_L0y) {\includegraphics[width=0.88cm]{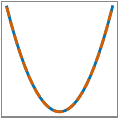}};

            \node[nd] at (s4_out) {};
            \node[nd] at (s4_ha) {};
            \node[nd, label={[font=\scriptsize]below:$x$}] at (s4_xin) {};
            \node[nd, label={[font=\scriptsize]below:$y$}] at (s4_yin) {};

            \node[smallnote, anchor=north] at (0,-0.24)
                {reach machine precision};
        \end{scope}

        \node[steplbl, anchor=south] at (-6.50,5.38)
            {Step 1: train\\with sparsification};
        \node[steplbl, anchor=south, text width=1.55cm] at (0.40,5.38)
            {Step 2:\\prune};
        \node[steplbl, anchor=south, text width=2.85cm] at (5.65,5.38)
            {Step 3: set\\symbolic funcs};

        \node[steplbl, anchor=south, text width=3.05cm] at (5.65,-0.88)
            {Step 4: train\\affine parameters};

        \draw[flow] (-2.80,5.11) -- (-1.35,5.11);
        \draw[flow] ( 2.15,5.11) -- ( 3.95,5.11);

        \draw[flow] (7.72,1.28)
            .. controls (8.80,0.60) and (8.80,-0.55) ..
            (7.05,-1.05);

        \draw[flow] (4.03,-2.55)
            .. controls (1.65,0.10) and (-2.55,-0.10) ..
            (-4.85,-1.30);

        \node[steplbl, anchor=south] at (-6.50,-1.80)
            {Step 5: output\\symbolic formula};
        \node[font=\Large, anchor=north] at (-6.50,-2.00)
            {$1.0\,e^{\,1.0y^2 + 1.0\sin(3.14x)}$};

        \draw[finalflow] (-6.50,-2.95) -- (-6.50,-4.44);
        \node[steplbl, anchor=east] at (-7.10,-4.20) {Step 6:\\number snap};
        \node[snapbox, anchor=north] at (-6.50,-4.64)
            {$\displaystyle e^{\,y^2+\sin(\pi x)}$};
    \end{tikzpicture}
    }
    \caption{
        \textbf{Full symbolic-regression workflow for QKAN on $f(x,y)=\exp(\sin(\pi x)+y^2)$.}
        Reproduction on QKAN of all six steps of Fig.~2.4 in~\citet{liu2024kan}. Small panels show the learned QKAN edge activation (blue) with the best-fit symbolic primitive (orange dashed) overlaid; Step~3 panels show only the symbolic curve (orange).
        \textbf{Step~1}: train QKAN$([2,5,1])$ with L1+entropy sparsification and incremental DARUAN layer extension $r{:}3{\to}24$ (test MSE $9.6\times 10^{-5}$).
        \textbf{Step~2}: the most-important hidden neuron is kept and a fresh QKAN$([2,1,1])$ is refit from scratch with layer extension $r{:}3{\to}48$; the resulting $3$-edge subnet reaches test MSE $3.7\times 10^{-6}$.
        \textbf{Step~3}: \texttt{fit\_params} identifies the three surviving edges as $\sin$, $x^2$, and $\exp$ respectively, all with $R^2=1.000$.
        \textbf{Step~4}: LBFGS refines the per-edge affine parameters $(a,b,c,d)$ to machine precision.
        \textbf{Step~5}: \texttt{sympy} composes the three edges into $0.999\cdot\exp\!\bigl(0.0613\,(4.04y)^2 + 1.001\,\sin(3.142 x)\bigr) + 5.9{\times}10^{-4}$.
        \textbf{Step~6}: after expanding inner products and rounding near-integer, near-$\pi$, and near-small-rational floats, the expression snaps \emph{exactly} to $\exp(y^2+\sin(\pi x))$ --- recovering the ground-truth target end-to-end. The per-edge symbolic table is \cref{tab:qkan_symbolic}.
    }
    \label{fig:qkan_symbolic}
\end{figure}

\cref{fig:qkan_symbolic} and \cref{tab:qkan_symbolic} summarise the per-edge fits after Steps~2--4. The three surviving edges of the pruned QKAN$([2,1,1])$ cleanly recover the three symbolic primitives that appear in $f(x,y)=\exp(\sin(\pi x)+y^2)$:
\begin{itemize}
    \item Layer-$0$ edge-$0$ on the $x$ channel: $\sin$ with snapped pre-activation frequency $a\approx\pi$ ($R^2=1.000$).
    \item Layer-$0$ edge-$1$ on the $y$ channel: $x^2$ ($R^2=1.000$).
    \item Layer-$1$ edge-$0$ from hidden to output: $\exp(\cdot)$ ($R^2=1.000$).
\end{itemize}
Composing the three refined edges yields $0.999\cdot\exp\!\bigl(0.0613\,(4.04y)^2 + 1.001\,\sin(3.142 x)\bigr) + 5.9{\times}10^{-4}$, and after number snap (expanding $0.0613\times 4.04^2\to 1$, $1.001\to 1$, $3.142\to \pi$, and dropping the $5.9{\times}10^{-4}$ bias) this collapses \emph{exactly} to $\exp(y^2+\sin(\pi x))$, i.e.\ the ground-truth target. The recovery requires two QKAN-specific ingredients: (i)~training the $[2,1,1]$ subnet from scratch with incremental DARUAN layer extension ($r{:}3{\to}48$) rather than simply pruning the $[2,5,1]$ model, and (ii)~initialising preact weights so that each edge starts in a different frequency band. Without these, na\"ive pruning of the sparsified $[2,5,1]$ leaves MSE $>10^{-1}$ on the $[2,1,1]$ subnet; with them, the single hidden neuron carries sin/$x^2$/exp at machine precision and the full symbolic workflow of KAN Fig.~2.4 applies end-to-end.

\begin{table}[t!]
\centering

\caption{Per-edge best-fit symbolic forms for the pruned QKAN$([2,1,1])$ trained on $f(x,y)=\exp(\sin(\pi x)+y^2)$. Each surviving edge is parameterised as $\phi(z) = c\,\mathrm{fun}(a z + b) + d$; after LBFGS refinement (Step~4) all three edges fit their matched primitive at $R^2 = 1.000$. Pre-symbolic test MSE: $3.66\times 10^{-6}$.}
\label{tab:qkan_symbolic}
\begin{tabular}{lllcrrrrc}
\toprule
Layer & Edge $j$ & Input & Symbolic key & $a$ & $b$ & $c$ & $d$ & $R^2$ \\
\midrule
$L_0$ & $0$ & $x$    & $\sin$     & $+3.142$ & $\approx 0$ & $+0.937$ & $+0.221$ & $1.000$ \\
$L_0$ & $1$ & $y$    & $x^2$      & $+4.039$ & $\approx 0$ & $+0.057$ & $-0.821$ & $1.000$ \\
$L_1$ & $0$ & $h_0$  & $\exp$     & $+1.068$ & $-0.101$    & $+2.097$ & $\approx 0$ & $1.000$ \\
\bottomrule
\end{tabular}

\end{table}

\section{Additional Baseline Comparisons and Ablation Studies}\label{sec:sup_baselines}

This section presents additional experiments on baseline comparisons, ablation studies, runtime analysis, and the nature of QKAN's advantage over classical alternatives.
The regression baseline and ablation experiments in \crefrange{sec:sup_fourierkan}{sec:sup_layerext} use 10\% label noise and report mean $\pm$ standard deviation over 5 random seeds. Runtime and precision analyses use the task-specific protocols stated in their respective subsections.

\subsection{Fourier-based KAN (FourierKAN) baseline comparison}\label{sec:sup_fourierkan}

To address whether QKAN is merely a Fourier series generator, we compare QKAN against a Fourier-based KAN (FourierKAN), a KAN variant that replaces B-spline activations with truncated Fourier series ($\sin$/$\cos$ basis).
FourierKAN is the most natural classical counterpart to DARUAN's trigonometric spectral structure.
QKAN uses reps$=3$; FourierKAN is tested at grid sizes (Fourier series size) $g \in \{1, 5, 15\}$.
Results are shown in \cref{tab:fourierkan}.

\begin{table}[t!]
\centering

\caption{QKAN vs FourierKAN: Test RMSE (mean $\pm$ std, 5 seeds). Best in bold. QKAN wins all 10 equations.}
\label{tab:fourierkan}
\small
\begin{tabular}{lcccc}
\toprule
Equation & QKAN ($r$=3) & FK ($g$=1) & FK ($g$=5) & FK ($g$=15) \\
\midrule
$1+x\sin y$                         & \textbf{0.122$\pm$0.001} & 0.128$\pm$0.003 & 0.190$\pm$0.052 & 0.280$\pm$0.078 \\
$\sqrt{1{+}x^2{+}2x\cos(y{-}z)}$   & \textbf{0.147$\pm$0.001} & 0.153$\pm$0.004 & 0.213$\pm$0.061 & 0.475$\pm$0.144 \\
$xe^{-y}$                           & \textbf{0.038$\pm$0.010} & 0.087$\pm$0.075 & 0.222$\pm$0.016 & 0.238$\pm$0.026 \\
$\cos x + y\cos^2 x$                & \textbf{0.119$\pm$0.001} & 0.124$\pm$0.004 & 0.309$\pm$0.067 & 0.410$\pm$0.023 \\
$(x{-}1)y$                          & \textbf{0.030$\pm$0.003} & 0.080$\pm$0.042 & 0.043$\pm$0.011 & 0.116$\pm$0.085 \\
$\tfrac{z\sqrt{x^2+y^2}}{4\pi}$     & \textbf{0.008$\pm$0.004} & 0.034$\pm$0.034 & 0.040$\pm$0.029 & 0.040$\pm$0.016 \\
$\tfrac{x}{2\cosh y}$               & \textbf{0.026$\pm$0.003} & 0.061$\pm$0.057 & 0.071$\pm$0.042 & 0.104$\pm$0.061 \\
$x+yz$                              & \textbf{0.089$\pm$0.016} & 0.106$\pm$0.024 & 0.096$\pm$0.015 & 0.591$\pm$0.364 \\
$\sqrt{1{+}x^2{+}y^2}$              & \textbf{0.134$\pm$0.016} & 0.157$\pm$0.033 & 0.169$\pm$0.037 & 0.695$\pm$1.108 \\
$x(1{+}y\cos z)$                    & \textbf{0.074$\pm$0.001} & 0.091$\pm$0.017 & 0.083$\pm$0.005 & 0.303$\pm$0.220 \\
\bottomrule
\end{tabular}
\end{table}

In this benchmark, increasing the FourierKAN grid size does not improve performance and often worsens generalization under noisy labels. This supports the interpretation that QKAN's advantage is not merely the presence of Fourier-type modes, but also the structured coupling of coefficients through shared rotation parameters, which can act as an implicit regularizer.

\subsection{GeometricFourierKAN comparison}\label{sec:sup_geofourier}

To test whether QKAN's advantage comes from geometric frequency allocation rather than the quantum parameterization itself, we implement a GeometricFourierKAN that uses the same geometric frequencies concept as DARUAN ($\{1, 2, 4, \ldots, 2^{K-1}\}$) instead of standard arithmetic frequencies ($\{1, 2, 3, \ldots, K\}$).
Results are shown in \cref{tab:geofourier}.

\begin{table}[t!]
\centering

\caption{QKAN vs GeometricFourierKAN vs FourierKAN: Test RMSE (mean $\pm$ std, 5 seeds). Lowest or tied-lowest values are in bold.}
\label{tab:geofourier}
\small
\begin{tabular}{lccc}
\toprule
Equation & QKAN ($r$=3) & GeoFK ($K$=3) & FK ($g$=5) \\
\midrule
$1+x\sin y$                         & 0.123$\pm$0.004 & 0.123$\pm$0.004 & 0.135$\pm$0.018 \\
$\sqrt{1{+}x^2{+}2x\cos(y{-}z)}$   & \textbf{0.148$\pm$0.005} & 0.155$\pm$0.007 & 0.220$\pm$0.059 \\
$xe^{-y}$                           & \textbf{0.032$\pm$0.001} & 0.246$\pm$0.043 & 0.343$\pm$0.132 \\
$\cos x + y\cos^2 x$                & \textbf{0.120$\pm$0.004} & 0.127$\pm$0.006 & 0.131$\pm$0.017 \\
$(x{-}1)y$                          & 0.025$\pm$0.001 & 0.025$\pm$0.001 & 0.025$\pm$0.001 \\
$\tfrac{z\sqrt{x^2+y^2}}{4\pi}$     & \textbf{0.003$\pm$0.000} & 0.010$\pm$0.007 & 0.007$\pm$0.005 \\
$\tfrac{x}{2\cosh y}$               & \textbf{0.022$\pm$0.001} & 0.023$\pm$0.001 & 0.051$\pm$0.026 \\
$x+yz$                              & 0.075$\pm$0.002 & 0.075$\pm$0.002 & 0.111$\pm$0.052 \\
$\sqrt{1{+}x^2{+}y^2}$              & \textbf{0.127$\pm$0.004} & 0.138$\pm$0.018 & 0.137$\pm$0.021 \\
$x(1{+}y\cos z)$                    & \textbf{0.071$\pm$0.002} & 0.074$\pm$0.002 & 0.073$\pm$0.003 \\
\bottomrule
\end{tabular}
\end{table}

QKAN is lowest or tied-lowest in mean RMSE across the reported equations, and neither GeometricFourierKAN nor FourierKAN obtains a lower mean RMSE in this comparison. Thus the observed advantage is not explained by geometric frequency allocation alone. The remaining distinction is the DARUAN parameterization, in which the Fourier amplitudes are jointly determined by shared single-qubit rotation parameters rather than independently assigned coefficients.

\subsection{DARUAN contribution ablation (\texorpdfstring{$w_b$ vs $w_d$}{wb vs wd})}\label{sec:sup_ablation}

The QKAN activation on each edge combines a classical base activation and the DARUAN output:
$\phi(x) = w_b \cdot \sigma_{\text{base}}(x) + w_d \cdot \text{DARUAN}(x)$.
To verify that DARUAN is the primary contributor, we compare four variants across 5 benchmark equations shown in \cref{tab:ablation}.

\begin{table}[t!]
\centering

\caption{Ablation: RMSE by variant (mean $\pm$ std, 5 seeds). Removing DARUAN ($w_d=0$) causes 2--10$\times$ degradation.}
\label{tab:ablation}
\small
\begin{tabular}{lcccc}
\toprule
Equation & Full QKAN & No Base ($w_b{=}0$) & No DARUAN ($w_d{=}0$) & No Preact \\
\midrule
$1{+}x\sin y$          & 0.122$\pm$0.001 & 0.127$\pm$0.005 & 0.448 & 0.123$\pm$0.002 \\
$xe^{-y}$              & 0.038$\pm$0.010 & 0.037$\pm$0.004 & 0.152 & 0.038$\pm$0.009 \\
$(x{-}1)y$             & 0.030$\pm$0.003 & 0.029$\pm$0.002 & 0.320 & 0.035$\pm$0.005 \\
$\cos x{+}y\cos^2 x$   & 0.119$\pm$0.001 & 0.119$\pm$0.002 & 0.640 & 0.122$\pm$0.001 \\
$x(1{+}y\cos z)$       & 0.074$\pm$0.001 & 0.076$\pm$0.001 & 0.111 & 0.076$\pm$0.002 \\
\bottomrule
\end{tabular}
\end{table}

Across all equations, the learned DARUAN contribution fraction $|w_d|/(|w_b|+|w_d|)$ is consistently 63--69\%.
Removing the base activation ($w_b=0$) has only a small effect in this benchmark, whereas removing DARUAN ($w_d=0$) substantially worsens performance, giving a 2--10$\times$ RMSE increase.
The base activation serves primarily as a residual connection that stabilizes training.

\subsection{Layer extension ablation}\label{sec:sup_layerext}

We compare progressive layer extension (reps schedule $\{1, 5, 10, 15, 20\}$, warm-starting from previous model, 100 steps per stage) against training the full model from scratch (reps$=20$, 500 total steps).
Results over 5 seeds are shown in \cref{tab:layerext}.

\begin{table}[t!]
\centering

\caption{Layer extension ablation: test MSE (mean $\pm$ std, 5 seeds). Extension wins all reported equations.}
\label{tab:layerext}
\small
\begin{tabular}{lccc}
\toprule
Equation & With Extension & Without (scratch) & Improvement \\
\midrule
$1+x\sin y$                         & $\bm{1.50\times10^{-2}\pm9\times10^{-4}}$ & $1.79\times10^{-2}\pm3\times10^{-3}$ & $1.2\times$ \\
$\sqrt{1{+}x^2{+}2x\cos(y{-}z)}$    & $\bm{2.21\times10^{-2}\pm1\times10^{-3}}$ & $1.11\times10^{-1}\pm2\times10^{-2}$ & $5.0\times$ \\
$xe^{-y}$                           & $\bm{1.04\times10^{-3}\pm5\times10^{-5}}$ & $8.60\times10^{-2}\pm6\times10^{-2}$ & $83\times$ \\
$(x{-}1)y$                          & $\bm{6.07\times10^{-4}\pm6\times10^{-5}}$ & $1.34\times10^{-3}\pm1\times10^{-3}$ & $2.2\times$ \\
$\tfrac{z\sqrt{x^2+y^2}}{4\pi}$     & $\bm{9.71\times10^{-6}\pm7\times10^{-7}}$ & $4.45\times10^{-4}\pm2\times10^{-4}$ & $46\times$ \\
$\tfrac{x}{2\cosh y}$               & $\bm{4.78\times10^{-4}\pm3\times10^{-5}}$ & $1.34\times10^{-3}\pm8\times10^{-4}$ & $2.8\times$ \\
$x+yz$                              & $\bm{5.60\times10^{-3}\pm3\times10^{-4}}$ & $1.00\times10^{-2}\pm1\times10^{-3}$ & $1.8\times$ \\
$x(1{+}y\cos z)$                    & $\bm{5.18\times10^{-3}\pm3\times10^{-4}}$ & $6.28\times10^{-3}\pm7\times10^{-4}$ & $1.2\times$ \\
\bottomrule
\end{tabular}
\end{table}

Layer extension achieves up to 83$\times$ improvement on $xe^{-y}$ and 46$\times$ on $\tfrac{z\sqrt{x^2+y^2}}{4\pi}$.
Training the full large-$r$ model from scratch often converges to poorer solutions with higher variance, confirming that layer extension is an important optimization strategy for large-repetition QKANs.

\subsection{Runtime analysis}\label{sec:sup_runtime}

We use this section to isolate the cost of FlashQKAN, our optimized single-qubit QKAN solver family, because the main-text scalability claim depends on whether single-qubit DARUAN evaluations can be executed as ordinary dense GPU kernels rather than as expensive generic quantum simulation calls. FlashQKAN contains both Triton-based and CUTLASS CuTe-based backends. In the benchmark below, the CuTe backend is used for forward and training-step timing because it is optimized for this NVIDIA GPU path, while the Triton backend is used in \cref{sec:sup_precision} to test reduced-precision state-vector evaluation. We compare the QKAN bottleneck block against corresponding FourierKAN and B-spline-based KAN (SplineKAN) bottleneck blocks under the same architecture.

We benchmark QKAN's computational cost using the CuTe backend of FlashQKAN (\texttt{solver="cute"}), which implements fused kernels with \texttt{\_\_sincosf} intrinsics and warp-shuffle reductions.
The benchmark uses batch size 1000 and the bottleneck replacement block Linear$(100,10)\rightarrow$KAN-family$([10,10])\rightarrow$Linear$(10,100)$, matching the HQKAN design used for scalable model replacements.
All rows use \texttt{bfloat16} tensors where supported. QKAN uses the CuTe backend with \texttt{bfloat16} datatype; FourierKAN and SplineKAN are evaluated in the corresponding PyTorch \texttt{bfloat16} mode.
We report mean $\pm$ standard deviation over 100 CUDA-event timed runs after 10 warmup runs on an NVIDIA RTX~5090. The training-step timing includes zeroing gradients, the forward pass, mean-squared-error backpropagation, and one Adam optimizer update.
Results are shown in \cref{tab:runtime}.

\begin{table}[ht!]
\centering

\caption{Forward and training-step time (ms per batch) for the architecture-matched HQKAN-style Linear$(100,10)\rightarrow$KAN-family$([10,10])\rightarrow$Linear$(10,100)$ bottleneck block at batch size 1000 on an NVIDIA RTX~5090. Training step includes zeroing gradients, forward pass, MSE backward pass, and Adam update; values are mean $\pm$ standard deviation over 100 timed runs after 10 warmups.}
\label{tab:runtime}

\begin{tabular}{lccc}
\toprule
Model & \# Params & Forward (ms) & Training step (ms) \\
\midrule
QKAN & 3{,}110 & $0.072\,\pm\,0.005$ & $0.507\,\pm\,0.042$ \\
FourierKAN & 3{,}310 & $0.079\,\pm\,0.011$ & $0.519\,\pm\,0.014$ \\
SplineKAN & 3{,}110 & $0.200\,\pm\,0.011$ & $0.808\,\pm\,0.013$ \\
\bottomrule
\end{tabular}
\end{table}

In this architecture-matched replacement benchmark, the FlashQKAN bottleneck block gives the lowest mean latency among the compared KAN-family blocks.
The QKAN row is substantially faster than SplineKAN ($0.072\pm0.005$ vs. $0.200\pm0.011$\,ms forward, and $0.507\pm0.042$ vs. $0.808\pm0.013$\,ms per training step) and has lower mean latency than FourierKAN.
Together with the accuracy and parameter-efficiency comparisons in \crefrange{sec:sup_fourierkan}{sec:sup_layerext}, this supports the main advantage claimed for QKAN: a compact DARUAN-based replacement module whose optimized single-qubit circuit evaluation remains practical on standard GPU hardware.

\subsection{Parameter and simulation precision sensitivity}\label{sec:sup_precision}

We evaluate two distinct precision issues: (i) classical parameter precision (quantization) at inference time, and (ii) the simulation precision used to evaluate the quantum circuit.

\paragraph{(i) Parameter quantization.}
We train each model at full precision and then quantize its parameters to 32- and 4-bit representations.
\Cref{tab:precision} reports the degradation ratio, defined as the 4-bit test RMSE divided by the 32-bit test RMSE.

\begin{table}[ht!]
\centering

\caption{4-bit / 32-bit RMSE degradation ratio. Lower values indicate greater robustness to parameter quantization.}
\label{tab:precision}
\small
\begin{tabular}{lccc}
\toprule
Equation & QKAN & FourierKAN & MLP \\
\midrule
$1+x\sin y$     & 2.2$\times$ & 2.4$\times$ & 5.9$\times$ \\
$xe^{-y}$        & 1.3$\times$ & 1.0$\times$ & 2.0$\times$ \\
$x(1+y\cos z)$  & 2.9$\times$ & 2.2$\times$ & 3.5$\times$ \\
\midrule
Average          & 2.2$\times$ & 1.9$\times$ & 3.8$\times$ \\
\bottomrule
\end{tabular}
\end{table}

MLP degrades most severely at 4-bit (avg 3.8$\times$), while QKAN (2.2$\times$) and FourierKAN (1.9$\times$) are more robust.
QKAN's rotation angles are bounded in $[0, 2\pi)$, limiting quantization perturbations.

\paragraph{(ii) Simulation precision.}
We additionally sweep the compute dtype used for state-vector simulation, evaluating the \emph{same} trained QKAN weights through the Triton~\citep{tillet2019triton} backend of FlashQKAN (fused kernels) at each of \{\texttt{float64}, \texttt{float32}, \texttt{bfloat16}, \texttt{float8\_e4m3fn}\}.
Especially, the \texttt{float8\_e4m3fn} is done with pre-scaled reduced-precision state checkpoints.
Results over 5 seeds on the same three representative equations are in \cref{tab:sim_precision}.

\begin{table}[ht!]
\centering
\caption{QKAN simulation-precision sweep under the Triton backend of FlashQKAN: test RMSE (5-seed mean) at each reduced-precision $c\_dtype$ for the same trained weights. The max-degradation column reports the worst-case ratio to the \texttt{float32} reference.}
\label{tab:sim_precision}
\small
\begin{tabular}{l c c c c c}
\toprule
Equation & \texttt{float64} & \texttt{float32} & \texttt{bfloat16} & \texttt{float8\_e4m3fn} & max deg. \\
\midrule
$1+x\sin y$           & 0.124 & 0.124 & 0.124 & 0.124 & 1.00$\times$ \\
$xe^{-y}$             & 0.199 & 0.199 & 0.201 & 0.201 & 1.01$\times$ \\
$x(1{+}y\cos z)$      & 0.078 & 0.078 & 0.078 & 0.078 & 1.00$\times$ \\
\bottomrule
\end{tabular}
\end{table}
Two findings:
(a) \textbf{fp64 is not needed.} \texttt{float64} and \texttt{float32} agree to six decimals on all three equations, so the single-qubit state vector does not require fp64 intermediate precision.
(b) \textbf{FlashQKAN makes QKAN essentially invariant to reduced precision.} Under the Triton backend with pre-scaled state checkpoints, QKAN at \texttt{bfloat16} and \texttt{float8\_e4m3fn} both match the \texttt{float32} reference to within 1.01$\times$ on every equation tested, so quantum-state simulation for single-qubit DARUAN fits cleanly into standard mixed-precision deployment pipelines and ``high-precision simulation'' is not a prerequisite for QKAN inference.

\section{Experimental Details}\label{sec:sup_experimental_details}

\subsection{Software implementation.}\label{sec:sup_software_impl}

To enable efficient numerical simulation of QKANs on CUDA-enabled devices \citep{cuda}, we build our implementation using PyTorch \citep{Ansel_PyTorch_2_Faster_2024}, extending its tensor operations with custom routines for quantum state evolution.

The quantum state is represented as a complex-valued tensor with shape $(B, N, M, 2)$, where $B$ denotes the batch size, $N$ is the number of post-nodes, $M$ is the number of pre-nodes, and the final dimension encodes the amplitudes of the 2-level quantum system (i.e., a single qubit).

Quantum gates are encoded as complex tensors with shape $(N, M, 2, 2)$, where the last two dimensions represent the $2 \times 2$ matrix structure of a single-qubit unitary gate, and the first two dimensions index the interaction between input and output nodes.

We adopt the Pauli-$Z$ operator as the observable for measurement. The data encoding blocks and trainable unitaries in the data re-uploading ansatz are implemented as parameterized single-qubit rotation gates. 

To initialize the quantum state, a Hadamard gate is applied to place the qubit into an equal superposition of basis states, which empirically improves the stability and convergence of training.

Building on this PyTorch representation, the simulator exposes FlashQKAN, an optimized GPU solver family for the DARUAN layer. FlashQKAN includes two backends: a Triton-based backend and a CUTLASS CuTe-based backend. The Triton backend fuses the data-encoding rotations, trainable rotations, and Pauli-$Z$ readout into real-valued kernels and supports mixed-precision evaluation.
The CuTe backend implements the same single-qubit recurrence in custom CUDA/CuTe kernels with \texttt{\_\_sincosf} intrinsics and warp-level reductions; this backend is used for the runtime benchmark in \cref{sec:sup_runtime}. Both FlashQKAN backends exploit the fact that each DARUAN activation evolves only a two-amplitude state, so the computation is mapped to batched GPU tensor operations rather than a generic multi-qubit state-vector simulator.

\subsection{Software and system information.}\label{sec:sup_system_info}

For full reproducibility, \cref{tab:version} summarizes the software versions, dependencies, and hardware system specifications used in the main-text experiments.
For the additional FlashQKAN related experiments in supplementary materials, the software stack used the QKAN 0.2.2 codebase with Python~3.13.2, PyTorch~2.11.0+cu130, Triton~3.6.0~\citep{tillet2019triton}, the \texttt{cuda-toolkit} package~13.0.2, and an NVIDIA RTX~5090. The CuTe JIT build used Ninja~1.13.0 and CUDA compilation tools~12.8.93 with CUTLASS/CuTe headers from the CUTLASS checkout identified as v4.4.0-22-g4ca61d06. Here, \texttt{solver="cute"} and the optional \texttt{qkan[cute]} install extra refer to QKAN's CUTLASS/CuTe-header CUDA backend; the extra provides build support rather than an independent \texttt{cute} or CuTe DSL Python package. The backend is compiled from CUTLASS/CuTe headers during extension build, or through the JIT fallback when those headers are available.

\begin{table}[pthb]
\scriptsize
\centering
\caption{\textbf{Software and System Information for the main-text experiments.}
}
\label{tab:version}
\begin{tabular}{ll}
\toprule
Software & Version \\
\midrule
matplotlib & 3.6.2\\
wandb & 0.16.6\\
tqdm & 4.66.2\\
h5py & 3.11.0\\
numpy & 1.24.4\\
scikit\_learn & 1.1.3\\
setuptools & 65.5.0\\
sympy & 1.11.1\\
pandas & 2.0.3\\
requests & 2.31.0\\
transformers & 4.40.1\\
PennyLane & 0.37.0\\
PennyLane\_Lightning & 0.37.0\\
torch & 2.4.0\\
torchaudio & 2.4.0\\
torchvision & 2.4.0\\
pykan & 0.0.5\\
\midrule
\multicolumn{2}{l}{RTX 4090 Personal Computer}\\
\midrule
Parameter & Value \\
\midrule
Python version & 3.11.5 \\
Python compiler & GCC 11.2.0 \\
Python build & main, Sep 11 2023 13:54:46 \\
OS & Linux \\
CPUs & 1 \\
CPUs Memory (GB) & 64 \\
GPUs & 1 (NVIDIA GeForce RTX 4090) \\
GPUs Memory (GB) & 24 \\
\midrule
\multicolumn{2}{c}{Mon Sep 16 08:47:24 2024 UTC} \\
\midrule
\multicolumn{2}{l}{Tesla V100S Single Node Cluster}\\
\midrule
Parameter & Value \\
\midrule
Python version & 3.11.11 \\
Python compiler & GCC 11.2.0 \\
Python build & main, Dec 11 2024, 16:28:39 \\
OS & Linux \\
GPUs & 4 (Tesla V100S-PCIE-32GB) \\
GPUs Memory (GB) & 128 \\
\midrule
\multicolumn{2}{c}{Fri Mar 29 18:32:15 2025 UTC} \\
\midrule
\multicolumn{2}{l}{NVIDIA H100 GPUs \& NVIDIA H200 GPUs Cluster}\\
\midrule
Parameter & Value \\
\midrule
Python version & 3.11.13 \\
Python compiler & GCC 11.2.0 \\
Python build & main, Jun  5 2025, 13:12:00 \\
OS & Linux \\
Scheduling System & Slurm Workload Manager\\
GPUs on a H100 node & 8 (NVIDIA H100 PCIe) \\
GPUs Memory (GB)  on a H100 node & 640\\
GPUs on a H200 node  & 8 (NVIDIA H200 PCIe)\\
GPUs Memory (GB) on a H200 node  & 1128\\
Cross-node Interconnector & 8 InfiniBand NDR 400G network ports\\
\midrule
\multicolumn{2}{c}{Wed Jul 2 23:10:24 2025 UTC} \\
\bottomrule
\end{tabular}
\end{table}

\end{document}